\documentclass[12pt,a4paper]{article}
\usepackage[dvips]{epsfig}

\begin{document}
\author{\bf Yu.A. Markov$\!\,$\thanks{e-mail:markov@icc.ru}
$\,$, M.A. Markova$^*$,\\
\bf and A.N. Vall$\!\,$\thanks{e-mail:vall@irk.ru}}
\title{Nonlinear dynamics of soft boson\\
collective excitations in hot QCD plasma II:\\
plasmon -- hard-particle scattering and
energy losses}
\date{\it $^{\ast}\!$Institute of System Dynamics and Control Theory\\
Siberian Branch of Academy of Sciences of Russia,\\
P.O. Box 1233, 664033 Irkutsk, Russia\\
\vspace{0.4cm}
$^{\dagger}\!$Irkutsk State
University, Department of Theoretical Physics,\\ 664003, Gagarin blrd,
20, Irkutsk, Russia}
\thispagestyle{empty}
\maketitle{}


\def\theequation{\arabic{section}.\arabic{equation}}

\[
{\bf Abstract}
\]
In a general line with our first work [1], within hard thermal
loop (HTL) approximation a general theory of the scattering for an arbitrary
number of colorless plasmons off hard thermal particles of hot
QCD-medium is considered. Using generalized Tsytovich correspondence
principle, a connection between matrix elements of the scattering
processes and a certain effective current, generating these processes is
established.  The iterative procedure of calculation of these matrix
elements is defined, and a problem of their gauge-invariance is
discussed. An application of developed theory to a problem of calculating
energy losses of energetic color particle propagating through QCD-medium
is considered. It is shown that for limiting value of the plasmon
occupation number ($\sim 1/g^2$, where $g$ is a strong coupling) energy
losses caused by spontaneous scattering process of energetic particle off
soft-gluon waves is of the same order in the coupling as other known losses
type: collision and radiation ones. The Fokker-Planck equation, describing
decceleration (acceleration) and diffusion in momentum space of beam of
energetic color particles scattering off soft excitations of quark-gluon
plasma (QGP), is derived.

\section{Introduction}
\setcounter{equation}{0}

In the second part of our work we carry on with analysis of
dynamics of boson excitations in hot QCD-medium at the soft momentum scale,
started in \cite{mark1} (to be reffered to as ``Paper I'' throughout this
text) in the framework of HTL effective theory \cite{bra1}.
Here, we focuss our research on the studies of the scattering processes of
soft-gluon plasma waves off hard particles of higher order within a real time
formalism based on a Boltzmann type kinetic equation for soft modes. The
nonlinear Landau damping process studied before \cite{mark2} is simple example
of this type of scattering processes. In a similar case of plasmon-plasmon
scattering \cite{mark1}, for sufficiently high energy level of the soft plasma
excitations (exact estimation will be given in Section 5 below) all higher
processes of plasmon -- hard-partical scattering will give a contribution of
the same order to the right-hand side of the Boltzmann equation.

Our approach is based on the system of dynamical equations derived by
Blaizot and Iancu \cite{bla1} complemented by famous Wong equation
\cite{wong} that describes the precession of the classical color charge
$Q = (Q^a),\,a=1,\ldots,N_c^2-1$ for hard particle in a field of incident
soft-gluon plasma wave. Using the so-called
Tsytovich correspondence principle introduced in Paper I with necessary
minimal generalization  for relevant problem, we establish a link between
matrix elements of studied scattering processes and a certain effective
currents, generating these processes. These effective currents appear in solving
combined equation system of Blaizot-Iancu and Wong's equations in the form
of expansion in powers of free gauge field $A_{\mu}^{(0) a}$ and initial value
of a color charge $Q_0^a$. The coefficient functions in this expansion
determines matrix elements of investigated scattering processes. The appearance
of classical soft-gluon loop corrections to the scattering processes,
that in leading order in coupling can be formally presented in the form of
tree diagramm with vertices and propagators in HTL or eikonal
approximations, here, is a new interest ingredient.

We apply the current approach to study of the propagation of high energy color
parton (gluon or quark) through a hot QCD-medium and energy losses
associated with this motion. Research of the energy losses of energetic partons in
QGP at present is of a great interest with respect to {\it jet quenching}
phenomenon \cite{gyul, wang1, lok1, bai1}
and its related high-$p_t$ azimuthal asymmetry \cite{wang2, lok2},
large-$p_t$ $\pi^0$ suppression \cite{levai, hira} etc., that have observed
in ultrarelativistic heavy ion collisions at RHIC (see, review \cite{gyul1}).
Throughout the past twenty years by the efforts of many authors
several possible mechanisms of energy losses have been analyzed:
(1)elastic small distance collisional losses due to the final state
interaction of high energy parton with medium constituent (Bjorken \cite{bjor},
Braaten and Thoma \cite{bra2});
(2)polarization losses or losses caused by large distance
collision\footnote{Here, energy loss can be considered as a work
performed by color charge polarizing of QCD-medium by own field.}
(Thoma and Gyulassy, Mr\'owczy\'nski, Koike and Matsui \cite{thom},
Braaten and Thoma \cite{bra2});
(3)losses due to gluon bremsstrahlung induced by multiple scattering
(Ryskin \cite{rysk}; Gyulassy, Wang and Pl\"ummer \cite{gyul2}, Baier
at al. \cite{bai2, bai3, bai4}, Zakharov \cite{zakh}, Levin \cite{lev}, 
Wiedemann \cite{wied}, Kovner and Wiedemann \cite{kov} etc.). The first two 
mechanisms often combined into one collision mechanism of energy loss, but it is
convenient for our purpose to separated them.

It was shown \cite{gyul} that the elastic and polarization energy losses of
partons in QCD plasma turned out to be too small for jet extinction, while
induced radiative energy losses prove to be sufficient large to make itself
evident in jet quenching in collisions of heavy nuclei\footnote{As was mentioned
in \cite{lok3} that although on theoretical estimations of the radiative energy
loss of a hard parton much large than the elastic energy loss, a direct
experimental verification of this phenomenon remains an open problem.}. For
this reason recent theoretical stydies of parton energy loss have concentrated
on gluon radiation \cite{gyul2}\,--\,\cite{kov}.

However these `traditional' approaches have difficulties
accounting for the large jet losses reported at RHIC \cite{gyul1}. This is the
motivation for more detail analysis of mechanisms of energy losses already
studied or their alternative formulations (in particulary, radiation losses)
as well as the considered novel mechanisms. To the first case it can be
assigned the works of Gyulassy, Levai and Vitev on construction an algebraic
reaction operator formalism \cite{gyul3}, twist expansion approach developed
by Wang et al. \cite{guo}. The work of Shuryak and Zahed \cite{shur1}
considered synchrotron-like radiation in QCD by generalized Schwinger's
treatment of quantum synchrotron radiation in QED, a work of E. Wang
and X.-N. Wang \cite{wang3} connected with allowing for stimulated gluon
emission and thermal absorption by the propagating parton in dense QGP,
and also mechanism of coherent final state interaction proposed by Zakharov
\cite{zakh1}, can be assigned to the second case.

In this work we would like to turn to study of just one more mechanism of loss
(or gain) energy, the special and simplest case of which is polarisation loss.
The approximation within of which polarization loss is calculated, valids
only for extremely low level of excitations for soft fields of medium. This is
circumstance on which it was indicated by Mr\'owczy\'nski in Ref.\,\cite{thom}. 
The expression for polarization part of energy loss derived in \cite{thom}
up to color factors is exactly coincident with expression obtained early in
the theory of usual plasma \cite{akh} within standard linear response theory.
However, we can expect that for ultrarelativistic heavy ion collisions
generated QGP will be far from equilibrium in highly excited state. It can
be valid at least for the subsystem of soft plasmons (Section 2, Paper I),
when interaction by soft waves preponderate over effects of particle collisions.
One could study the influence of such off-equilibrium effects on parton
propagation and radiation for unbounded QGP. For existence of intensity
of soft radiation in medium, additional mechanism of deceleration
(or acceleration) of energetic color particle connected with spontaneous and
stimulated scattering processes of this particle off soft gluon excitations
arises. In a limiting case of a strong gauge field
$\vert A_{\mu}^a(X)\vert\sim T$, where $T$ is a temperature of a system,
this type
of energy losses becomes of the same order in $g$ as above mechanisms of energy
losses and therefore can give quite appreciable contribution to total losses
balance (this problem is discussed in more detail in Section 7 and 8 of
present work). It is precisely here non-Abelian character of interaction of
hard color particle with soft-gluon field in QGP is to be completely
manifested. The first work, where an attempt
was made to accounting for interaction of energetic (massive) colored particle
with stochastic background chromoelectric field, using the semiclassical
equations of motion, is the work of Leonidov \cite{leon}. In this work
an approach developed to the problem of stochastic deceleration and acceleration
of cosmic rays in usual plasma \cite{stur} has been used.
However here, no account has been taken the fact that in the case of dense
medium, what is QGP, in the scattering process, plasma surrounding traveling
color charge not remains indifferent towards. In QGP the nonlinear polarization
currents arise, essentially varying a physical picture. In the case of Abelian
plasma this was shown by Gailitis and Tsytovich in \cite{gai}. To account for
polarization currents it is necessary to use a kinetic equation, that were
not made in above-mentioned work by Leonidov.

A more close to the subject of our research on a conceptional plane
is the work of E. Wang and X.-N. Wang \cite{wang3} have been already
cited above. Here, at the first a problem of influence on energy losses
of stimulated gluon emission and thermal absorption by the propagating
parton because of the presence of hard thermal gluons in the hot QCD-medium,
was posed. This mechanism of energy losses is important in
a medium with large initial gluon density (such it is proportional to gluon
density) that it is really possible takes place by virtue of 
a strong suppression of high transverse momentum hadron spectra observed by 
experiments at RHIC \cite{gyul1}.  However, in Ref \cite{wang3} only gluon 
emission and absorption of hard thermal gluons (whose energy is of order $T$) with 
the use of the methods of perturbative QCD, is considered. For large density of 
soft gluon radiation consideration of contribution to energy loss of soft emission
and absorption by the propagating parton, also becomes important. These processes
(proportional to soft-gluon number density) by virtue of large occupation number
of soft excitations adequately describe by using quasiclassical methods
based on HTL approaches.

Returning to the problem stated in this work, it is necessary run ahead note the
following important circumstance: effective currents, determining the processes
of spontaneous and stimulated scattering posses remarkable pecularity that
they not explicitly dependent  on mass of hard particle (in HTL-approximation),
and therefore developed theory is suitable equally for light color particle as
well as massive ones. This in particular is reflected in that losses caused for
instance by spontaneous scattering of energetic particle off soft excitations
will be connected  with not varying momentum of particle, but with rotation of
its (classical) color charge in effective color space, governing by Wong equation.
Here, we have a principal distinction from energy loss of charge particles in
usual plasma. In Abelian case as known \cite{akh}, a rate of energy loss
(or gain) for interaction with stochastic plasma field is inverse
propotional to mass of hard particle, and therefore it is essential for light
particles (electron, positron) and suppressed for heavy ones (proton, ion),
although their polarization losses are practically identical.
In non-Abelian plasma, at least within HTL-approximation the mass dependence
can enter only through integration limits and a value of velocity of energetic
parton, and therefore explicit suppression by mass of this energy loss type,
is not arisen.
In this connection note that study of propagation of heavy partons
($c$ or $b$ quarks) through QGP is of great independent interest. The first
estimation of energy loss of heavy quark will be made by Svetitsky \cite{svet}
for study of the diffusion process of charmed quark in QGP, and then the
estimation will be given also in the works by Braaten and Thoma \cite{bra2}
Mustafa et al., Walton and Rafelski \cite{must1}(collision losses),
and Mustafa et al. \cite{must2},
Dokshitzer and Kharzeev \cite{dok}(radiation losses). The energy losses of
heavy quarks are important not only with jet quenching phenomenon, but also
for other important one for diagnostic of QGP: the modification of the
high mass dimuon spectra from semileptonic $B$ and $D$ meson decays
(Shuryak, Lin, Vogt and Wang \cite{shur2}, K\"ampfer et al.,
Lokhtin and Snigirev \cite{kamp}).

Paper II is organized as follows. In Section 2 preliminary comments with regard
to derivation of the Boltzmann equation taking into account the scattering
processes of an arbitrary number of colorless plasmons off hard thermal
particles, are given. Section 3 presents a detailed consideration of the
kinetic equation for the nonlinear Landau damping process, and here, also
appropriate extension of the Tsytovich correspondence principle is presented.
In Section 4 a complete algorithm of the succesive calculation of the certain
effective currents, generating considered scattering processes, is given.
In Section 5 on the based of the correspondence principle these effective
currents are used for calculation of the proper matrix elements. Here,
we estimate the typical value of plasmon occupation numbers, wherein one can
restricted the consideration to acounting for only the contribution from the
nonlinear Landau damping process or all higher scattering processes should be
considered. In Section 6 we discuss a problem of a gauge independence of matrix
elements defined in previous Section. Section 7 is concerned with definition of
general quasiclassical expressions for energy loss of energetic color particle
for its scattering off soft plasma excitations. In Section 8 the energy loss
caused by spontaneous scattering off colorless plasmons in lower order in powers
of plasmon number density is analyzed in detail, and in Section 9 some problems
connected with scattering of energetic particle by soft-gluon excitations lying off
mass-shell, is discussed. Finally in Section 10 the Fokker-Planck equation,
describing an evolution of a distribution function for a beam of energetic
color partons scattering off plasmons, is derived. In the closing section we
briefly discuss some interest and important moments concerned with mechanism
of energy loss studied in this work, and remaining to be considered.

\section{\bf Preliminares}
\setcounter{equation}{0}

As in Paper I we consider a pure gluon plasma with no quarks,
where soft longitudinal excitations is propagated. Here, we restrict
our cosideration to soft colorless exsitations, i.e. we assume that a
localized number density of plasmons
$N^l ({\bf p}, x)\equiv(N_{\bf p}^{l\,ab})$ is diagonal in color space
\[
N_{\bf p}^{l \, ab} = \delta^{ab} N_{\bf p}^l ,
\]
where $a,b = 1, \ldots , N_c^2 - 1$ for $SU(N_c)$ gauge group. We
consider in Paper II the change of the number density of the colorless plasmons
$N_{\bf p}^l$ as result of their scattering off hard thermal gluons.
We expect in this case that the time-space evolution of scalar function
$N_{\bf p}^l$ will be described by
\begin{equation}
\frac{\partial N_{\bf p}^l}{\partial t} +
{\bf v}_{\bf p}^l\cdot\frac{\partial N_{\bf p}^l}{\partial {\bf x}} =
- N_{\bf p}^l \Gamma_{\rm d} [ N_{\bf p}^l ] + ( 1 + N_{\bf p}^l )
\Gamma_{\rm i}[ N_{\bf p}^l]\equiv -C[N_{\bf p}^l],
\label{eq:2q}
\end{equation}
where ${\bf v}_{\bf p}^l = \partial \omega_{\bf p}^l / \partial {\bf p}$
is a group velocity of the longitudinal oscillations and $\omega_{\bf
p}^l \equiv \omega^l({\bf p})$ is the dispersion relation for plasmons.
As it usually is, a functional dependence is denoted by argument of a
function in square brackets.

The more general expressions for generalized decay rate $\Gamma_{\rm d}$
and regenerating rate $\Gamma_{\rm i}$ can be written in the form of
functional expansion in powers of the plasmon number density
\begin{equation}
\Gamma_{\rm d} [N_{\bf p}^l] = \sum_{n = 0}^{\infty}
\Gamma_{\rm d}^{(2n + 1)} [N_{\bf p}^l] , \; \;
\Gamma_{\rm i} [N_{\bf p}^l] = \sum_{n = 0}^{\infty}
\Gamma_{\rm i}^{(2n + 1)} [N_{\bf p}^l],
\label{eq:2w}
\end{equation}
where
\begin{equation}
\Gamma_{\rm d}^{(2n + 1)} [N_{\bf p}^l] =
\int\!\frac{d {\bf k}}{(2\pi)^3}\!\int\!{\rm d}{\cal T}^{(2n + 1)}\,
{\it w}_{2n + 2}({\bf k}\vert\,{\bf p}, {\bf p}_1, \ldots ,{\bf p}_n;
{\bf p}_{n + 1}, \ldots , {\bf p}_{2n + 1})
\label{eq:2e}
\end{equation}
\[
\times N_{{\bf p}_1}^l \ldots N_{{\bf p}_n}^l(1 + N_{{\bf p}_{n + 1}}^l) \ldots
(1 + N_{{\bf p}_{2n + 1}}^l)f_{\bf k}[1+f_{{\bf k}^\prime}],
\]
\begin{equation}
\Gamma_{\rm i}^{(2n + 1)} [N_{\bf p}^l] =
\int\!\frac{d {\bf k}}{(2\pi)^3}\!\int\!{\rm d}{\cal T}^{(2n + 1)} \,
{\it w}_{2n + 2}({\bf k}\vert\,{\bf p}, {\bf p}_1, \ldots ,{\bf p}_n;
{\bf p}_{n + 1}, \ldots , {\bf p}_{2n + 1})
\label{eq:2r}
\end{equation}
\[
\times (1 +  N_{{\bf p}_1}^l) \ldots (1 + N_{{\bf p}_n}^l)
N_{{\bf p}_{n + 1}}^l \ldots N_{{\bf p}_{2n + 1}}^l
f_{{\bf k}^\prime}[1+f_{\bf k}].
\]
Here, $f_{\bf k}\equiv f({\bf k},x)$ is the disribution function of hard
thermal gluons, and the phase-space integration is
\begin{equation}
\int\!{\rm d} {\cal T}^{(2n + 1)}\!\equiv\!
\int (2 \pi)
\delta(E_{\bf k}+{\cal E}_{\rm in}
- E_{{\bf k}+{\bf p}_{\rm in}-{\bf p}_{\rm out}} - {\cal E}_{\rm out})\!
\prod_{i =1}^{2n+1}\!\frac{d{\bf p}_{i}}{(2\pi)^3},
\label{eq:2t}
\end{equation}
where
\[
{\cal E}_{\rm in} = \omega^l_{\bf p}+\omega^l_{{\bf p}_1} + \ldots
+\omega^l_{{\bf p}_n},\quad
{\bf p}_{\rm in} = {\bf p} + {\bf p}_1 +\ldots +{\bf p}_n,
\hspace{0.3cm}
\]
\[
{\cal E}_{\rm out} = \omega^l_{{\bf p}_{n+1}} +\ldots
+\omega^l_{{\bf p}_{2n+1}},\quad
{\bf p}_{\rm out} ={\bf p}_{n+1} +\ldots +{\bf p}_{2n+1},
\]
are total energies and momenta of incoming and outgoing external
plasmon legs, respectively, and $E_{\bf k}=\vert{\bf k}\vert$ for massless
hard gluons.

The $\delta$-function in Eq.\,(\ref{eq:2t}) expresses the energy conservation
of the processes of stimulated emission and absorption of the plasmons.
The function
\begin{equation}
{\it w}_{2n + 2}={\it w}_{2n + 2}({\bf k}\vert\,{\bf p},
{\bf p}_1,\ldots ,{\bf p}_n; {\bf p}_{n + 1}, \ldots , {\bf p}_{2n + 1})
\label{eq:2tt}
\end{equation}
is a probability of absorption of $n+1$ plasmons with frequencies
$\omega_{\bf p}^l,\,\omega_{{\bf p}_1}^l,\ldots,\omega_{{\bf p}_n}^l$
(and appropriate wavevectors ${\bf p},\,{\bf p}_1,\ldots,{\bf p}_n$)
by a hard thermal gluon\footnote{In the subsequent discussion a hard thermal
particle of plasma, for which we consider the scattering processes of soft
waves, we will call also a {\it test particle}.}
carrying momentum ${\bf k}$ with consequent radiation of $n+1$ plasmons with
frequencies $\omega_{{\bf p}_{n+1}},\ldots\omega_{{\bf p}_{2n+1}}$
(and the wavevectors ${\bf p}_{n+1},\ldots,{\bf p}_{2n+1}$). We note that
for generalized decay and regenerated rates (\ref{eq:2e}) and
(\ref{eq:2r}) it was assumed that the scattering probability (\ref{eq:2tt})
satisfies the symmetry relation over permutation of incoming and outgoing
soft plasmon momenta
\begin{equation}
{\it w}_{2n + 2}({\bf k}\vert\,{\bf p}, {\bf p}_1,
\ldots ,{\bf p}_n; {\bf p}_{n + 1}, \ldots ,
{\bf p}_{2n + 1}) = {\it w}_{2n + 2}({\bf k}\vert\,{\bf p}_{n+1},\ldots
,{\bf p}_{2n+1}; {\bf p}, {\bf p}_1,\ldots, {\bf p}_n).
\label{eq:2ttt}
\end{equation}
This relation is a consequence of more general relation for exact
scattering probability depending on initial and final values of momentum
of hard particle, namely,
\[
{\it w}_{2n + 2}({\bf k},{\bf k}^{\prime}\vert\,{\bf p},
{\bf p}_1,\ldots ,{\bf p}_n; {\bf p}_{n + 1}, \ldots , {\bf p}_{2n + 1})
=
{\it w}_{2n + 2}({\bf k}^{\prime},{\bf k}\vert
\,{\bf p}_{n+1},\ldots ,{\bf p}_{2n+1};{\bf p}, {\bf p}_1,\ldots,
{\bf p}_n).
\]
It expresses detailed balancing principle in scattering processes, and in
this sence it is exact.  The scattering probability (\ref{eq:2tt}) is
obtained by integrating of total probability over ${\bf k}^{\prime}$ with
regard to momentum conservation law:
\[
{\it w}_{2n + 2}
({\bf k}\vert\,{\bf p}, {\bf p}_1,\ldots ,{\bf p}_n;
{\bf p}_{n + 1}, \ldots,
{\bf p}_{2n + 1}) \,2\pi\delta(E_{\bf k}+{\cal E}_{\rm in} -
E_{{\bf k}+{\bf p}_{\rm in}-{\bf p}_{\rm out}} - {\cal E}_{\rm out})
\]
\[
=\int\!
{\it w}_{2n + 2}({\bf k},{\bf k}^{\prime}\vert\,{\bf p},
{\bf p}_1,\ldots ,{\bf p}_n; {\bf p}_{n + 1}, \ldots , {\bf p}_{2n + 1})
\,2\pi
\delta(E_{\bf k}+{\cal E}_{\rm in}-E_{{\bf k}^{\prime}}-{\cal E}_{\rm out})
\]
\[
\times\,(2\pi)^3
\delta({\bf k}+{\bf p}_{\rm in}-{\bf k}^{\prime}-{\bf p}_{\rm out})
\,\frac{d{\bf k}^{\prime}}{(2\pi)^3}.
\]
Such obtained scattering probability satisfies the relation (\ref{eq:2ttt})
in a limit of interest to us, i.e. when we neglect by (quantum) recoil of test
particle. In the general case the expression (\ref{eq:2ttt}) is replaced by
more complicated one (see Section 10).

In Eqs.\,(\ref{eq:2e}) and (\ref{eq:2r}) as in the case of pure
plasmon-plasmon interaction (Paper I), we take into account the scattering processes
only with equal number of the plasmons prior to interaction and upon it,
i.e. the scattering processes of the following ''elastic`` type
\[
{\rm g}^{\ast} + {\rm G} \rightleftharpoons
{\rm g}_1^{\ast} + {\rm G}^{\prime},\quad {\rm for}\;n =0,
\]
\begin{equation}
{\rm g}^{\ast} + {\rm g}_1^{\ast} + {\rm G}\rightleftharpoons
{\rm g}_2^{\ast} + {\rm g}_3^{\ast} + {\rm G}^{\prime},\quad {\rm for}\;n = 1,
\label{eq:2y}
\end{equation}
\[
\ldots,
\]
where ${\rm g}^{\ast}, \, {\rm g}_1^{\ast}, \ldots$ are plasmon collective
excitations and ${\rm G},\, {\rm G}^{\prime}$ are excitations with
characteristic momenta of order $T$. The scattering processes with
odd number of the plasmons are kinematically forbidden by the conservation
laws. Finally the scattering processes with
unequal even number of incoming and outgoing soft external legs, i.e.
the processes of ``inelastic'' type
\[
{\rm G}\rightleftharpoons {\rm g}^{\ast} + {\rm g}_1^{\ast}
+ {\rm G}^{\prime},\quad {\rm for}\;n=0,
\]
\[
{\rm g}^{\ast} + {\rm g}_1^{\ast} + {\rm G}\rightleftharpoons
{\rm g}_2^{\ast} + {\rm g}_3^{\ast} + {\rm g}_4^{\ast}
+ {\rm g}_5^{\ast} + {\rm G}^{\prime},\quad{\rm for}\;n = 2,
\]
etc., have kinematic regions of momentum variables accessible by conservation
laws not coincident with kinematic regions of the corresponding processes
of elastic type (\ref{eq:2y}), and we suppose that a contribution of the last
processes to the nonlinear plasmon dynamics of the order of our interest is not
important.

The scattering process for $n=0$ (Eq.\,(\ref{eq:2y})) known as the process of
nonlinear Landau damping \cite{akh}, in the case of a quark-gluon plasma
was studied in detail in Ref.\,\cite{mark2} (we will consider this process
in the following section in the somewhat different context). In the present
work we would like to extend the approach developed in \cite{mark2} to
the scattering processes involving arbitrary number of plasmons.

By using the fact that $\vert{\bf k}\vert
\gg\vert{\bf p}\vert,\,\vert{\bf p}_1\vert,
\ldots,\vert{\bf p}_{2n+1}\vert$, the energy conservation law can be
represented in the form of the following ``generalized'' resonance condition
\begin{equation}
{\cal E}_{\rm in} - {\cal E}_{\rm out} -
{\bf v}\cdot({\bf p}_{\rm in} -{\bf p}_{\rm out}) =0,\quad
{\bf v}={\bf k}/\vert{\bf k}\vert.
\label{eq:2u}
\end{equation}
In particular for $n=0$ we have resonance
condition \begin{equation} \omega_{\bf p}^l-\omega_{{\bf p}_1}^l -{\bf
v}\cdot({\bf p}-{\bf p}_1)=0, \label{eq:2i} \end{equation} defining the
nonlinear Landau damping process. Furthermore, one can approximate the
distribution function of hard thermal gluons on the right-hand side of
Eqs.\,(\ref{eq:2e}) and (\ref{eq:2r})
\begin{equation}
f_{{\bf k}^{\prime}}\simeq f_{\bf k} + ({\bf p}_{\rm in} - {\bf p}_{\rm out})
\cdot\frac{\partial f_{\bf k}}{\partial{\bf k}},
\label{eq:2o}
\end{equation}
and set $1+f_{\bf k}\simeq 1+f_{{\bf k}^{\prime}}\simeq 1$ by virtue of
$f_{\bf k},\;f_{{\bf k}^{\prime}}\ll1$.

We introduce the following assumption. We suppose that characteristic time
for nonlinear relaxation of the soft oscillations is a small quantity
compared with the time of relaxation of the distribution of hard gluons
$f_{\bf k}$. In other words the intensity of soft plasma excitations
are sufficiently small and they cannot essentially change such
`crude' equilibrium parameters of plasma as particle density,
temperature and thermal energy. Therefore, we neglect by a space-time
change of the distribution function $f_{\bf k}$, assuming that this function
is specified and describe the global equilibrium state of non-Abelian
plasma
\[
f_{\bf k} = 2\,\frac{1}{{\rm e}^{E_{\bf k}/T}-1}.
\]
Here, the coefficient 2 takes into account that the hard gluon has two
helicity states. In the context of this assumption (as it will be further
shown) the scattering probability
${\it w}_{2n+2}({\bf k}\vert\,{\bf p},{\bf p}_1,\ldots{\bf p}_{2n+1})$
depends upon the velocity ${\bf v}$ (a unit vector), but not upon
the magnitude $\vert{\bf k}\vert$ of the hard momentum. This enables us
somewhat simplify expression for collision term $C[N_{\bf p}^l]$.

We use the fact that the occupation numbers $N^l_{{\bf p}_i}$ are more large
than one, i.e. $1+N_{{\bf p}_i}^l\simeq N_{{\bf p}_i}^l$.
Furthermore we present the integration measure as
\[
\int\!\frac{d{\bf k}}{(2\pi)^3} =
\int\!\frac{d\vert{\bf k}\vert}{2\pi^2}\,\vert{\bf k}\vert^2
\!\int\!\frac{d\Omega_{\bf v}}{4\pi},
\]
where the solid integral is over the directions of unit vector ${\bf v}$.
Taking into account above-mentioned, considering (\ref{eq:2o}), the
collision term can be approximated by following expression
\begin{equation}
C[N_{\bf p}^l]\simeq\bigg(\int\!\frac{d\vert{\bf k}\vert}{2\pi^2}
\,\vert{\bf k}\vert^2\frac{\partial f_{\bf k}}
{\partial \vert{\bf k}\vert}
\bigg)
\sum_{n=0}^{\infty}
\int\!\frac{d\Omega_{\bf v}}{4\pi}
\int\!d{\cal T}^{(2n+1)}
\label{eq:2p}
\end{equation}
\[
\times({\cal E}_{\rm in} - {\cal E}_{\rm out})
{\it w}_{2n + 2}({\bf v}\vert\,{\bf p}, {\bf p}_1, \ldots ,{\bf p}_n;
{\bf p}_{n + 1},\ldots,{\bf p}_{2n + 1})
N_{\bf p}^l N_{{\bf p}_1}^l \ldots N_{{\bf p}_{2n+1}}^l.
\]

To derive the scattering probability ${\it w}_{2n+2}$ in Section 5 we
need somewhat different approximation of collision term. Setting
$1+N_{{\bf p}_i}^l \simeq N_{{\bf p}_i}^l$ and $f_{{\bf k}^{\prime}}
\simeq f_{\bf k}$, in the limit of a small intensity $N_{\bf p}^l
\rightarrow 0$ we have the following expression for collision term
\begin{equation}
C[N_{\bf p}^l]\vert_{N_{\bf p}^l\rightarrow 0}
\simeq\biggl(\int\!\frac{d\vert{\bf k}\vert}{2\pi^2}
\,\vert{\bf k}\vert^2 f_{\bf k}\biggr)
\sum_{n=0}^{\infty}
\int\!\frac{d\Omega_{\bf v}}{4\pi}
\int\!d{\cal T}^{(2n+1)}
\label{eq:2a}
\end{equation}
\[
\times{\it w}_{2n + 2}({\bf v}\vert\,{\bf p},{\bf p}_1,\ldots,{\bf p}_n;
{\bf p}_{n + 1},\ldots,{\bf p}_{2n + 1})
N_{{\bf p}_1}^l \ldots N_{{\bf p}_{2n+1}}^l.
\]
The kinetic equation (\ref{eq:2q}) with collision term in the form of
(\ref{eq:2a}) defines a change of plasmons number, caused by processes
of spontaneous plasmon scattering off hard test gluon only.

\section{\bf  Nonlinear Landau damping process}
\setcounter{equation}{0}

In this section, we review the main features of the scattering probability
for nonlinear Landau damping process derived in Ref.\,\cite{mark2}.
This will be done by another way (that has already used
in Paper I) using Tsytovich correspondence principle \cite{gai}, \cite{tsyt},
admitting a direct extension to calculation of scattering probabilities
for the processes of a higher order. We preliminary discuss a basic
equation for a soft gauge field, that will play a main role in our subsequent
research. We have already written out this equation in Paper I
(Eq.\,(I.3.8))\footnote{References to formulas in \cite{mark1} are
prefixed by the roman number I.}. Here, it should be correspondingly
extented to take into account the presence of a current caused by test
particle passing through hot gluon plasma.

One expects the word lines of the hard modes to obey classical trajectories in
the manner of Wong \cite{wong} since their coupling to the soft modes is weak
at very high temperature. Considering this circumstance, we
add the color current of color point charge
\begin{equation}
j_Q^{a\mu}(x)=gv^{\mu}Q^a{\delta}^{(3)}({\bf x}-{\bf v}t)
\label{eq:3q}
\end{equation}
to the right-hand side of the basic field equation.
Here, $Q^a$ is a color classical charge. Considering $Q^a$ as a constant
quantity
we lead to the nonlinear integral equation for gauge potential $A_{\mu}$
instead of Eq.\,(I.3.8)
\begin{equation}
\,^{\ast}\tilde{\cal D}^{-1 \, \mu \nu}(p) A^{a}_{\nu}(p) =
-J^{a\mu}_{NL}[A](p) - j_Q^{a\mu}(p),
\label{eq:3w}
\end{equation}
where
\begin{equation}
J^{a\mu}_{NL}[A](p)=\sum_{s=2}^{\infty} J^{(s)a\mu}(A,\ldots,A),
\label{eq:3e}
\end{equation}
\[
J^{(s)a}_{\mu}(A,\ldots,A)
=  \frac{1}{s!}\,g^{s-1}\!\!\int\!\!
\,^{\ast}\Gamma^{a a_1\ldots a_s}_{\mu\mu_1\ldots\mu_s}(p,-p_{1},\ldots,-p_{s})
A^{a_1\mu_1}(p_{1})A^{a_2\mu_2}(p_{2})\ldots A^{a_s\mu_s}(p_{s})
\]
\[
\times{\delta}^{(4)}(p - \sum_{i=1}^{s}p_{i})\prod_{i=1}^{s}dp_{i},
\]
and
\begin{equation}
j_Q^{a\mu}(p)=\frac{g}{(2\pi)^3}\,v^{\mu}Q^a\delta(v\cdot p).
\label{eq:3r}
\end{equation}
Here, we recall that the coefficient functions
$\,^{\ast}\Gamma^{a a_1\ldots a_s}_{\mu\mu_1\ldots\mu_s}$ are usual
HTL-amplitudes and $\,^{\ast}\tilde{\cal D}^{\mu \nu}(p)$ is a medium
modified (retarded) gluon propagator in a temporal gauge defined by
Eqs.\,(I.3.10)\,--\,(I.3.12).

As in Paper I (Section 5) we consider a solution of the nonlinear
integral equation (\ref{eq:3w}) by the approximation scheme method.
Discarding the nonlinear terms in $A_{\mu}^a$ on the right-hand side, we
obtain in the first approximation
\[
\,^{\ast}\tilde{\cal D}^{-1\,\mu \nu}(p)A_{\nu}^a(p)=
-j_Q^{a\mu}(p).
\]
The general solution of the last equation is
\begin{equation}
A_{\mu}^a(p) = A_{\mu}^{(0)a}(p) -
\,^{\ast}\tilde{\cal D}_{\mu \nu}(p)j_Q^{a\nu}(p),
\label{eq:3t}
\end{equation}
where $A_{\mu}^{(0)a}(p)$ is a solution of homogeneous equation (a free field),
and the last term on the right-hand side represents a gauge field induced
by test parton in medium.

Furthermore, we keep the term quadratic in field on the right-hand side
of Eq.\,({\ref{eq:3w}). Substituting derived solution (\ref{eq:3t})
into the right-hand
side, we obtain the following correction to the interacting field
\[
A_{\mu}^{(1)a}(p) =
-\!\,^{\ast}\tilde{\cal D}_{\mu \nu}(p)J^{(2)a\nu}(A^{(0)},A^{(0)})
\,-\!\,^{\ast}\tilde{\cal D}_{\mu \nu}(p)
\{J^{(2)a\nu}(A^{(0)},-\!\,^{\ast}\tilde{\cal D}j_Q)
+J^{(2)a\nu}(-\!\,^{\ast}\tilde{\cal D}j_Q,A^{(0)})\}
\]
\[
-\!\,^{\ast}\tilde{\cal D}_{\mu \nu}(p)
J^{(2)a\nu}(-\!\,^{\ast}\tilde{\cal D}j_Q,-\!\,^{\ast}\tilde{\cal D}j_Q)\}.
\]
The last term on the right-hand side of this expression equals
zero\footnote{It is obvious that this term will be not equal to zero if
the charges $Q^b$ and $Q^c$ are refered to two different hard test
particles.  It will define contribution to effective current connected
with the process of plasmon bremsstrahlung, that is a subject of research
in our next paper \cite{mark3}.} by virtue of $f^{abc}Q^bQ^c=0$. The first
term is associated with a pure plasmon-plasmon interaction and was
analyzed previously in Paper I.  Therefore now we have concentrated on
new nontrivial terms putting in braces. Using explicit definition of the
nonlinear current of the second order (Eq.\,(\ref{eq:3e})) it can be
shown that a second term in braces equals the first one.

We determine thus a new effective current (more exactly, the first term
in the expansion over free field) that is a correction to ``starting''
current (\ref{eq:3r}), caused by interaction of a medium with color test 
particle
\begin{equation}
J_{Q\mu}^{(1)a}[A^{(0)}](p) =
2J^{(2)a\nu}(A^{(0)},-\!\,^{\ast}\tilde{\cal D}j_Q)
\label{eq:3y}
\end{equation}
\[
= -\,\frac{g^2}{(2\pi)^3}\,Q^{a_2}\!
\int\!^\ast\Gamma_{\mu\mu_1\mu_2}^{aa_1a_2}(p,-p_1,-p+p_1)
A^{(0)a_1\mu_1}(p_1)\,^\ast\!\tilde{\cal D}^{\mu_2\mu_2^{\prime}}(p-p_1)
v_{\mu_2^{\prime}}\delta(v\cdot(p-p_1))dp_1.
\]
Here, in integrand the typical $\delta$-function, determining the resonance
condition (\ref{eq:2i}) of the nonlinear Landau damping process arises.
Using an explicit expression for effective current (\ref{eq:3y}) one can
define a scattering probability of plasmons off hard test particle. For
this purpose according to Tsytovich correspondence principle it should
be substituted $J_{Q\mu}^{(1)a}[A^{(0)}]$ into expression defining the
emitted radiant power of the longitudinal waves ${\cal I}^{\,l}$ Eq.\,(I.4.4).
However as in the case of the field equation (I.3.8) it needs to be preliminary
carried out a minimal extension of an expression (I.4.4) taking into account
a specific of considered problem.

To the procedure of the ensemble average in Eq.\,(I.4.4) we add
an integration over the colors $Q^a$ with a measure
\[
dQ=\prod_{a=1}^{d_A} dQ^a\delta(Q^aQ^a-C_A),\quad
d_A = N_c^2 - 1,
\]
with the gluon Casimir $C_A = N_c$ normalized such that $\int\!dQ=1$,
and thus
\begin{equation}
\int\!dQ\,Q^aQ^b = \frac{C_A}{d_A}\,\delta^{ab}.
\label{eq:3yy}
\end{equation}
Besides it should be added an averaging over distribution of hard particles
in thermal equilibrium: $\int\!d{\bf k}/(2\pi)^3f_{\bf k}\ldots$ .
Taking into account above-mentioned we will use a following expression for the
emitted radiant power ${\cal I}^l$ instead of (I.4.4)
\begin{equation}
{\cal I}^l = -\pi\!\lim\limits_{\tau\rightarrow\infty}
\frac{(2\pi)^4}{\tau}
\biggl(\int\frac{d\vert{\bf k}\vert}{2\pi^2}\,{\bf k}^2 f_{\bf k}\biggr)
\frac{1}{N_c}\int\!dQ\int\!\frac{d\Omega_{\bf v}}{4\pi}
\label{eq:3u}
\end{equation}
\[
\times\!\int\!dp\,\omega\,{\rm sign}(\omega)\tilde{Q}^{\mu\mu^{\prime}}(p)
\,\langle J^{\ast a}_{Q\,\mu}({\bf v},p)J^a_{Q\,\mu^{\prime}}({\bf v},p)\rangle
\delta({\rm Re}\,^{\ast}\!\Delta^{\!-1\,l}(p)).
\]
Here, as distinct from (I.4.4) we remove an averaging over volume of a system,
since in the definition of integration measure (\ref{eq:2t}) a momentum
conservation law is explicitly considered. Besides in notation (\ref{eq:3u})
we take into account that for global equilibrium system an effective
current (more exactly, the part caused by test particle) depends on
momentum ${\bf k}$ only trough velocity ${\bf v}={\bf k}/\vert{\bf k}\vert$.
According to corresponding principle (Paper I, Section 4) for definition
of the scattering probability ${\it w}_{2n+2}({\bf v}\vert\,{\bf p},{\bf p}_1,
\ldots,{\bf p}_{2n+1})$ it should be compared an expression obtained from
(\ref{eq:3u}), with the expression determining the change of energy of
the longitudinal excitations, caused by spontaneus processes of plasmon
emission only
\begin{equation}
\left( \frac{d{\cal E}}{dt} \right)^{\!{\rm sp}}\!=
\frac{d}{dt}\!\left(
\int\!\frac{d{\bf p}}{(2 \pi)^3} \;
\omega^l_{\bf p} N^l_{\bf p}\right)=
\left(\int\frac{d\vert{\bf k}\vert}{2\pi^2}\,{\bf k}^2 f_{\bf k}\right)
\label{eq:3i}
\end{equation}
\[
\times\sum_{n=0}^{\infty}\int\!\frac{d\Omega_{\bf v}}{4\pi}
\int\!\frac{d{\bf p}}{(2 \pi)^3}
\int\!d{\cal T}^{(2n + 1)}\,\omega_{\bf p}^l{\it w}_{2n + 2}
({\bf v}\vert\,{\bf p}, {\bf p}_1, \ldots ,{\bf p}_n;
{\bf p}_{n + 1}, \ldots , {\bf p}_{2n + 1})
N_{{\bf p}_1}^l \ldots N_{{\bf p}_{2n + 1}}^l.
\]
In derivation of the last equality the kinetic equation (\ref{eq:2q}) with
collision term in the limit of a small intensity $N_{\bf p}^l\rightarrow 0$,
(Eq.\,(\ref{eq:2a})), is used.

With all required formulas in hand now one can define the scattering
probability for nonlinear Landau damping process. Substituting an effective
current (\ref{eq:3y}) into (\ref{eq:3u}) and comparing the obtained
expression with the first term ($n=0$) in the expansion on the right-hand
side of (\ref{eq:3i}), we obtain desired elastic scattering probability of
plasmon off hard thermal particle. However such an obtained expression
${\it w}_2({\bf v}\vert\,{\bf p};{\bf p}_1)$ will not be a gauge invariant.
This is associated with the fact that an expression for a first correction
$J_{Q\,\mu}^{(1)a}$ to ``starting'' current is not complete. As we have shown
in Ref.\,\cite{mark2}, the effective current (\ref{eq:3y}) defines the scattering
amplitude of a longitudinal wave off dressing the `cloud' of a test particle.
Diagrammatically this process of scattering is depicted in Fig.\,\ref{FIG1}.
\begin{figure}[hbtp]
\begin{center}
\includegraphics*[height=7cm]{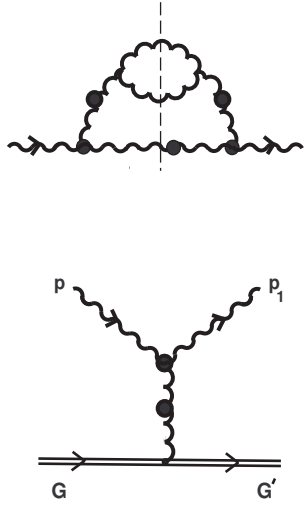}
\end{center}
\caption{\small The process of the stimulated scattering soft boson
excitation off test gluon through a resummed gluon propagator
$\,^{\ast}\tilde{\cal D}$, where a vertex of a three-soft-wave interaction
is induced by $\,^{\ast}\Gamma^{(3)}$. The blob stands for HTL resummation
an the double line denotes hard test particle.}
\label{FIG1}
\end{figure}
Here, the upper figure indicates what the Feynman diagram defines this
scattering process -- cutting of the effective self-energy graph before
inserting a hard bubble along the gluon line.

There is a further process given a contribution of the same order as
above-considered. This contribution on the classic language represents the normal
Thomson scattering of a wave off thermal particle: a wave with the original
frequency $\omega_{\bf p}^l$ is a set in oscillatory particle motion, and an
oscillating particle radiates a wave with modified frequency
$\omega_{{\bf p}_1}^l$. Diagrammatically this process of scattering is depicted in
Fig.\,\ref{FIG2}.
\begin{figure}[hbtp]
\begin{center}
\includegraphics*[height=9cm]{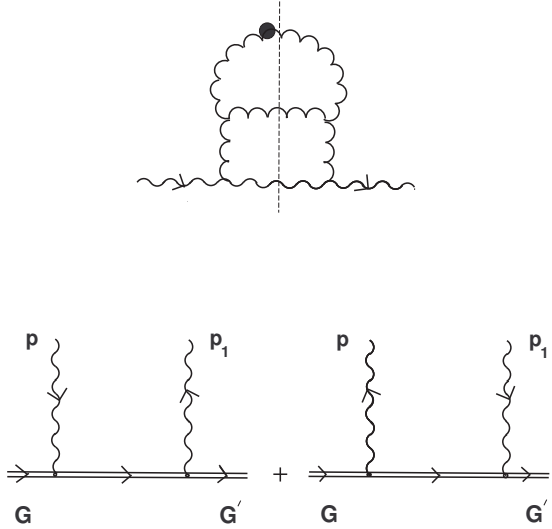}
\end{center}
\caption{\small The Compton scattering of soft boson excitations
off a test gluon. The upper diagram represents cutting of the effective
`tadpole' graph defining this scattering process.}
\label{FIG2}
\end{figure}
To define contribution of this scattering process to an effective current,
account must be taken of influence of the soft random field on a state of
color test particle itself. The particle motion in a wave field is described by
the system of equations
\begin{equation}
m\frac{d^2x^{\mu}}{d \tau^2} =
gQ^aF^{a\mu\nu}(x)\frac{dx_{\nu}}{d\tau},
\label{eq:3o}
\end{equation}
\[
\frac{dQ^a}{d\tau} =
-gf^{abc}\frac{dx^{\mu}}{d\tau}A_{\mu}^b(x)Q^c.
\]
Above, $\tau$ is the proper time. The second equation is known Wong
equation. Instead of the expression (\ref{eq:3q}) for the color current
of a test charge now we must use more general expression
\[
j_Q^{a\mu}=g\,
\frac{dx^{\mu}}{d\tau}Q^a(\tau)\delta^{(4)}(x-x(\tau)).
\]

The system of equations (\ref{eq:3o}) is in a general case very complicated since
it describes both the Abelian contribution to radiation connected with the
change of trajectory and momentum of the test particle due to interaction with
the soft fluctuating gauge field, and the non-Abelian part of the radiation
induced by precession of a color spin in the field of the incident wave.
Our interest is only with a leading HTL-contribution
over coupling constant in considered
theory. This makes it possible to simplify the treatment and consider the hard
test gluon as moving along the straight line, with constant velocity.
In HTL-approximation the QED-like part of the Compton scattering is suppressed,
and only the dominant specific non-Abelian contribution survives\footnote{
Let us note a similarity of this fact with cases which arize in considering
of the problem of a different kind. In deriving energy loss of fast parton
due to bremsstrahlung processes (radiation losses) it was shown that the
QED-like part of the induced radiation associated with change of path, is
suppressed in coupling constant as comparison with non-Abelian contribution,
proportional to the commutator of color generators \cite{gyul2, bai2}.}.
The processes of emission
and absorption of plasmon is defined by a `rotation' of color vector
$Q^a=Q^a(\tau).$ Thus, in the limit of the accepted accuracy of calculation
we need only a minimal extension of the color current of hard test particle
\begin{equation}
j_Q^{a\mu}=gv^{\mu}Q^a(t){\delta}^{(3)}({\bf x}-{\bf v}t),\quad
v^{\mu}=(1,{\bf v}),
\label{eq:3p}
\end{equation}
where a color charge satisfies the Wong equation
\begin{equation}
\frac{dQ^a(t)}{dt} = gf^{abc}(v\cdot A^b(x))Q^c.
\label{eq:3a}
\end{equation}
Here, we turn to a coordinate time.

To derive the oscillations of a color particle, excited by a soft random field
and being linear in amplitude of field on the right-hand side
of Eq.\,(\ref{eq:3a}) we set the color charge equal to its initial value:
$Q^c(t_0)\equiv Q_0^c$, and we replace a field $A_{\mu}^b(x)$ by a free field
\[
A_{\mu}^b(x)\rightarrow A_{\mu}^{(0)b}(x) =A_{\mu}^{(0)b}(t,{\bf v}t)=
\int\!{\rm e}^{-i(v\cdot p)t}A_{\mu}^{(0)b}(p)dp.
\]
In this approximation the solution of the Wong equation (\ref{eq:3a}) has
the form
\begin{equation}
Q^a(t) = Q_0^a + igf^{abc}Q_0^c\int\!\frac{1}{v\cdot p}
\,[\,{\rm e}^{-i(v\cdot p)t} - {\rm e}^{-i(v\cdot p)t_0}]
(v\cdot A^{(0)b}(p))dp.
\label{eq:3s}
\end{equation}
Substituting this expression into (\ref{eq:3p}) and turn to
Fourier-transformation we obtain additional to (\ref{eq:3y}) correction to
``starting'' current of test particle
\[
j_Q^{(1)a\mu}(p)=
i\,\frac{g^2}{(2\pi)^3}\,f^{abc}Q_0^c
\int\!\frac{1}{v\cdot p_1}\,[\delta(v\cdot (p-p_1)) - \delta(v\cdot p)
{\rm e}^{-i(v\cdot p_1)t_0}]
(v\cdot A^{(0)b}(p_1))dp_1.
\]
Here, the first term in square brackets in integrand defines a resonance
condition for the process of the nonlinear Landau damping. The second term
is associated with Cherenkov radiation that is absent in our case.
In the subsequent discussion all terms in the expressions of (\ref{eq:3s})
type, containing initial time will be dropped, because they
not contribute to the scattering processes of interest to us.

Adding an obtained current $j_Q^{(1)a\mu}(p)$ with current (\ref{eq:3y})
we derive complete expression for the first term in the expansion of an
effective current in powers of a free field in the leading HTL-approximation
\begin{equation}
\tilde{J}_{Q\mu}^{(1)a}(p) = \frac{g^2}{(2\pi)^3}
\!\int\!\!K_{\mu\mu_1}^{aa_1}({\bf v}\vert\,p,-p_1)A^{(0)\,a_1\mu_1}(p_1)
\delta(v\cdot(p-p_1))dp_1,
\label{eq:3d}
\end{equation}
where the coefficient function in integrand is defined as
\begin{equation}
K_{\mu\mu_1}^{aa_1}({\bf v}\vert\,p,-p_1)=
K_{\mu\mu_1}^{aa_1b}({\bf v}\vert\,p,-p_1)Q_0^b \equiv
if^{aa_1b}Q_0^b
\stackrel{\scriptscriptstyle{\,(1)}}{\displaystyle{K}}_{\mu\mu_1}
\!({\bf v}\vert\,p,-p_1),
\label{eq:3f}
\end{equation}
where in turn
\begin{equation}
\stackrel{\scriptscriptstyle{\,(1)}}{\displaystyle{K}}_{\mu\mu_1}
\!({\bf v}\vert\,p,-p_1)\equiv
\frac{v_{\mu}v_{{\mu}_1}}{v\cdot p_1} +
\!\,^\ast\Gamma_{\mu\mu_1\nu}(p,-p_1,-p+p_1)
\,^\ast\!\tilde{\cal D}^{\nu\nu^{\prime}}\!(p-p_1)v_{\nu^{\prime}}.
\label{eq:3g}
\end{equation}
The denominator $v\cdot p_1$ in Eq.\,(\ref{eq:3g}) is eikonal, that was
expected in an approximation to the small-angle scattering of a high-energy
particle.  The sense of entering a symbol $\footnotesize{`(1)'}$ over
$K_{\mu\mu_1}\!({\bf v}\vert\,p,-p_1)$ will be clear from the context
below.

We use the expression of an effective current (\ref{eq:3d}) for deriving
of desired scattering probability ${\it w}_2({\bf v}\vert\,{\bf p},-{\bf p}_1)$.
The procedure of a calculation of the scattering probabilities with the use of
Eqs.\,(\ref{eq:3u}) and (\ref{eq:3i}) will be given for general case in
Section 6. Here, we present only the final result
\begin{equation}
{\it w}_2({\bf v}\vert\,{\bf p},{\bf p}_1) =
N_c\,\vert T({\bf v}\vert\,{\bf p},-{\bf p}_1)\vert^2\equiv
N_c\,\vert T_{\bf v}({\bf p},-{\bf p}_1)\vert^2,
\label{eq:3h}
\end{equation}
where the function
\begin{equation}
T_{\bf v}({\bf p},-{\bf p}_1)
= g^{2}\!
\left[\left(\frac{{\rm Z}_l({\bf p})}{2\omega_{\bf p}^l}\right)^{\!1/2}
\!\!\frac{\tilde{u}^{\mu}(p)}{\sqrt{\bar{u}^2(p)}}\right]
\!\!\left[\left(\frac{{\rm Z}_l({\bf p}_1)}{2\omega_{{\bf p}_1}^l}\right)^{\!1/2}
\!\!\frac{\tilde{u}^{\mu_1}(p_1)}{\sqrt{\bar{u}^2(p_1)}}\right]
\!\stackrel{\scriptscriptstyle{\,(1)}}{\displaystyle{K}}_{\mu\mu_1}
\!({\bf v}\vert\,p,-p_1)
\Big|_{\rm on-shell}
\label{eq:3j}
\end{equation}
represents the scattering amplitude of the nonlinear Landau damping process.
The expression (\ref{eq:3h}) was obtained in Ref.\,\cite{mark2} by a different
method, that will be mentioned below.

\section{\bf The higher coefficient functions}
\setcounter{equation}{0}

In previous Section we consider a simpler process of scattering of plasmons off
hard thermal particle -- the nonlinear Landau damping process. As in pure
plasmon-plasmon interaction \cite{mark1}, a calculation of the scattering probability
here, reduces to computation of some effective current (Eq.\,(\ref{eq:3d}))
generating this process. We have shown that this effective current apears in the
solution of the nonlinear integral field equation (\ref{eq:3w}), which defines
interacting soft-gluon field $A_{\mu}$ in the form of an expansion in a free
field $A_{\mu}^{(0)}=(A_{\mu}^{(0)a})$ and also initial value of a color charge
$Q_0=(Q_0^a)$. The last circumstance is a new feature of considered problem
different from pure plasmon nonlinear dynamics.

Based on results of Paper I and previous section we can rewrite now more
general structure of an effective current in the form of a functional expansion
in a free field $A_{\mu}^{(0)}$ (and generally speaking, color charge $Q_0$)
generating the plasmon decay processes and the scattering processes of
arbitrary number of plasmons\footnote{Here, for concreteness we consider
the scattering processes with longitudinal oscillation, but this refer also
to transverse soft gluon excitations.} by hard thermal particle
\begin{equation}
\tilde{J}_{\mu}^{({\rm tot})a}[A^{(0)}](p) =
\frac{g}{(2\pi)^3}\,Q_0^av_{\mu}\delta(v\cdot p) +
\sum_{s=1}^{\infty}
\tilde{J}_{Q\mu}^{(s)a}[A^{(0)}]({\bf v},p)+
\sum_{s=2}^{\infty}
\tilde{J}_{\mu}^{(s)a}[A^{(0)}](p),
\label{eq:4q}
\end{equation}
where
\[
\tilde{J}_{Q\mu}^{(s)a}[A^{(0)}]({\bf v},p)=
\frac{1}{s!}\,\frac{g^{s+1}}{(2\pi)^3}\!\int\!
K_{\mu\mu_1\ldots\mu_s}^{aa_1\ldots a_s}({\bf v}\vert\,p,-p_1,\ldots,-p_s)
A^{(0)a_1\mu_1}(p_1)\ldots A^{(0)a_s\mu_s}(p_s)
\]
\[
\times\delta(v\cdot(p-\sum_{i=1}^{s}p_i))\prod_{i=1}^{s}dp_i,
\]
\[
\tilde{J}^{(s)a}_{\mu}[A^{(0)}](p) =
\frac{1}{s!}\;g^{s-1}\!\!\int\!\!
\,^{\ast}\tilde{\Gamma}^{a a_1\ldots a_s}_{\mu\mu_1\ldots\mu_s}
(p,-p_{1},\ldots,-p_{s})
A^{(0)a_1\mu_1}(p_{1})\ldots A^{(0)a_s\mu_s}(p_{s})
\]
\[
\times\delta (p - \sum_{i=1}^{s}p_{i})\prod_{i=1}^{s}dp_{i}.
\]
The last sum on the right-hand side of Eq.\,(\ref{eq:4q}) represents an
effective current generating plasmon decay processes. The current was studied
in detail in Paper I, where a complete algorithm of succesive calculation
of coefficient functions
$\,^\ast\tilde{\Gamma}_{\mu\mu_1\ldots\mu_s}^{aa_1\ldots a_s}$,
was proposed. Now our problem is in constraction of a similar algorithm for
calculation of coefficient function $K_{\mu\mu_1\ldots\mu_s}^{aa_1\ldots a_s}$.
In general the structure of these functions is more complicated and tangled,
as compared with $\,^\ast\tilde{\Gamma}_{\mu\mu_1\ldots\mu_s}^{aa_1\ldots a_s}$.
This follows from the
fact that they represent infinite series in expansion over color charge
$Q_0$, i.e.
\begin{equation}
K_{\mu\mu_1\ldots\mu_s}^{aa_1\ldots a_s}
({\bf v}\vert\,p,-p_1,\ldots,-p_s)=
\sum_{m=0}^{\infty}
K_{\mu\mu_1\ldots\mu_s}^{aa_1\ldots a_s\,bb_1\ldots b_m}
({\bf v}\vert\,p,-p_1,\ldots,-p_s)
Q_0^bQ_0^{b_1}\ldots Q_0^{b_m},
\label{eq:4w}
\end{equation}
where
\[
K_{\mu\mu_1\ldots\mu_s}^{aa_1\ldots a_s\,bb_1\ldots b_m}=
\left.\frac{\delta^m K^{aa_1\ldots a_s}_{\mu\mu_1\ldots\mu_s}}
{\delta Q_0^b \delta Q_0^{b_1}\ldots\delta Q_0^{b_m}}\right\vert_{Q_0=0}.
\]
Below we discuss a physical meaning of coefficients of this expansion
on the example
of particular computation. Here, we note that primary consideration will be
focussed on computation of the terms linear in $Q_0$ in the expansion
(\ref{eq:4w}) that give leading contribution to the scattering processes of our
interest.

The more direct and explicit way of calculation of required
coefficient functions $K^{(s)}(\equiv K^{aa_1\dots a_s}_{\mu\mu_1\ldots\mu_s})$
was presented in previous section. It is based on usual procedure of computing
perturbative solutions of the classical Yang-Mills equation. However such
a direct approach for determination of an explicit form of the higher cofficient
functions $K^{(s)},\,s>1$, becomes very complicated and as consequence,
ineffective. For deriving $K^{(s)}$ we use the approach suggested in Paper I,
which was applied to obtain $\!\,^{\ast}\tilde{\Gamma}^{(s)}$ with
appropriate modification.  Remind that this approach is based on simple
fact, namely, that the total color current $J^{({\rm
tot})\,a\mu}[A]\equiv J^{a\mu}_{NL}[A] + j_Q^{a\mu}[A]$, entering into
the right-hand side of the field equation (\ref{eq:3w}), has two
representation: by means of free and interacting fields, which must be
equal each other. Thus in representation of free field $A^{(0)}_{\mu}$
the current is defined by Eq.\,(\ref{eq:4q}).  To rewrite explicit
expression for total current in representation of interacting field we
need to define the expansion for current $j_Q^{a\mu}[A](p)$ similar to
expansion (\ref{eq:3e}). For this purpose we write the solution of
Eq.\,(\ref{eq:3a}) in the form \[ Q^a(t)=U^{ab}(t,t_0)Q_0^b, \] where \[
U(t,t_0) = {\rm T}\exp\{-ig\!\int_{t_0}^t\!
(v\cdot A^a(\tau,{\bf v}\tau))T^ad\tau\}=
\]
\[
1+\sum_{s=1}^{\infty}(-ig)^s\!\int_{t_0}^{t}\!d\tau_1\int_{t_0}^{\tau_1}\!
d\tau_2\ldots\int_{t_0}^{\tau_{s-1}}\!d\tau_s
\,(v\cdot A^{a_1}(\tau_1))(v\cdot A^{a_2}(\tau_2))\dots
(v\cdot A^{a_s}(\tau_s))T^{a_1}T^{a_2}\dots T^{a_s}
\]
is an evolution operator accounting for the color precession along the parton
trajectory. Here, $(T^a)^{bc}=-if^{abc}$, and in the last line we set
$A^a_{\mu}({\tau})\equiv A^a_{\mu}(\tau,{\bf v}\tau)$.
Using this form of the solution of the Wong equation,
we can present a current $j_Q^{a\mu}[A](p)$ in the form
\begin{equation}
j_Q^{a\mu}[A](p)=\frac{g}{(2\pi)^3}\,Q_0^a v^{\mu}\delta (v\cdot p) +
\frac{g}{(2\pi)^3}\,v^{\mu}\!\!\int\!{\rm e}^{i(v\cdot p)t}[U(t,t_0)-1]^{ab} Q_0^b
\,\frac{dt}{2\pi}.
\label{eq:4e}
\end{equation}
Adding the obtained expression with nonlinear current $(\ref{eq:3e})$, we
derive a total expression for current in the representation of an interacting
field
\[
J_{\mu}^{({\rm tot})a}[A](p) = J_{NL\,\mu}^a[A](p) + j_{Q\,\mu}^a[A](p)
\]
\begin{equation}
=\sum_{s=2}^{\infty}
\frac{1}{s!}\,g^{s-1}\!\!\int\!
\,^{\ast}\Gamma^{a a_1\ldots a_s}_{\mu\mu_1\ldots\mu_s}(p,-p_{1},\ldots,-p_{s})
A^{a_1\mu_1}(p_{1})A^{a_2\mu_2}(p_{2})\ldots A^{a_s\mu_s}(p_{s})
\delta (p - \sum_{i=1}^{s}p_{i})\prod_{i=1}^{s}dp_{i}
\label{eq:4r}
\end{equation}
\[
+\,\frac{g}{(2\pi)^3}\,Q_0^av_{\mu}\delta(v\cdot p) + v_{\mu}\sum_{s=1}^{\infty}
\frac{g^{s+1}}{(2\pi)^3}\int\!\frac{1}
{(v\cdot (p_1 +\ldots +p_s))(v\cdot (p_2 + \ldots + p_s))\ldots (v\cdot p_s)}
\]
\[
\times (v\cdot A^{a_1}(p_1))\ldots (v\cdot A^{a_s}(p_s))
\delta (v\cdot (p-\sum_{i=1}^{s}p_i))\prod_{i=1}^s dp_i
\,(T^{a_1}\ldots T^{a_s})^{ab}Q_0^b.
\]
Here, in derivation of two last lines in Eq.\,(\ref{eq:4r}) we drop all terms
containing initial time $t_0$. The interacting fields on the right-hand side
of Eq.\,(\ref{eq:4r}) are defined by expansion
\begin{equation}
A^{a\mu}(p) = A^{(0)\,a\mu}(p) -\!
\,^{\ast}\tilde{\cal D}^{\mu\mu^{\prime}}(p)
\tilde{J}^{({\rm tot})\,a}_{\mu^{\prime}}[A^{(0)}](p),
\label{eq:4t}
\end{equation}
where the current $\tilde{J}^{({\rm tot})a}_{\mu^{\prime}}$ is defined by
Eq.\,(\ref{eq:4q}). Thus we have two different representations for total color
current: Eqs.\,(\ref{eq:4q}) and (\ref{eq:4r}), which must be equal each other
\begin{equation}
\tilde{J}^{({\rm tot})\,a}_{\mu}[A^{(0)}](p)=
J^{({\rm tot})\,a}_{\mu}[A](p).
\label{eq:4y}
\end{equation}
Substitution of Eq.\,(\ref{eq:4t}) into the right-hand side of
Eq.\,(\ref{eq:4y}) turns this equation into identity. As for pure
plasmon-plasmon scattering \cite{mark1}, for derivation of required coefficient
functions $K^{(s)}$ it should be differentiated left- and right-hand sides
of equality (\ref{eq:4y}) with respect to free field $A_{\mu}^{(0)}$
considering Eq.\,(\ref{eq:4r}) for differentiation on the right-hand side,
and set $A_{\mu}^{(0)}=0$ after all calculations. However in this case it is
necessary to add differentiation with respect to initial color charge $Q_0$
to differentiation with respect to free field $A_{\mu}^{(0)}$,
and it should be added the condition $Q_0=0$
to condition $A_{\mu}^{(0)}=0$ after all calculations.
Below we shall give a few examples.

The second differentiation of a total current yields
\begin{equation}
\left.\frac{\delta^2\!J^{({\rm tot})\,a}_{\mu}[A](p)}
{\delta Q_0^b\,\delta A^{(0)a_1\mu_1}(p_1)}\right|_{A^{(0)}=0,\,Q_0=0}
=\left.\frac{\delta K_{\mu\mu_1}^{aa_1}({\bf v}\vert\,p,-p_1)}{\delta Q_0^b}
\right|_{Q_0=0}\!\delta(v\cdot(p-p_1))
\label{eq:4u}
\end{equation}
\[
=(T^{a_1})^{ab}
\stackrel{\scriptscriptstyle{\,(1)}}{\displaystyle{K}}_{\mu\mu_1}
({\bf v}\vert\,p,-p_1)\,\delta (v\cdot (p-p_1)),
\]
where Lorentz tensor
$\stackrel{\scriptscriptstyle{\,(1)}}{\displaystyle{K}}_{\mu\mu_1}$
is defined by Eq.\,(\ref{eq:3g})
and represents a sum of two different contribution depicted in 
Figs.\,\ref{FIG1} and \ref{FIG2}.

A more nontrivial example arises in calculation of the next derivative. 
It defines
the process of nonlinear interaction of three wave with hard test particle
\begin{equation}
\left.\frac{\delta^3\!J^{({\rm tot})\,a}_{\mu}[A](p)}
{\delta Q_0^b\,
\delta A^{(0)a_1\mu_1}(p_1)
\delta A^{(0)a_2\mu_2}(p_2)}
\right|_{A^{(0)}=0,\,Q_0=0}
=\left.\frac{\delta K_{\mu\mu_1\mu_2}^{aa_1a_2}({\bf v}\vert\,p,-p_1,-p_2)}
{\delta Q_0^b}
\right|_{Q_0=0}\!\delta(v\cdot(p-p_1-p_2))
\label{eq:4i}
\end{equation}
\[
=[(T^{a_1}T^{a_2})^{ab}
\stackrel{\scriptscriptstyle{\,(1)}}{\displaystyle{K}}_{\mu\mu_1\mu_2}
\!({\bf v}\vert\,p,-p_1,-p_2) +
(T^{a_2}T^{a_1})^{ab}
\stackrel{\scriptscriptstyle{\,(1)}}{\displaystyle{K}}_{\mu\mu_2\mu_1}
\!({\bf v}\vert\,p,-p_2,-p_1)]\,\delta(v\cdot(p-p_1-p_2)).
\]
The color factor on the right-hand side of Eq.\,(\ref{eq:4i}) are multiplied
by pure kinematical coefficients, which we will call {\it partial coefficient
functions}, and are defined as follows
\[
\stackrel{\scriptscriptstyle{\,(1)}}{\displaystyle{K}}_{\mu\mu_1\mu_2}
\!({\bf v}\vert\,p,-p_1,-p_2) =
\frac{v_\mu v_{\mu_1} v_{\mu_2}}{(v\cdot p_2)(v\cdot (p_1 + p_2))}
\]
\begin{equation}
+\,\frac{v_\mu v_\nu}{v\cdot (p_1 + p_2)}
\,^{\ast}\tilde{\cal D}^{\nu\nu^{\prime}}(p_1 + p_2)
\,^{\ast}\Gamma_{\nu^{\prime}\mu_1\mu_2}(p_1+p_2,-p_1,-p_2)
\label{eq:4o}
\end{equation}
\[
-\,^{\ast}\Gamma_{\mu\mu_1\mu_2\nu}(p,-p_1,-p_2,-p+p_1+p_2)\,^{\ast}
\tilde{\cal D}^{\nu\nu^{\prime}}(p-p_1-p_2)v_{{\nu}^{\prime}}
\]
\[
+\,^{\ast}\Gamma_{\mu\mu_1\nu}(p,-p_1,-p+p_1)\,^{\ast}
\tilde{\cal D}^{\nu\nu^{\prime}}(p-p_1)
\stackrel{\scriptscriptstyle{\,(1)}}{\displaystyle{K}}_{\nu^{\prime}\mu_2}
\!({\bf v}\vert\,p-p_1,-p_2)
\]
\[
-\,^{\ast}\Gamma_{\mu\nu\lambda}(p,-p+p_1+p_2,-p_1-p_2)
\!\,^{\ast}\tilde{\cal D}^{\nu\nu^{\prime}}(p-p_1-p_2)v_{{\nu}^{\prime}}
\!\,^{\ast}\tilde{\cal D}^{\lambda\lambda^{\prime}}(p_1+p_2)
\!\,^{\ast}\Gamma_{\lambda^{\prime}\mu_1\mu_2}(p_1+p_2,-p_1,-p_2).
\]
Here, the fuction $\stackrel{\scriptscriptstyle{\,(1)}}
{\displaystyle{K}}_{\mu\mu_1\mu_2}$ consists of five terms different in
structure\footnote{We note that with regard to an explicit expression
$\stackrel{\scriptscriptstyle{\,(1)}}{\displaystyle{K}}_{\mu\mu_1}$
(Eq.\,(\ref{eq:3g})) the last three terms in Eq.\,(\ref{eq:4o}) can be also
presented in another useful form
\[
\,^{\ast}\Gamma_{\mu\mu_1\nu}(p,-p_1,-p+p_1)\,^{\ast}
\tilde{\cal D}^{\nu\nu^{\prime}}(p-p_1)v_{\nu^{\prime}}\,\frac{v_{\mu_2}}
{v\cdot p_2}
-\!\,^{\ast}\tilde{\Gamma}_{\mu\mu_1\mu_2\nu}(p,-p_1,-p_2,-p+p_1+p_2)\,^{\ast}
\tilde{\cal D}^{\nu\nu^{\prime}}(p-p_1-p_2)v_{{\nu}^{\prime}},
\]
where $\!\,^{\ast}\tilde{\Gamma}^{(4)}$ is defined by Eq.\,(I.5.6).},
whose diagram interpretation was presented on Fig.\,\ref{FIG3}.
\begin{figure}[hbtp]
\begin{center}
\includegraphics*[height=9cm]{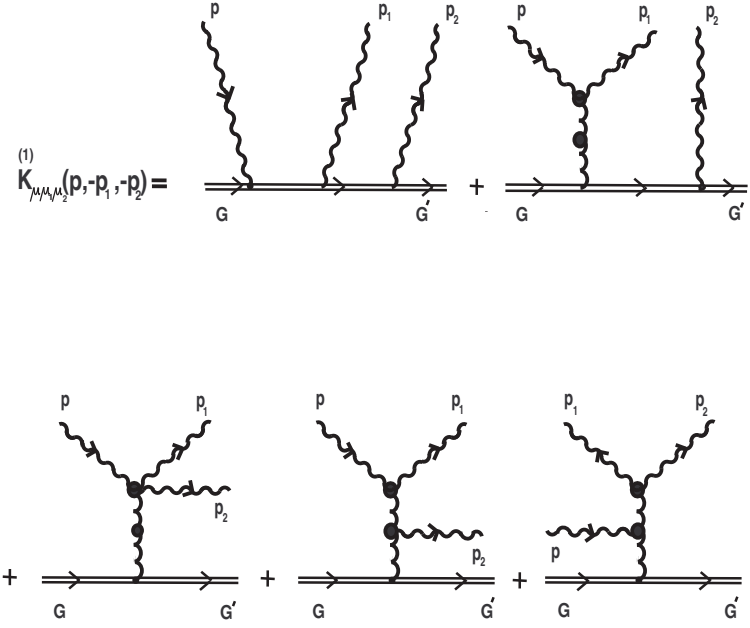}
\end{center}
\caption{\small The process of the stimulated scattering of third soft boson
excitations off hard test parton.}
\label{FIG3}
\end{figure}
Above-mentioned two examples suggest on that coefficient functions
$(\delta K_{\mu\mu_1\ldots\mu_s}^{aa_1\ldots a_s}/\delta Q_0^b)_{Q_0=0}$
have a color structure, similar to color structure of effective amplitudes
$\,^{\ast}\tilde{\Gamma}^{aa_1\ldots a_{s+1}}_{\mu\mu_1\ldots\mu_{s+1}}$,
and thus coincide with color structure of usual HTL-amplitudes, first
proposed by Braaten and Pisarski in Ref.\,\cite{bra1} for an arbitrary number 
of external soft-gluon legs. To test this assumption we calculate a next
derivative with respect to free field $A^{(0)}_{\mu}$:
\begin{equation}
\left.\frac{\delta^4\!J^{({\rm tot})\,a}_{\mu}[A](p)}
{\delta Q_0^b\,
\delta A^{(0)a_1\mu_1}(p_1)
\delta A^{(0)a_2\mu_2}(p_2)
\delta A^{(0)a_3\mu_3}(p_3)}
\right|_{A^{(0)}=0,\,Q_0=0}
\label{eq:4p}
\end{equation}
\[
=\left.\frac{\delta K_{\mu\mu_1\mu_2\mu_3}^{aa_1a_2a_3}
({\bf v}\vert\,p,-p_1,-p_2,-p_3)}
{\delta Q_0^b}\right|_{Q_0=0}\!\delta(v\cdot(p-p_1-p_2-p_3))
\]
\[
=[(T^{a_1}T^{a_2}T^{a_3})^{ab}
\stackrel{\scriptscriptstyle{\,(1)}}{\displaystyle{K}}_{\mu\mu_1\mu_2\mu_3}
\!({\bf v}\vert\,p,-p_1,-p_2,-p_3) +
({\rm perm.}\,1,2,3)]\,\delta(v\cdot(p-p_1-p_2-p_3)),
\]
where
\[
\stackrel{\scriptscriptstyle{\,(1)}}
{\displaystyle{K}}_{\mu\mu_1\mu_2\mu_3}\!({\bf v}\vert\,p,-p_1,-p_2,-p_3) =
\frac{v_\mu}{v\cdot (p_1 + p_2 + p_3)}
\left\{
\frac{v_{\mu_1} v_{\mu_2} v_{\mu_3}}{(v\cdot p_2)(v\cdot (p_2 + p_3))}
\right.
\]
\[
+\,
v_{\nu}\!\,^{\ast}\tilde{\cal D}^{\nu\nu^{\prime}}\!(p_1 + p_2 + p_3)
\,^{\ast}\Gamma_{\nu^{\prime}\mu_1\mu_3\mu_2}(p_1+p_2+p_3,-p_1,-p_3,-p_2)
\]
\[
+\,\frac{1}{v\cdot p_2}\,
v_{\nu}\!\,^{\ast}\tilde{\cal D}^{\nu\nu^{\prime}}\!(p_1 + p_3)
\,^{\ast}\Gamma_{\nu^{\prime}\mu_1\mu_3}(p_1+p_3,-p_1,-p_3)v_{\mu_2}
\]
\[
\left.
-\,\frac{1}{v\cdot (p_2+p_3)}\,
v_{\nu}\!\,^{\ast}\tilde{\cal D}^{\nu\nu^{\prime}}\!(p_2 + p_3)
\,^{\ast}\Gamma_{\nu^{\prime}\mu_2\mu_3}(p_2+p_3,-p_2,-p_3)v_{\mu_1}
\right\}
\]
\[
-\,\Bigl\{
\!\,^{\ast}\Gamma_{\mu\mu_1\nu}(p,-p_1,-p+p_1)
\,^{\ast}\tilde{\cal D}^{\nu\nu^{\prime}}\!(p-p_1)
\stackrel{\scriptscriptstyle{\,(1)}}
{\displaystyle{K}}_{\nu^{\prime}\mu_2\mu_3}
\!({\bf v}\vert\,p-p_1,-p_2,-p_3)
\]
\[
+\!\,^{\ast}\Gamma_{\mu\nu\lambda}(p,\!-p_3,\!-p+p_3)
\!\,^{\ast}\!\tilde{\cal D}^{\nu\nu^{\prime}}\!(p_3)
\!\!\stackrel{\scriptscriptstyle{\,(1)}}
{\displaystyle{K}}_{\nu^{\prime}\mu_3}
\!\!({\bf v}\vert\,p\!-p_1\!-p_2,\!-p_3)
\!\,^{\ast}\!\tilde{\cal D}^{\lambda\lambda^{\prime}}\!(p_1+p_2)
\!\,^{\ast}\Gamma_{\lambda^{\prime}\mu_1\mu_2}(p_1+p_2,\!-p_1,\!-p_2)
\]
\[
+\,^{\ast}\Gamma_{\mu\nu\lambda}(p,-p+p_1+p_2+p_3,-p_1-p_2-p_3)
\!\,^{\ast}\tilde{\cal D}^{\nu\nu^{\prime}}\!(p-p_1-p_2-p_3)v_{{\nu}^{\prime}}
\]
\[
\times
\!\,^{\ast}\tilde{\cal D}^{\lambda\lambda^{\prime}}\!(p_1+p_2+p_3)
\!\,^{\ast}\Gamma_{\lambda^{\prime}\mu_1\mu_2\mu_3}
(p_1+p_2+p_3,-p_1,-p_2,-p_3)
\Bigr\}
\]
\[
-\,\Bigl\{
\!\,^{\ast}\Gamma_{\mu\mu_1\mu_2\nu}(p,-p_1,-p_2,-p+p_1+p_2)\,^{\ast}
\tilde{\cal D}^{\nu\nu^{\prime}}\!(p-p_1-p_2)
\!\stackrel{\scriptscriptstyle{\,(1)}}
{\displaystyle{K}}_{\nu^{\prime}\mu_3}
\!({\bf v}\vert\,p-p_1-p_2,-p_3)
\]
\[
+\,^{\ast}\Gamma_{\mu\lambda\mu_3\nu}(p,-p_1-p_2,-p_3,-p+p_1+p_2+p_3)
\!\,^{\ast}\tilde{\cal D}^{\nu\nu^{\prime}}\!(p-p_1-p_2-p_3)v_{{\nu}^{\prime}}
\]
\[
\times
\!\,^{\ast}\tilde{\cal D}^{\lambda\lambda^{\prime}}\!(p_1+p_2)
\!\,^{\ast}\Gamma_{\lambda^{\prime}\mu_1\mu_2}(p_1+p_2,-p_1,-p_2)
\]
\[
+\,^{\ast}\Gamma_{\mu\mu_1\lambda\nu}(p,-p_1,-p_2-p_3,-p+p_1+p_2+p_3)
\!\,^{\ast}\tilde{\cal D}^{\nu\nu^{\prime}}\!(p-p_1-p_2-p_3)v_{{\nu}^{\prime}}
\]
\[
\times
\!\,^{\ast}\tilde{\cal D}^{\lambda\lambda^{\prime}}\!(p_2+p_3)
\!\,^{\ast}\Gamma_{\lambda^{\prime}\mu_2\mu_3}(p_2+p_3,-p_2,-p_3)
\Bigr\}
\]
\[
-\,^{\ast}\Gamma_{\mu\mu_1\mu_2\mu_3\nu}(p,-p_1,-p_2,-p_3,-p+p_1+p_2+p_3)
\,^{\ast}
\tilde{\cal D}^{\nu\nu^{\prime}}\!(p-p_1-p_2-p_3)v_{{\nu}^{\prime}}.
\]
The right-hand side of this expression is represented for convenience
as a sum of four groups of terms defined by derivation of currents
$j_{Q\mu}^a,\,J_{\mu}^{(2)a},\,J_{\mu}^{(3)a}$ and $J_{\mu}^{(4)a}$
(Eqs.\,(\ref{eq:3w}), (\ref{eq:3e})), respectively. Thus one can
state that a color structure  of the first term in the expansion of
arbitrary coefficient function (\ref{eq:4w})
linear over color charge $Q_0^b$, entirely coincides with the color
structure of usual $(s+2)$-gluon HTL-amplitudes.

In closing we consider the problem on physical meaning of
higher power of color charge $Q_0$ in the expansion of coefficient
functions $K_{\mu\mu_1\ldots\mu_s}^{aa_1\ldots a_s}$ (Eq.\,(\ref{eq:4w})).
For this purpose we calculate the second derivative of the expression
(\ref{eq:4u}) with respect to $Q_0$
\[
\left.\frac{\delta^3\!J^{({\rm tot})\,a}_{\mu}[A](p)}
{\delta Q_0^b\,\delta Q_0^{b_1}\,
\delta A^{(0)a_1\mu_1}(p_1)}\right|_{A^{(0)}=0,\,Q_0=0}
=\frac{g^2}{(2\pi)^3}
\left.\frac{\delta^2 K_{\mu\mu_1}^{aa_1}({\bf v}\vert\,p,-p_1)}
{\delta Q_0^b\,\delta Q_0^{b_1}}
\right|_{Q_0=0}\!\delta(v\cdot(p-p_1))
\]
\[
=\frac{g^2}{(2\pi)^3}\,\{T^b,T^{b_1}\}^{aa_1}
\stackrel{\scriptscriptstyle{\,(2)}}
{\displaystyle{K}}_{\mu\mu_1}\!({\bf v}\vert\,p,-p_1)\,
\delta(v\cdot(p-p_1)),
\]
where
\[
\stackrel{\scriptscriptstyle{\,(2)}}
{\displaystyle{K}}_{\mu\mu_1}\!({\bf v}\vert\,p,-p_1) =
\int\!\bigg\{
\frac{v_\mu v_{\mu_1}}{(v\cdot p_1)(v\cdot (p_1 + p_1^{\prime}))}
\,(v_{\nu}\!\,^{\ast}\tilde{\cal D}^{\nu\nu^{\prime}}\!(p_1^{\prime})
v_{\nu^{\prime}})
\]
\begin{equation}
-\,\frac{v_\mu v_\nu}{v\cdot (p_1 + p_1^{\prime})}
\,^{\ast}\tilde{\cal D}^{\nu\nu^{\prime}}\!(p_1 + p_1^{\prime})
\!\stackrel{\scriptscriptstyle{\,(1)}}
{\displaystyle{K}}_{{\nu}^\prime\mu_1}\!({\bf v}\vert\,p_1+p_1^{\prime},-p_1)
\label{eq:4a}
\end{equation}
\[
+\,^{\ast}\Gamma_{\mu\nu_1\nu_2\mu_1}
(p,-p_1^{\prime},-p+p_1+p_1^{\prime},-p_1)
\,^{\ast}\tilde{\cal D}^{\nu_1\nu_1^{\prime}}
(p_1^{\prime})v_{{\nu_1}^{\prime}}
\!\,^{\ast}\tilde{\cal D}^{\nu_2\nu_2^{\prime}}
(p-p_1-p_1^{\prime})v_{{\nu_2}^{\prime}}
\]
\[
+\,^{\ast}\Gamma_{\mu\nu_1\nu_2}(p,-p_1^{\prime},-p+p_1^{\prime})
\,^{\ast}\tilde{\cal D}^{\nu_1\nu_1^{\prime}}(p_1^{\prime})v_{{\nu}_1^{\prime}}
\!\,^{\ast}\tilde{\cal D}^{\nu_2\nu_2^{\prime}}(p-p_1^{\prime})
\!\stackrel{\scriptscriptstyle{\,(1)}}
{\displaystyle{K}}_{{\nu}^\prime_2\mu_1}\!({\bf v}\vert\,p-p_1^{\prime},-p_1)
\bigg\}\,
\delta(v\cdot p_1^{\prime}) dp_1^{\prime}.
\]
It is easy to check by direct calculation that partial coefficient function
$\stackrel{\scriptscriptstyle{\,(2)}}
{\displaystyle{K}}_{\mu\mu_1}\!({\bf v}\vert\,p,-p_1)$ is symmetric relative
to permutation $p\leftrightarrow -p_1$, as it must be.
The diagrammatic interpretation of different terms in the right-hand side
of Eq.\,(\ref{eq:4a}) is presented on Fig.\,\ref{FIG4}.
\begin{figure}[hbtp]
\begin{center}
\includegraphics*[height=9.5cm]{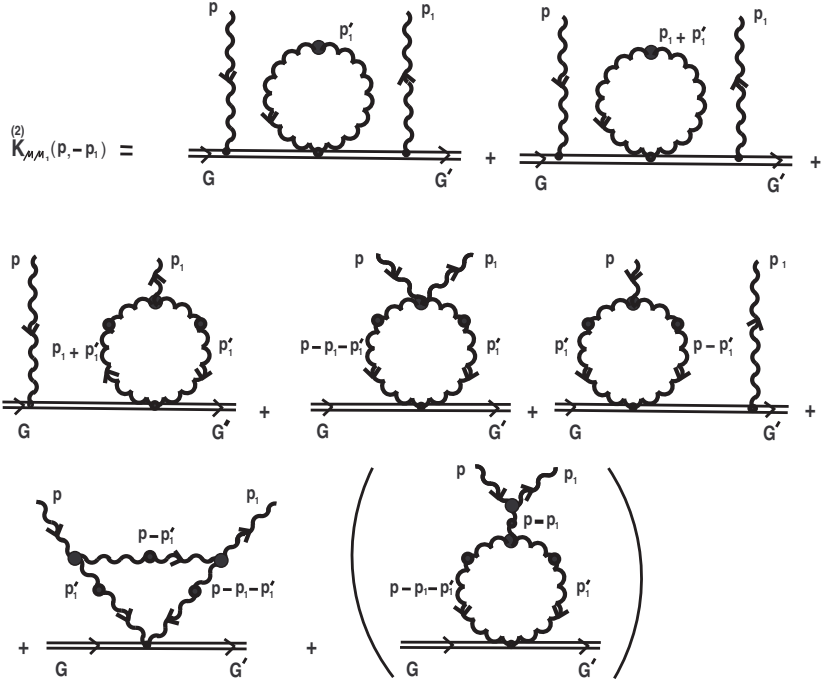}
\end{center}
\caption{\small The soft one-loop corrections to nonlinear Landau
damping process.}
\label{FIG4}
\end{figure}
The presence of resonance factor $\delta(v\cdot(p-p_1))$ and integration over
loop momentum $p_1^{\prime}$ points to the fact that here, we are
concerned with soft one-loop correction to nonlinear Landau damping process,
that is suppressed by power of $g^2$ as compared with tree approximation
(\ref{eq:3f}), (\ref{eq:3g}). We note specially that the region of integration
in loops is restricted by cone $v\cdot p_1^{\prime}=0$. The contribution
\[
\int\!\!\,^{\ast}\Gamma_{\mu_1\mu_2\nu}(p,-p_1,-p+p_1)
\,^{\ast}\tilde{\cal D}^{\nu\nu^{\prime}}\!(p-p_1)
\,^{\ast}\Gamma_{\nu\lambda\sigma}(p-p_1,-p_1^{\prime},-p+p_1+p_1^{\prime})
\]
\[
\times
\,^{\ast}\tilde{\cal D}^{\lambda\lambda^{\prime}}\!(p_1^{\prime})
v_{{\lambda}^{\prime}}
\,^{\ast}\tilde{\cal D}^{\sigma\sigma^{\prime}}\!(p-p_1-p_1^{\prime})
v_{\sigma^{\prime}}
\,\delta(v\cdot p_1^{\prime})\,\delta(v\cdot(p-p_1))dp_1^{\prime}
\]
corresponds to the last term (in parenthesis) in Fig.\,\ref{FIG4}. 
By virtue of a property of three-gluon HTL-amplitude:
$\!\,^{\ast}\Gamma_{\mu\mu_1\mu_2}(p,-p_1,-p_2)=-
\!\,^{\ast}\Gamma_{\mu\mu_2\mu_1}(p,-p_2,-p_1)$,
the integrand is odd under the interchange of
$p_1^{\prime}\rightarrow p - p_1 - p_1^{\prime}$ and integrates to zero.

It is clear that higher powers in $Q_0$ in the expansion (\ref{eq:4w}) is
associated with soft loop corrections of higher orders, and they are suppressed
relative to linear term
$(\delta\!K_{\mu\mu_1}^{aa_1}/\delta Q_0^b)_{Q_0=0}$ as $(g^2)^2$ etc.
The partial coefficients functions  $\stackrel{\scriptscriptstyle{\,(r)}}
{\displaystyle{K}}_{\mu\mu_1}\!({\bf v}\vert\,p,-p_1),\,r\geq 2$
(more precisely, an effective
current connected with them) can be interpretated as ``dressing''
of initial current of hard test color particle or simple of a particle caused
by interaction of this current with hot bath. For derivation of total
probability of the process of the nonlinear Landau damping (process of elastic
scattering for $n = 0$ in Eq. (\ref{eq:2y})), we must, generally speaking,
summarize a series in expansion in a color charge degree $Q_0$. It can be
considered as a replacement of initial test particle by some effective
quasiparticle on which in fact the scattering process of soft-gluon wave
takes place\footnote{Besides it can be interpretated classical correction
loops as change of state not test color particle, but properties of a medium
by the action of color field of test particle and interaction of initial
``no dressed'' current of particle with such modified medium.}.
It is successfully that even in the case of highly excited gluon plasma
(see next section) soft loop corrections are suppressed and therefore in
leading order we can neglect by it.

Finally we point to one interest relation, which exists between partial
coefficient functions $\stackrel{\scriptscriptstyle{\,(2)}}
{\displaystyle{K}}_{\mu\mu_1}\!({\bf v}\vert\,p,-p_1),\,
\stackrel{\scriptscriptstyle{\,(1)}}
{\displaystyle{K}}_{\mu\mu_1\mu_2}\!({\bf v}\vert\,p,-p_1,-p_2)$
and four\,--\,soft-gluon effective subamplitude
$\!\,^{\ast}\tilde{\Gamma}_{\mu \mu_1 \mu_2 \mu_3}(p,-p_1,-p_2,-p_3)$
defined by equation (I.5.6).
For its derivation we contract the expression (\ref{eq:4o}) with
$^{\ast} \tilde{\cal D}^{\mu_1 \mu_1^{\prime}} (p_1) v_{\mu_1^{\prime}}$,
integrate over $d p_1$ and replace variables $p_1 \rightarrow p_1^{\prime},
\,p_2 \rightarrow p_1$ and indices: $\mu_1 \rightarrow \nu,\,
\mu_2 \rightarrow \mu_1$ etc. After some algebraic transformations with use of
Eq. (I.5.12) and explicit form $\stackrel{\scriptscriptstyle{\,(2)}}
{\displaystyle{K}}_{\mu\mu_1}\!({\bf v}\vert\,p,-p_1)$ we obtain desired
relation
\[
\stackrel{\scriptscriptstyle{\,(2)}}
{\displaystyle{K}}_{\mu\mu_1}\!({\bf v}\vert\,p,-p_1) =
-\int\!
\frac{v_\mu v_{\mu_1}}{(v\cdot (p_1 + p_1^{\prime}))^2}
\,(v_{\nu}\!\,^{\ast}\tilde{\cal D}^{\nu\nu^{\prime}}\!(p+p_1^{\prime})
v_{\nu^{\prime}})\,
\delta(v\cdot p_1^{\prime}) dp_1^{\prime}
\]
\begin{equation}
+\int
\!\stackrel{\scriptscriptstyle{\,(1)}}
{\displaystyle{K}}_{\mu\nu\mu_1}\!({\bf v}\vert\,p,-p_1^{\prime},-p_1)
\,^{\ast}\tilde{\cal D}^{\nu\nu^{\prime}}\!(p_1^{\prime})
v_{\nu^{\prime}}\,
\delta(v\cdot p_1^{\prime})dp_1^{\prime}
\label{eq:4s}
\end{equation}
\[
+\,\frac{1}{2}\!\int\!\!
\,^{\ast}\tilde{\Gamma}_{\mu\nu\mu_1\lambda}
(p,-p_1^{\prime},-p_1,-p+p_1+p_1^{\prime})
\,^{\ast}\tilde{\cal D}^{\lambda\lambda^{\prime}}\!(p-p_1-p_1^{\prime})
v_{{\lambda}^{\prime}}
\,^{\ast}\tilde{\cal D}^{\nu\nu^{\prime}}\!(p_1^{\prime})
v_{\nu^{\prime}}
\,\delta(v\cdot p_1^{\prime}) dp_1^{\prime}.
\]
The existence of such relation is not difficult for understanding by using
diagramm representation of the functions entering in it.
Thus the second term on the right-hand
side of Eq.\,(\ref{eq:4s}) corresponds to procedure of closing one of soft
plasmon legs in Fig.\,\ref{FIG3} (in our case the legs with soft momentum $p_1$)
on hard test particle, as it depicted in Fig.\,\ref{FIG4}. The last term on the
right-hand side of (\ref{eq:4s}) is connected with closing two external soft legs
of diagramm on the left hand side in Fig.\,I.1 and attaching hard line of
test particle to close loop. The physical meaning of the first term on the
right-hand side of (\ref{eq:4s}) is less clear (see however below).
It is evident that relation of such a kind exists and for higher partical
coefficient function and effective subamplitudes. It can be supposed that they
are a consequence of gauge invariance (more precisely, covariance) of total
effective current (\ref{eq:4q}) with respect to gauge transformations
concerned both a field $A_{\mu}^{(0)a}(p)$ and a color charge $Q_{0}^{a}$.
We still return to analysis of relation (\ref{eq:4s}) in our next work
\cite{mark3} in consideration of the process of nonlinear Landau damping
with regard to rescattering of test particle off another hard test particle,
where a reason of appearing the first term on the right-hand side of
(\ref{eq:4s}) becomes more explicit.

\section{\bf Characteristic amplitudes of the soft gluon field}
\setcounter{equation}{0}

In this Section we shall estimate for which typical amplitude of the soft
gluon field, the contribution of the nonlinear Landau damping process to a
generalized decay and regeneration rates,
$\Gamma_{\rm d} [N_{\bf p}^l]$ and $\Gamma_{\rm i} [N_{\bf p}^l]$, will be
leading and for which value all terms in the expansions (\ref{eq:2w})
will be of the same order in $g$. As a preliminary we define a general
expression for scattering probability
${\it w}_{2n + 2}={\it w}_{2n + 2}({\bf v}\vert\,{\bf p},
{\bf p}_1,\ldots ,{\bf p}_n; {\bf p}_{n + 1}, \ldots , {\bf p}_{2n + 1})$.
Since the basic moments of computation of the scattering probability by a hard particle repeat
reasoning used for deriving the scattering probability in pure plasmon-plasmon
scattering in Paper I, Section 4, here, we restrict our consideration to
brief scheme of calculation of ${\it w}_{2n+2}$.

Initial expression for deriving ${\it w}_{2n+2}$ is Eq.\,(\ref{eq:3u}) defining
the emitted radiant power of the longitudinal waves ${\cal I}^l$. The current
entering into correlation function is given by
\begin{equation}
\tilde{J}_{Q\,\mu}^{a}[A^{(0)}](p) =
\frac{g}{(2\pi)^3}\,Q_0^av_{\mu}\delta(v\cdot p) +
\sum_{s=1}^{\infty}
\tilde{J}_{Q\mu}^{(s)a}[A^{(0)}]({\bf v},p),
\label{eq:5q}
\end{equation}
where
\[
\tilde{J}_{Q\mu}^{(s)a}[A^{(0)}]({\bf v},p)=
\frac{1}{s!}\,\frac{g^{s+1}}{(2\pi)^3}\int\!
K_{\mu\mu_1\ldots\mu_s}^{aa_1\ldots a_s}({\bf v}\vert\,p,-p_1,\ldots,-p_s)
A^{(0)a_1\mu_1}(p_1)\ldots A^{(0)a_s\mu_s}(p_s)
\]
\[
\times\delta(v\cdot(p-\sum_{i=1}^{s}p_i))\prod_{i=1}^{s}dp_i.
\]
Here, in the coefficient functions $K^{(s)}$ we leave only the term leading
order in
$g$ in the expansion (\ref{eq:4w}), i.e. linear in color charge $Q_0$
\[
K_{\mu\mu_1\ldots\mu_s}^{aa_1\ldots a_s}
({\bf v}\vert\,p,-p_1,\ldots,-p_s)
\simeq K_{\mu\mu_1\ldots\mu_s}^{aa_1\ldots a_s\,b}
({\bf v}\vert\,p,-p_1,\ldots,-p_s)Q_0^b.
\]

In substitution of the current expansion (\ref{eq:5q}) into a correlation
function in integrand of Eq.\,(\ref{eq:3u}) we face the product of two series.
As in deriving probability of plasmon decays, in this product it is necessary
to leave only a sum of a product of terms having the same order in power
of a potential $A_{\mu}^{(0)}$. In this case only a desired $\delta$-function
entering to integration measure $\int d{\cal T}^{(2n+1)}$, defining the
generalized resonance condition (\ref{eq:2u}), arise. Thus the emitted
radiant power (\ref{eq:3u}), taking into account emission caused by the processes
of scattering plasmons off hard test particle, can be represented in the form
of expansion
\[
{\cal I}^l = {\cal I}^{l(0)} + \sum_{s=1}^{\infty}{\cal I}^{l(s)},
\]
where
\[
{\cal I}^{l(0)} =
-\,\frac{g^2}{2(2\pi)^2}\!
\left(
\int\!\frac{d\vert{\bf k}\vert}{2\pi^2}\vert {\bf k}\vert^2 f_{\bf k}\right)
\!\int\!\frac{d\Omega_{\bf v}}{4\pi}\!\int\!d\omega\,\vert\omega\vert
\!\int\!\!d{\bf p}\,(v_{\mu}\tilde{Q}^{\mu\mu^{\prime}}(p)v_{{\mu}^{\prime}})
\delta (v\cdot p)\delta({\rm Re}\,^{\ast}\!\Delta^{-1\,l}(p)),
\]
and
\begin{equation}
{\cal I}^{l\,(s)} = -\,\frac{1}{4\pi}
\left(
\int\!\frac{d\vert{\bf k}\vert}{2\pi^2}\vert {\bf k}\vert^2 f_{\bf k}\right)
\lim\limits_{\tau\rightarrow\infty}
\frac{1}{\tau}\,\frac{g^{2s+2}}{(s!)^2}
\left(\frac{1}{N_c}\!\int\!dQ\,Q_0^bQ_0^{b^{\prime}}\right)
\label{eq:5w}
\end{equation}
\[
\times\!\int\!\frac{d\Omega_{\bf v}}{4\pi}
\int\!dp\,\vert\omega\vert\,\delta^{a a^{\prime}}
\tilde{Q}^{\mu\mu^{\prime}}(p)
K^{\ast\, a a_1\ldots a_s b}_{\mu\mu_1\ldots\mu_s}
({\bf v}\vert\,p,-p_{1},\ldots,-p_{s})
K^{a^{\prime} a^{\prime}_1\ldots a^{\prime}_s
b^{\prime}}_{\mu^{\prime}\mu^{\prime}_1\ldots\mu^{\prime}_s}
({\bf v}\vert\,p,-p_{1}^{\prime},\ldots,-p_{s}^{\prime})
\]
\[
\times\langle A^{\ast(0)a_1\mu_1}(p_{1})\ldots A^{\ast(0) a_s\mu_s}(p_{s})
A^{(0)a^{\prime}_1\mu^{\prime}_1}(p^{\prime}_{1})\ldots
A^{(0)a^{\prime}_s\mu^{\prime}_s}(p^{\prime}_{s})\rangle
\delta(v\cdot (p - \sum_{i=1}^{s}p_{i}))
\]
\[
\times\delta(v\cdot (p - \sum_{i=1}^{s}p^{\prime}_{i}))
\delta({\rm Re}\,^{\ast}\!\Delta^{\!-1\,l}(p))\!
\prod_{i=1}^{s}dp_{i}dp_{i}^{\prime}.
\]
The first term ${\cal I}^{(0)}$ is connected with Cherenkov radiation
(the linear Landau damping) which is kinematically forbidden in hot gluon
plasma and therefore, this term can be setting zero. Let us consider
the remaining terms ${\cal I}^{l(s)},\,s\geq 1$.

The integration over color charge in (\ref{eq:5w}) is trivial. Furthermore
we write out decoupling of the $2s$th-order correlator on the right-hand side
of Eq.\,(\ref{eq:5w}) in terms of pair correlators by rule defined by
us in Paper I. Employing a condensed notion,
$A_1\equiv A_{\mu_1}^{(0)a_1}(p_1)$ etc, we have
\begin{equation}
\langle A^{\ast(0)}_{1}\ldots A^{\ast(0)}_{s}
A^{(0)}_{1^{\prime}}\ldots A^{(0)}_{s^{\prime}}\rangle =
(s!)^2
\langle A^{\ast(0)}_{1}A^{(0)}_{1^{\prime}}\rangle\ldots
\langle A^{\ast(0)}_{s}A^{(0)}_{s^{\prime}}\rangle
\label{eq:5e}
\end{equation}
\[
=(s!)^2\prod_{i=1}^{s}\delta^{a_i a_i^{\prime}}
\tilde{Q}^{\mu_i \mu_i^{\prime}}(p_i) I^l(p_i)\delta(p_i - p_i^{\prime})
\]
\[
\simeq
(s!)^2\prod_{i=1}^{s}\delta^{a_i a_i^{\prime}}
\tilde{Q}^{\mu_i \mu_i^{\prime}}(p_i)
\!\left(-\,\frac{1}{(2\pi)^3}
\frac{{\rm Z}_l({\bf p}_i)}{2\omega_{{\bf p}_i}^l}\right)\!
\{N_{{\bf p}_i}^l \delta (\omega_i - \omega_{{\bf p}_i}^l)
+ N_{-{\bf p}_i}^l \delta (\omega_i + \omega_{{\bf p}_i}^l)\}
\delta(p_i - p_i^{\prime}).
\]
Here, in the last line, within the accepted accuracy, we replace equilibrium
spectral densities $I^l(p_i)$ by off-equlibrium ones in the
Wigner form (I.3.15): $I^l(p_i)\rightarrow I^l(p_i,x_i)$, slowly depending
on $x$, futhermore, we take the functions $I^l(p_i,x_i)$ in the form of
{\it the quasiparticle approximation} (Eq.\,(I.4.12))
and pass from functions $I_{{\bf p}_i}^l$ to the plasmon number density
\[
N_{{\bf p}_i}^l = -\,(2\pi)^3\,2\omega_{{\bf p}_i}^l {\rm Z}_l^{-1}({\bf p}_i)
I_{{\bf p}_i}^l.
\]

Substituting expression (\ref{eq:5e}) into (\ref{eq:5w}), performing an
integration over $\prod_{i=1}^{s}dp_i^{\prime}$ and taking into account the
relation
\begin{equation}
\Bigr[\delta(v\cdot(p - \sum_{i=1}^{s}p_{i}))\Bigl]^2 =
\frac{1}{2\pi}\,\tau\delta(v\cdot(p-\sum_{i=1}^{s}p_{i})),
\label{eq:5ee}
\end{equation}
we obtain instead of (\ref{eq:5w})
\begin{equation}
{\cal I}^{l\,(s)} = \frac{(-1)^{s+1}}{2d_A}
\!\left(
\int\!\frac{d\vert{\bf k}\vert}{2\pi^2}\vert {\bf k}\vert^2 f_{\bf k}\right)
\!\int\!\frac{d\Omega_{\bf v}}{4\pi}\!
\int\!d\omega\,\vert\omega\vert
\!\int \prod_{i=1}^{s}\!d\omega_i\!
\int\!\frac{d{\bf p}}{(2\pi)^3}\!\int\prod_{i=1}^{s}
\frac{d{\bf p}_i}{(2\pi)^3}
\label{eq:5r}
\end{equation}
\[
\times\,2\pi\delta(v\cdot(p - \sum_{i=1}^s p_{i}))
{\rm T}^{\ast\,a a_1 \ldots a_s b}
({\bf v}\vert\,{\bf p}, -{\bf p}_1,\ldots,-{\bf p}_s)
{\rm T}^{a a_1 \ldots a_s b}
({\bf v}\vert\,{\bf p}, -{\bf p}_1,\ldots,-{\bf p}_s)
\]
\[
\times\{\delta (\omega - \omega_{{\bf p}}^l)
+ \delta (\omega + \omega_{{\bf p}}^l)\}
\prod_{i=1}^s\{N_{{\bf p}_i}^l \delta (\omega_i - \omega_{{\bf p}_i}^l)
+ N_{-{\bf p}_i}^l \delta (\omega_i + \omega_{{\bf p}_i}^l)\}.
\]
Here, we introduce the function
\begin{equation}
{\rm T}^{a a_1 \ldots a_s b}
({\bf v}\vert\,{\bf p}, -{\bf p}_1,\ldots,-{\bf p}_s)= g^{s+1}\!
\left(\frac{{\rm Z}_l({\bf p})}{2\omega_{\bf p}^l}\right)^{\!1/2}\!
\Biggl(\frac{\tilde{u}^{\mu}(p)}{\sqrt{\bar{u}^2(p)}}\Biggr)\times
\label{eq:5t}
\end{equation}
\[
\times\prod_{i=1}^{s}
\left(\frac{{\rm Z}_l({\bf p}_i)}{2\omega_{{\bf p}_i}^l}\right)^{\!1/2}\!
\Biggl(\frac{\tilde{u}^{\mu_i}(p_i)}{\sqrt{\bar{u}^2(p_i)}}\Biggr)
K^{a a_1\ldots a_s b}_{\mu\mu_1\ldots\mu_s}
({\bf v}\vert\,p,-p_{1},\ldots,-p_{s})\mid_{\rm on-shell}
\]
representing the interaction matrix element for the process of nonlinear
interaction of $s+1$ soft longitudinal oscillations with hard test particle.
Since external soft-gluon legs lie on the plasmon mass-shell, then account
must be taken of the fact that resonance conditions (\ref{eq:2u}) admit
scattering processes with even number of plasmons, i.e. it is necessary to set
in Eq.\,(\ref{eq:5r}) $s=2n+1,\,n=0,1,\ldots\,.$

Furthermore, we multiply out terms in curly brackets in integrand of
Eq.\,(\ref{eq:5r}) and use combinatorial transformation identical with
transformation in Paper I. Confronting such an obtained expression for
${\cal I}^{l\,(2n+1)}$ with corresponding terms in expansion (\ref{eq:3i}),
one identifies the required probability ${\it w}_{2n+2}$:
\begin{equation}
{\it w}_{2n+2}
({\bf v}\vert\,{\bf p},{\bf p}_1,\ldots,
{\bf p}_n;{\bf p}_{n+1},\ldots,{\bf p}_{2n+1})
\label{eq:5y}
\end{equation}
\[
=\frac{1}{d_A}\Big\{{\rm T}^{\ast \{a\}b}_{\bf v}{\rm T}_{\bf v}^{\{a\}b}
+\sum_{1\leq i_1\leq n}^{({\bf p}_{i_1})}
{\rm T}^{\ast\{a\}b}_{\bf v}{\rm T}_{\bf v}^{\{a\}b}
+\sum_{1\leq i_1<i_2\leq n}^{({\bf p}_{i_1},\,{\bf p}_{i_2})}
{\rm T}_{\bf v}^{\ast\{a\}b}{\rm T}_{\bf v}^{\{a\}b}+
\ldots+\sum^{({\bf p}_1,\ldots,{\bf p}_n)}
{\rm T}_{\bf v}^{\ast\{a\}b}{\rm T}_{\bf v}^{\{a\}b}\Big\},
\]
where ${\rm T}_{\bf v}^{\{a\}b}\equiv {\rm T}^{\{a\}b}
({\bf v}\vert\,{\bf p},{\bf p}_1,\ldots,{\bf p}_n,-{\bf p}_{n+1},
\ldots, -{\bf p}_{2n+1})$, and for brevity we enter multi-index notation
$\{a\}=(a,a_1,\dots,a_{2n+1})$.
The summing symbol $\sum\limits_{1\leq i_1\leq n}^{({\bf p}_{i_1})}$
denotes summing over all possible momentum interchange
${\bf p}_{i_1},\;1\leq i_1\leq n$, by momentum $(-{\bf p}_{j_1}),\;
n+1\leq j_1\leq 2n+1$. The symbol
$\sum_{1\leq i_1<i_2\leq n}^{({\bf p}_{i_1},{\bf p}_{i_2})}$ analogously
denotes a summing over all possible interchange of momenta pair
$({\bf p}_{i_1},{\bf p}_{i_2}),\;1\leq i_1\!<\!i_2\leq n$, by momenta pair
$(-{\bf p}_{j_1},-{\bf p}_{j_2})$, where $n+1\leq j_1\!<\!j_2\leq 2n+1$ etc.

With explicit expression for scattering probability (Eq.\,(\ref{eq:5y}))
in hand, now we can turn to the estimation of typical amplitudes
of the soft-gluon field. First of all we estimate an order of matrix element
${\rm T}^{aa_1\ldots_{2n+1}b}$. In the soft region of the momentum scale
the following estimation results from expression (\ref{eq:5t})
\[
{\rm T}^{a a_1 \ldots a_{2n + 1}b} \sim \frac{1}{(gT)^{(n+1)}} \,
K_{\mu\mu_1\ldots\mu_{2n+1}}^{a a_1\ldots a_{2n + 1}b}.
\]
The order of the coefficient function
$K_{\mu\mu_1\ldots\mu_{2n+1}}^{a a_1\ldots a_{2n + 1}b}$
can be estimated from arbitrary tree diagram
with amputate $2n+2$ soft external legs. Let us consider, for example,
the diagram drawn in Fig.\ref{FIG5}.
\begin{figure}[hbtp]
\centering
\includegraphics[height=3cm]{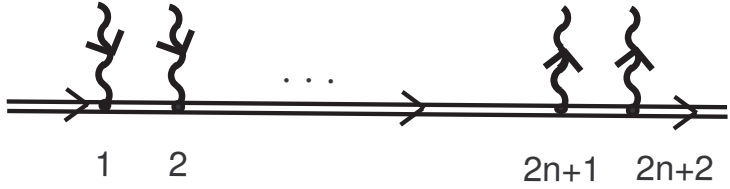}
\caption{\small
The typical tree-level Feynman diagram for scattering $(2n+2)$-plasmons
off hard test particle}
\label{FIG5}
\end{figure}
The factor $g$ is related to vertices, and eikonal propagator $1/(v\cdot p)
\sim 1/gT$ is related to an internal lines. From simple power counting of the
diagram it follows an estimation
\begin{equation}
K_{\mu\mu_1\ldots\mu_{2n+1}}^{a a_1\ldots a_{2n + 1}b}
\sim g^{2n+2}N_c^{(2n+1)/2}\,\frac{1}{(gT)^{2n+1}}\,d_A^{1/2},
\label{eq:5yy}
\end{equation}
and thus
\[
{\rm T}^{a a_1 \ldots a_{2n + 1}b}\sim g^{2n+2}N_c^{(2n+1)/2}
\,\frac{1}{(gT)^{3n+2}}\,d_A^{1/2}.
\]
Using this estimation we derive from Eq.\,(\ref{eq:5y})
\begin{equation}
{\it w}_{2n+2}\sim g^{4n+4}\frac{1}{(gT)^{6n+4}}\,N_c^{2n+1}.
\label{eq:5u}
\end{equation}
Furthermore integration measure has an estimation
\begin{equation}
{\rm d}{\cal T}^{(2n+1)}\sim(gT)^{6n+2}.
\label{eq:5i}
\end{equation}
Power counting of the decay and regenerating rates (\ref{eq:2e}) and
(\ref{eq:2r}) with regard to (\ref{eq:5u}) and (\ref{eq:5i})
gives the following estimation
\[
\Gamma_{\rm d,\,i}^{(2n+1)\,a a^{\prime}}\sim g^{4n+2}N_c^{2n+1}T
(N_{\bf p}^l)^{2n+1}.
\]
If we now set $N_{\bf p}^l\,\sim\,\displaystyle\frac{1}{g^{\rho}},\;
\rho\!>\!0,$ then from the last expression it follows
\begin{equation}
\Gamma_{\rm d,\,i}^{(2n+1)\,a a^{\prime}}\sim g^{(2n+1)(2-\rho)}
N_c^{2n+1}T.
\label{eq:5o}
\end{equation}
For small value of oscillation amplitude (Eq.\,(I.6.5)) we have
\[
\Bigl(\Gamma_{\rm d,\,i}^{(2n+1)\,a a^{\prime}}\Bigr)_{A\sim\sqrt{g}T}
\sim g^{2n+1}N_c^{2n+1}T,\quad n=0,1,\dots\,.
\]
From this estimation it can be seen that each subsequent term in the
functional expansions (\ref{eq:2w}) is suppressed by more power of $g^2$.
Thus for low excited state of plasma, corresponding to level of thermal
fluctuations at soft scale \cite{bla2}, we can only restrict
ourselves to first leading term in the expansions (\ref{eq:2w})
\begin{equation}
\Bigl(\Gamma_{\rm d,\,i}^{(1)\,a a^{\prime}}\Bigr)_{A\sim\sqrt{g}T}
\sim gN_cT,
\label{eq:5p}
\end{equation}
describing the nonlinear Landau damping processes.

In the other case of a strong field, $\vert A_{\mu}(X)\vert\sim T$, for $\rho=2$
from estimation (\ref{eq:5o}) it follows
\begin{equation}
\Bigl(\Gamma_{\rm d,\,i}^{(2n+1)\,a a^{\prime}}\Bigr)_{A\sim T}
\sim N_c^{2n+1}T,\quad n=0,1,\dots\,
\label{eq:5a}
\end{equation}
i.e. the generalized rates are independent of $g$. All terms in the
expansions (\ref{eq:2w}) become of the same order in magnitude, and the problem
of resummation of all relevant contributions arises.

In closing this Section we compare estimations (\ref{eq:5p}) and (\ref{eq:5a})
with similar ones for generalized velocities in the case of pure plasmon-plasmon
interactions. For low excited state of plasma, as shown in Paper I, the first
leading terms in functional expansions of $\Gamma_{\rm d,\,i}$ (according to
Eq.\,(I.6.6)) have following estimation
\[
\Bigl(\Gamma_{\rm d,\,i}^{(3)\,a a^{\prime}}\Bigr)_{A\sim\sqrt{g}T}
\sim g^2N_c^2T.
\]
From this estimation and (\ref{eq:5p}) we see that four-plasmon decay process is
suppresed by $g$. Such, when the soft-gluon fields are thermal fluctuations,
the nonlinear Landau damping process is a basic process determining
damping of plasma waves.

A situation is qualitatively changed, when a system is highly excited.
In this case
we have an estimation for generalized rates, determining $2n+2$-plasmon decay
processes, from Eq.\,(I.6.6)
\begin{equation}
\Bigl(\Gamma_{\rm d,\,i}^{(2n+1)\,a a^{\prime}}\Bigr)_{A\sim T}
\sim \frac{1}{g}\,N_c^{2n}T,\quad n=1,2,\dots\,.
\label{eq:5s}
\end{equation}
Thus, from estimations (\ref{eq:5a}) and (\ref{eq:5s}) for sufficiently large
intensity of plasma excitations, the processes of pure plasmon-plasmon
interactions becomes dominant.
There is certain intermediate level of excitations, when these two
different scattering processes give the same contribution to damping of
plasmon excitations. Considering the plasmos number density as
$N_{\bf p}^l\sim 1/g^{\rho}$ and comparing estimations (\ref{eq:5o})
with (I.6.6), one can estimate roughly the value $\alpha$, for which the
orders of generalized velocities for these processes are comparable
\[
\rho=\rho^{\ast}\simeq\frac{3}{2}+\frac{1}{2}
\left(\frac{\ln N_c}{\ln g}\right),\quad g\ll 1,
\]
this corresponds to $\vert A_{\mu}(X)\vert\sim(g/N_c)^{1/4}T$.

\section{\bf Gauge invariance of the matrix elements
${\rm T}^{aa_1\ldots a_sb}$}
\setcounter{equation}{0}

Let us consider the problem of a gauge invariance of the interaction of matrix
elements defined in previous section. From Eq.\,(\ref{eq:5t}) and expansions
(\ref{eq:4u}), (\ref{eq:4i}), (\ref{eq:4p}) and (\ref{eq:4a})
with respect to antisymmetric
structural constant we see that a proof of gauge invariance is reduced to
proof of the gauge invariance of functions presenting convolution of projector
$\tilde{u}_{\mu}(p) = p^2(u_{\mu}(p\cdot u)-p_{\mu})/(p\cdot u)$
with partial coefficient functions taken on mass-shell\footnote{In Paper I
we have used not entirely correct definition of a projector
$\tilde{u}_{\mu}(p)$. It distincts from above-written by a sign.
This in turn led to mistaken statement in Section 8, Paper I, on
non-physical nature of
matrix elements for plasmon decays with regard to odd number of
plasmons. These matrix elements are related to actual physical
processes, which would arise if they not kinematically forbidden by laws
of conservation of energy and momentum by virtue of specific of spectrum
of longitudinal oscillations in hot QCD plasma.}, i.e.
\begin{equation}
\stackrel{\scriptscriptstyle{\,(r)}}{\displaystyle{K}}
\!({\bf v}\vert\,p,-p_1,\ldots,-p_s)\equiv\,
\stackrel{\scriptscriptstyle{\,(r)}}{\displaystyle{K}}_{\mu\mu_1\dots\mu_s}
\!({\bf v}\vert\,p,-p_1,\ldots,-p_s)
\tilde{u}^{\mu}(p)\tilde{u}^{\mu_1}(p_1)\ldots
\tilde{u}^{\mu_s}(p_s)\vert_{\rm on-shell}.
\label{eq:6q}
\end{equation}
We will show below that these functions are reduced to simple forms of
(I.8.1) and (I.8.3) types, obtained in calculation of similar convolution with
effective subamplitudes
$\,^{\ast}\tilde{\Gamma}_{\mu\mu_1\ldots\mu_s}(p,-p_1,\ldots,-p_s)$.

Let us consider the function (\ref{eq:6q}) for $s=1$
in linear approximation over color charge $Q_0$, i.e. for $r=1$.
Lorentz tensor
$\stackrel{\scriptscriptstyle{\,(1)}}{\displaystyle{K}}_{\mu\mu_1}$
is defined by Eq.\,(\ref{eq:3g}).
Using effective Ward identities for HTL-amplitudes \cite{bra1} and
mass-shell condition, we derive
\begin{equation}
\stackrel{\scriptscriptstyle{\,(1)}}{\displaystyle{K}}\!({\bf v}\vert\,p,-p_1)
= p^2p_1^2\biggl\{
\frac{1}{v\cdot p_1} +
\,^\ast\Gamma_{00\nu}(p,-p_1,-p+p_1)
\,^\ast\!\tilde{\cal D}^{\nu\nu^{\prime}}(p-p_1)v_{\nu^{\prime}}
\!\!\left.\biggr\}\right|_{\rm on-shell}.
\label{eq:6w}
\end{equation}
It is easy to check that the term with a gauge parameter in Eq.\,(\ref{eq:6w})
vanishes on mass-shell. Now we consider a function
$\stackrel{\scriptscriptstyle{\,(1)}}{\displaystyle{K}}\!({\bf v}\vert\,p,-p_1)$
in covariant gauge. For this purpose we perform the replacements (I.8.2) of
projector and propagator on the left-hand side of Eq.\,(\ref{eq:6w}). Then
after analogous computations we lead to the same expression on the
right-hand side of Eq.\,(\ref{eq:6w}). Thus, we have shown that at least in
the class of temporal and covariant gauges the matrix element for the
nonlinear Landau damping process is gauge-invariant.

We can assume that a similar reduction holds for an arbitrary partial
coefficient function. However, as in the case of a pure plasmon-plasmon
interaction a proof of this statement is imposible in general case by
reason of absence of general expression
$\stackrel{\scriptscriptstyle{\,(r)}}{\displaystyle{K}}_{\mu\mu_1\ldots\mu_s}$.
Similar to example in Paper I (Section 8) the only thing that we can make is to
consider a contraction (\ref{eq:6q}) for three-point partial coefficient
function, exact form of which (\ref{eq:4o}) is known. Slightly cumbersome,
but not complicated computations with the use of the effective Ward identities
and mass-shell condition lead to the following expression
\[
\stackrel{\scriptscriptstyle{\,(1)}}{\displaystyle{K}}
\!({\bf v}\vert\,p,-p_1,-p_2) = -\,p^2p^2_1p^2_2\biggl\{
\frac{1}{(v\cdot p_2)(v\cdot (p_1 + p_2))}
\]
\begin{equation}
+\,\frac{1}{v\cdot (p_1 + p_2)}
\,^{\ast}\Gamma_{00\nu}(-p_1,-p_2,p_1+p_2)
\,^{\ast}\tilde{\cal D}^{\nu\nu^{\prime}}(p_1 + p_2)v_{\nu^{\prime}}
\label{eq:6r}
\end{equation}
\[
-\,^{\ast}\Gamma_{000\nu}(p,-p_1,-p_2,-p+p_1+p_2)\,^{\ast}
\tilde{\cal D}^{\nu\nu^{\prime}}(p-p_1-p_2)v_{{\nu}^{\prime}}
\]
\[
+\,^{\ast}\Gamma_{00\nu}(p,-p_1,-p+p_1)\,^{\ast}
\tilde{\cal D}^{\nu\nu^{\prime}}(p-p_1)
\!\stackrel{\scriptscriptstyle{\,(1)}}
{\displaystyle{K}}_{\nu^{\prime}0}\!\!({\bf v}\vert\,p-p_1,-p_2)
\]
\[
-\!\,^{\ast}\Gamma_{0\nu\lambda}(p,-p+p_1+p_2,-p_1-p_2)
\!\,^{\ast}\tilde{\cal D}^{\nu\nu^{\prime}}\!(p-p_1-p_2)v_{{\nu}^{\prime}}
\!\!\,^{\ast}\tilde{\cal D}^{\lambda\lambda^{\prime}}\!(p_1+p_2)\!
\,^{\ast}\Gamma_{\lambda^{\prime}00}(p_1+p_2,-p_1,-p_2)\!
\left.\!\biggr\}\right|_{\rm on-shell}.
\]
Here, it is also easy to check that all terms with a gauge parameter vanish
on mass-shell. In deriving (\ref{eq:6r}) we use relation
$p^{\mu}\!\!\stackrel{\scriptscriptstyle{\,(1)}}
{\displaystyle{K}}_{\mu\mu_1}\!({\bf v}\vert\,p,-p_1)\vert_{\rm on-shell}=0$,
that directly follows from (\ref{eq:3g}).

If we now perform a replacements (I.8.2) on the left-hand side (\ref{eq:6r}),
then after analogous computations we lead to the same expression on the
right-hand side of (\ref{eq:6r}). The function (\ref{eq:6r}) by virtue of
Eqs.\,(\ref{eq:4i}) and (\ref{eq:5t}) defines the matrix element for
scattering process including three soft-gluon waves and hard test particle.
This process is kinematically forbidden, i.e. the $\delta$-function
\[
\delta(\omega-\omega_1-\omega_2 - {\bf v}\cdot ({\bf p}-{\bf p}_1-{\bf p}_2))
\]
has no support on the plasmon mass-shell. This suggest that all scattering
processes involving odd number of plasmons are kinematically forbidden.
In an expansion of the color current (\ref{eq:5q}) all effective currents
$\tilde{J}_{Q\,\mu}^{(s)a}$ with odd number of free fields $A_{\mu}^{(0)}$
by $\delta$-functions are equal to zero on mass-shell. The reflection
of this fact is, in particular, a choice of the structure for generalized rates
$\Gamma_{\rm d}$ and $\Gamma_{\rm i}$ in collision term of kinetic equation
(\ref{eq:2q}).

As in Paper I one can note, that in spite of the fact that result
(\ref{eq:6r}) is of only pure methodological meaning, nevertheless two
examples (\ref{eq:6w}) and (\ref{eq:6r}) provide a reason to use
considerably simple expressions of (\ref{eq:6w}) type for all
$(2n + 2)$-matrix elements ${\rm T}^{a a_1\ldots a_{2n+1}b}$
in particular calculations.

All above-mentioned reasoning is concerned with a part of coefficient functions
linear over color charge $Q_0$. Genereally speaking, far from
obviosly, that a reduction of (\ref{eq:6w}) and (\ref{eq:6r}) types takes
place and for partial coefficient functions relating to expansion terms
(\ref{eq:4w}) for arbitrary order in $Q_0$. Let us consider, for example,
partial coefficient function (\ref{eq:4a}), that is connected with term in
the expansion of $K_{\mu\mu_1}^{aa_1}({\bf v}\vert\,p,-p_1)$ quadratic over
color charge. Using effective Ward identities
for HTL-amplitudes and mass-shell conditions, we derive from (\ref{eq:4a})
\[
\stackrel{\scriptscriptstyle{\,(2)}}{\displaystyle{K}}_{\mu\mu_1}
\!({\bf v}\vert\,p,-p_1)
\tilde{u}^{\mu}(p)\tilde{u}^{\mu_1}(p_1)
\Big\vert_{\rm on-shell}
= p^2p_1^2\!\int\!\bigg\{
\frac{1}{(v\cdot p_1)(v\cdot (p_1 + p_1^{\prime}))}
\,(v_{\nu}\!\,^{\ast}\tilde{\cal D}^{\nu\nu^{\prime}}\!(p_1^{\prime})
v_{\nu^{\prime}})
\]
\begin{equation}
-\,\frac{1}{v\cdot (p_1 + p_1^{\prime})}\,v_{\nu}
\!\,^{\ast}\tilde{\cal D}^{\nu\nu^{\prime}}\!(p_1 + p_1^{\prime})
\!\stackrel{\scriptscriptstyle{\,(1)}}
{\displaystyle{K}}_{{\nu}^\prime 0}\!({\bf v}\vert\,p_1+p_1^{\prime},-p_1)
\label{eq:6t}
\end{equation}
\[
+\,^{\ast}\Gamma_{0\nu_1\nu_2 0}
(p,-p_1^{\prime},-p+p_1+p_1^{\prime},-p_1)
\,^{\ast}\tilde{\cal D}^{\nu_1\nu_1^{\prime}}(p_1^{\prime})v_{{\nu_1}^{\prime}}
\!\,^{\ast}\tilde{\cal D}^{\nu_2\nu_2^{\prime}}
(p-p_1-p_1^{\prime})v_{{\nu_2}^{\prime}}
\]
\[
-\,^{\ast}\Gamma_{0\nu_1\nu_2}(p,-p_1^{\prime},-p+p_1^{\prime})
\!\,^{\ast}\tilde{\cal D}^{\nu_1\nu_1^{\prime}}
(p_1^{\prime})v_{{\nu}_1^{\prime}}
\!\,^{\ast}\tilde{\cal D}^{\nu_2\nu_2^{\prime}}(p-p_1^{\prime})
\!\stackrel{\scriptscriptstyle{\,(1)}}
{\displaystyle{K}}_{{\nu}^\prime_2 0}\!({\bf v}\vert\,p-p_1^{\prime},-p_1)
\bigg\}\,
\delta(v\cdot p_1^{\prime}) dp_1^{\prime}.
\]
In deriving (\ref{eq:6t}) we use a replacement of integration variable:
$p_1^{\prime}\rightarrow p-p_1-p_1^{\prime}$ and resonance condition
$v\cdot(p-p_1)=0$. It is not difficult to check that all terms in integrand
(\ref{eq:6t}) with a gauge parameter vanish on mass-shell. Performing a
replacement (I.8.2) on the left-hand side, after anologous computations
we lead to the same expression on the right-hand side of Eq.\,(\ref{eq:6t}).
Above-mentioned examples (\ref{eq:6w}), (\ref{eq:6r}) and (\ref{eq:6t})
suggest that all terms
$\stackrel{\scriptscriptstyle{\,(r)}}{\displaystyle{K}}_{\mu\mu_1},\,r>2$
in the expansion of $K_{\mu\mu_1}^{aa_1}$ and higher
coefficient functions
$K_{\mu\mu_1\dots\mu_{2n+1}}^{aa_1\ldots a_{2n+1}},\,n>0$ cause
gauge-invariant convolution in the form (\ref{eq:6q}).

At the end of this section we would like to discuss a possible existence of very
nontrivial connections between effective subamplitudes
$\,^{\ast}\tilde{\Gamma}_{\mu\mu_1\ldots\mu_{4n+3}}$,
defined in Paper I and partial coefficient functions linear over
color charge $\stackrel{\scriptscriptstyle{\,(1)}}
{\displaystyle{K}}_{\mu\mu_1\ldots\mu_{2n+1}}$. As was mentioned at the end of
Section 3 the matrix
element for nonlinear Landau damping process was obtained in \cite{mark2}
by an alternative method distinct from the method proposed in Sections 3 and 4
of present work. In \cite{mark2} it was shown that the obtained nonlinear
Landau damping rate for longitudinal modes
\[
\gamma^l({\bf p}) = -\,2g^2N_c\!
\int\!\frac{d{\bf p}_1}{(2\pi)^3}\,N_{{\bf p}_1}^l\!
\left(\frac{{\rm Z}_l({\bf p})}{2\omega_{\bf p}^l\bar{u}^2(p)}
\frac{{\rm Z}_l({\bf p}_1)}{2\omega_{{\bf p}_1}^l\bar{u}^2(p_1)}
\right)_{\rm on-shell}\!
{\rm Im}\,^{\ast}\tilde{\Gamma}(p,p_1,-p,-p_1),
\]
where
\[
\,^{\ast}\tilde{\Gamma} (p, p_1,-p_2,-p_3) \equiv
\,^{\ast}\tilde{\Gamma}^{\mu\mu_1\mu_2\mu_3} (p, p_1,-p_2,-p_3)
\tilde{u}_{\mu}(p)\tilde{u}_{{\mu}_1}(p_1)\tilde{u}_{{\mu}_2}(p_2)
\tilde{u}_{{\mu}_3}(p_3)\vert_{\rm on-shell},
\]
and the function $\!\,^{\ast}\tilde{\Gamma}^{\mu\mu_1\mu_2\mu_3}$ is defined
from Eq.\,(I.5.6) by direct transformation of imaginary part from convolution
of $\,^{\ast}\tilde{\Gamma}(p,p_1,-p_2,-p_3)$, can be introduce also in the form
\[
\gamma^l({\bf p}) = 2\pi g^2N_c m_D^2\!
\int\!\frac{d{\bf p}_1}{(2\pi)^3}\,N_{{\bf p}_1}^l\!
\left(\frac{{\rm Z}_l({\bf p})}{2\omega_{\bf p}^l\bar{u}^2(p)}
\frac{{\rm Z}_l({\bf p}_1)}{2\omega_{{\bf p}_1}^l\bar{u}^2(p_1)}
\right)_{\rm on-shell}
\]
\[
\times (\omega_{\bf p}^l - \omega_{{\bf p}_1}^l)\!
\int\!\frac{d\Omega_{\bf v}}{4\pi}\,
\delta(\omega_{\bf p}^l - \omega_{{\bf p}_1}^l -{\bf v}\cdot({\bf p}-{\bf p}_1))
\vert\!\stackrel{\scriptscriptstyle{\,(1)}}{\displaystyle{K}}
({\bf v}\vert\,p,-p_1)\vert^2,
\]
where the function
\[
\stackrel{\scriptscriptstyle{\,(1)}}{\displaystyle{K}}
\!({\bf v}\vert\,p,-p_1)\equiv\,
\stackrel{\scriptscriptstyle{\,(1)}}{\displaystyle{K}}_{\mu\mu_1}
\!({\bf v}\vert\,p,-p_1)\tilde{u}^{\mu}(p)\tilde{u}^{\mu_1}(p_1)
\vert_{\rm on-shell}
\]
and $m_D$ is Debay screening mass. Thus an existence of two
different representations of nonlinear Landau damping rate is caused by
relation between the functions $\,^{\ast}\tilde{\Gamma}$ and
$\stackrel{\scriptscriptstyle{\,(1)}}{\displaystyle{K}}$:
\begin{equation}
{\rm Im}\,^{\ast}\tilde{\Gamma}(p,p_1,-p_2,-p_3)\vert_{p_2=p,\,p_3=p_1}
\label{eq:6y}
\end{equation}
\[
=-\,\pi m_D^2\,
(\omega_{\bf p}^l - \omega_{{\bf p}_1}^l)\!
\int\!\frac{d\Omega_{\bf v}}{4\pi}\,
\delta(\omega_{\bf p}^l - \omega_{{\bf p}_1}^l -{\bf v}\cdot({\bf p}-{\bf p}_1))
\,\vert\!
\stackrel{\scriptscriptstyle{\,(1)}}{\displaystyle{K}}
\!({\bf v}\vert\,p,-p_1)\vert^2.
\]
Based on obtained relation (\ref{eq:6y}) we can make speculation relative to
connection between functions $\,^{\ast}\tilde{\Gamma}$ and
$\stackrel{\scriptscriptstyle{\,(1)}}{\displaystyle{K}}$ of arbitrary
(even) order in external soft lines
\begin{equation}
{\rm Im}\,^{\ast}\tilde{\Gamma}(p,p_1,\dots,p_{2n+1},
-p_{2n+2},\ldots,-p_{4n+3})\vert_{p_{2n+2}=p,\,p_{2n+3}=p_1,\ldots,
p_{4n+3}=p_{2n+1}}=
\label{eq:6u}
\end{equation}
\[
-\pi m_D^2
({\cal E}_{\rm in} - {\cal E}_{\rm out})\!\!
\int\!\frac{d\Omega_{\bf v}}{4\pi}\,
\delta({\cal E}_{\rm in} - {\cal E}_{\rm out}
-{\bf v}\cdot({\bf p}_{\rm in}-{\bf p}_{\rm out}))\vert\!
\stackrel{\scriptscriptstyle{\,(1)}}{\displaystyle{K}}
\!({\bf v}\vert\,p,p_1,\ldots,p_n,
-p_{n+1},\ldots,-p_{2n+1}))\vert^2.
\]
Here,
\[
\,^{\ast}\tilde{\Gamma}(p,p_1,\dots,p_{2n+1},-p_{2n+2},\ldots,-p_{4n+3})
\]
\[
\equiv
\,^{\ast}\tilde{\Gamma}_{\mu\mu_1\ldots\mu_{4n+3}}(p, p_1,\ldots, p_{4n+3})
\tilde{u}^{\mu}(p)\tilde{u}^{{\mu}_1}(p_1)\ldots
\tilde{u}^{{\mu}_{4n+3}}(p_{4n+3})\vert_{\rm on-shell},
\]
and the function
$\stackrel{\scriptscriptstyle{\,(1)}}{\displaystyle{K}}
\!({\bf v}\vert\,p,p_1,\ldots,p_{2n+1})$ is defined by
Eq.\,(\ref{eq:6q}).

The equality (\ref{eq:6y}) is actually reflection of relation between the
imaginary part of self-energy $\Pi(\omega)$ and the interaction rate
$\Gamma(\omega)$ in the form proposed by Weldon \cite{weld}
\begin{equation}
{\rm Im}\,\Pi(\omega)=-\omega\Gamma(\omega).
\label{eq:6i}
\end{equation}
In our case we have
\[
\Pi(\omega)\!\sim\!\!\int\! d{\bf p}_1 N_{{\bf p}_1}^l
\,^{\ast}\tilde{\Gamma}(p,p_1,-p,-p_1)\dots,\,
\omega\Gamma(\omega)\!\sim\!\!\int\!
d{\bf p}_1
(\omega_{\bf p}^l - \omega_{{\bf p}_1}^l)\!
\int\!\frac{d\Omega_{\bf v}}{4\pi}\,\vert\!
\stackrel{\scriptscriptstyle{\,(1)}}{\displaystyle{K}}
\!({\bf v}\vert\,p,-p_1)\vert^2\ldots\,.
\]
We remind that formula (\ref{eq:6i}) was obtained by Weldon from analysis
of Feynman diagrams with the use of bare propagators and vertices. Therefore
nontrivial moment in relation (\ref{eq:6y}) is the fact that it involves
resummed propagators and vertices. The example of relation of such a kind can
be found in the work of Braaten and Thoma \cite{bra2}.

\section{Energy loss of energetic parton in HTL appro\-xi\-mation.
Initial equation}
\setcounter{equation}{0}

As an application of the theory developed in previous Sections we study a
problem of calculating energy loss of energetic color parton\footnote{
As energetic color parton (or simple parton) we take either energetic quark
or gluon.} traversing the hot gluon plasma, i.e. energy loss due to a
scattering process off soft boson excitations of medium in the framework
quasiclassical (HTL) approximation. As initial expression for energy loss
we will use a classical expression for parton energy loss per unit length
being a minimal extension to a color freedom degree of standard formula
for energy loss in ordinary plasma \cite{akh}
\begin{equation}
-\!\frac{dE}{dx} =
\frac{1}{\vert\check{\bf v}\vert}
\lim\limits_{\tau\rightarrow\infty}
\frac{1}{\tau}\!\int\limits_{-\tau/2}^{\tau/2}\!\int
\!d{\bf x}dt\int\!dQ\,{\rm Re}\,
\langle
\tilde{\bf J}^a_Q ({\bf x},t)\cdot
{\bf E}^a_Q ({\bf x},t)
\rangle
\label{eq:7q}
\end{equation}
\[
=\frac{1}{\vert\check{\bf v}\vert}
\lim\limits_{\tau\rightarrow\infty}
\frac{(2\pi)^4}{\tau}
\int\!d{\bf p}d\omega\!\int\!dQ\,
{\rm Re}\,
\langle
\tilde{\bf J}^{\ast a}_Q ({\bf p},\omega)
\cdot{\bf E}^a_Q ({\bf p},\omega)
\rangle.
\]
Here, $\check{\bf v}$ is the velocity of energetic
parton. Check from above is introduced to distinguish the velocity of energetic
parton (color particle which is external with respect to medium) from the
velocity ${\bf v}$ $(\vert {\bf v}\vert = 1)$ of hard thermal gluons. The sign on
the left-hand side of (\ref{eq:7q}) corresponds to the choice of a sign in
front of the current in the Yang-Mills equation (I.3.1). Chromoelectric
field ${\bf E}^a_Q ({\bf x},t)$ is one responsible for parton at the
site of its locating. In our accompanying paper \cite{mark3} the expression
(\ref{eq:7q}) will be extended also to a case of energy loss due to plasmon
radiation induced by scattering in a medium off hard thermal particles
(plasmon bremsstrahlung).

In the expression (\ref{eq:7q}) as a color current $\tilde{\bf J}^a_Q ({\bf x},t)$
one necessary takes an effective {\it dressed} current of energetic parton, that
arises as a result of a screening action of all thermal hard particles and interactions
with soft color field of plasma. As resulted from an expansion of coefficient function
(\ref{eq:4w}), this effective current represents infinite series in expansion over
color charge $Q_0^b$ of energetic parton
\[
\tilde{J}_{Q\mu}^a[A^{(0)}](p)=
\sum_{m=0}^{\infty}
\tilde{J}_{Q\mu}^{a\,bb_1\ldots b_m}
[A^{(0)}](p)
Q_0^bQ_0^{b_1}\ldots Q_0^{b_m}.
\]
Here, the term linear over color charge, i.e. the term with $m=0$ is leading in
coupling constant. Let us write once again the general expression of the effective
current $\tilde{J}_{Q\mu}^{ab}[A^{(0)}](p)$ for the convenience of further references
\begin{equation}
\tilde{J}_{Q\mu}^{ab}[A^{(0)}](p) =
\tilde{J}_{Q\mu}^{(0)ab}(p)
+ \sum_{s=1}^{\infty}
\tilde{J}_{Q\mu}^{(s)ab}[A^{(0)}](p),
\label{eq:7w}
\end{equation}
where $\tilde{J}_{Q\mu}^{(0)ab}(p)=
g/(2\pi)^3\delta^{ab}\,\check{v}_{\mu}\delta(\check{v}\cdot p),\,
\check{v}=(1,\check{\bf v})$ is initial current of energetic parton, and
\[
\tilde{J}_{Q\mu}^{(s)a}[A^{(0)}](p)=
\frac{1}{s!}\,\frac{g^{s+1}}{(2\pi)^3}\!\int\!
K_{\mu\mu_1\ldots\mu_s}^{aa_1\ldots a_s}(\check{\bf v}\vert\,p,-p_1,\ldots,-p_s)
A^{(0)a_1\mu_1}(p_1)\ldots A^{(0)a_s\mu_s}(p_s)
\]
\begin{equation}
\times\delta(\check{v}\cdot(p-\sum_{i=1}^{s}p_i))\prod_{i=1}^{s}dp_i.
\label{eq:7e}
\end{equation}
The chromoelectric field caused by the current (\ref{eq:7w})
is defined by the field equation in the temporal gauge
\[
E_Q^{ai}(p) = -i\omega\, ^\ast{\cal D}^{i\nu}(p)\tilde{J}_{Q\nu}^{ab}[A^{(0)}]
(p)Q_0^b,
\]
where the soft-gluon propagator in a given gauge reads
\begin{equation}
\,^{\ast}\tilde{\cal D}^{ij}(p) =
\left(\frac{p^2}{\omega^2}\right)
\frac{{\rm p}^i{\rm p}^j}{{\bf p}^2}
\,^{\ast}\!\Delta^{l}(p) +
\left(\delta^{ij} - \frac{{\rm p}^i{\rm p}^j}{{\bf p}^2}\right)
\!\!\,^{\ast}\!\Delta^{t}(p),
\label{eq:7r}
\end{equation}
\[
\,^{\ast}\tilde{\cal D}^{i0}(p) =
\,^{\ast}\tilde{\cal D}^{0i}(p) =
\,^{\ast}\tilde{\cal D}^{00}(p) = 0.
\hspace{3cm}
\]
Substituting this expression for field $E_Q^{ai}(p)$ in Eq.\,(\ref{eq:7q})
and integrating over color charge by using (\ref{eq:3u}) we lead to formula for
energy loss, instead of (\ref{eq:7q})
\begin{equation}
-\!\frac{dE}{dx} =
-\frac{1}{\vert\check{\bf v}\vert}\,
\frac{C_A}{d_A}
\lim\limits_{\tau\rightarrow\infty}
\frac{(2\pi)^4}{\tau}
\int\!d{\bf p}d\omega\,
{\rm Im}\,\langle
\tilde{J}^{\ast\,ab}_{Q\mu}(p)
\,^{\ast}\tilde{\cal D}^{\mu\nu}(p)
\tilde{J}^{ab}_{Q\nu}(p)
\rangle.
\label{eq:7t}
\end{equation}
Substituting the expansion (\ref{eq:7w}) into Eq.\,(\ref{eq:7t}) we will
have in general case the terms nonlinear in soft fluctuation field
$A^{(0)a}_{\mu}(p)$ (more exactly, their correlators). These nonlinear
terms define a mean change of parton energy connected with existence
of both spatially-time correlations between fluctuations of soft gauge field,
and correlations between fluctuations of a direction of a color vector of
energetic parton and fluctuation of soft field of system. The existence of
such correlations result in additional (apart from polarization
\cite{thom})
energy loss\footnote{This kind of loss sometimes is called
{\it fluctuation loss} (see, e.g. \cite{akh}).} of moving particle.

First of all we write the expression for energy loss connected with initial
current $\tilde{J}^{(0)ab}_{Q\mu}(p)$. Substituting this current into
(\ref{eq:7t}) and taking into account a structure of a propagator (\ref{eq:7r})
we obtain
\begin{equation}
\left(\!-\frac{dE^{(0)}}{dx}\right)^{\!{\rm pol}}\!
= -\frac{1}{\vert\check{\bf v}\vert}
\Biggl(\frac{N_c\alpha_s}{2{\pi}^2}\Biggr)\!
\int\! dp\,\frac{\omega}{{\bf p}^2}
\biggl\{{\rm Im}\,(p^2\!\,^{\ast}{\!\Delta}^l(p)) +
(\check{\bf v}\times {\bf p})^2
{\rm Im}\,(^{\ast}{\!\Delta}^t(p))\biggr\}
\delta(\check{v}\cdot p),
\label{eq:7y}
\end{equation}
where $\alpha_s = g^2/4\pi$. In derivation of (\ref{eq:7y}) we use a rule
(\ref{eq:5ee}). This expression defines polarization loss of energetic parton,
connected with large distance collisions. The conclusions about energy loss
in works \cite{thom} are valid, generally speaking, only for
equilibrium (nonturbulent) plasma. However for sufficiently high level of
plasma excitations (strong turbulent plasma) contributions to energy loss
connected with higher terms, $\tilde{J}^{(s)ab}_{Q\mu}, s\ge 1,$
in the expansion of effective current, become comparable with polarization loss
(\ref{eq:7y}) and therefore these contributions are necessary for
considering. This can
be seen from the level of effective current (\ref{eq:7w}). Really, let us write an
estimation of oscillation amplitude of soft field $A^{(0)a}_{\mu}(p)$ in the
form
\begin{equation}
\vert A^{(0)a}_{\mu}(p)\vert\sim
\frac{g^{\delta}}{g(gT)^3},\;\; \delta\ge 0.
\label{eq:7u}
\end{equation}
By estimations (I.6.4) and (I.6.5) a value of parameter $\delta = 0$ corresponds
to the case of highly excited state of gluon plasma, and the value $\delta =1/2$
corresponds to weakly excited state or level of thermal fluctuations at
the soft
momentum scale \cite{bla2}. From expression for currents (\ref{eq:7e}) and
estimation for coefficient functions (\ref{eq:5yy}) it is follows
\[
\tilde{J}^{(0)ab}_{Q\mu}(p)\sim\frac{1}{T},\quad
\tilde{J}^{(s)ab}_{Q\mu}[A^{(0)}]\sim g^{s\delta}
\,\frac{N_c^{s/2}}{T}.
\]
From these estimations it can be seen that for a small value of
oscillation
amplitude, i.e. for $\delta=1/2$, each subsequent term in the expansion
(\ref{eq:7w}) is suppressed by more power of $g^{1/2}$. Here we can restrict
ourselves to the first leading term $\tilde{J}_{Q\mu}^{(0)ab}(p)$ defining
polarization loss (\ref{eq:7y}), and higher order terms in expansion of the
effective current will give perturbative corrections to expression
(\ref{eq:7y}). In another limiting case of a strong field, when $\delta
\rightarrow 0$ from these estimations it follows that all terms in the expansion
(\ref{eq:7w}) become of the same order and correspondingly are comparable to
value with polarization one.

In what follows we will suppose that a value of parameter $\delta$ is
different from zero, but it is sufficiently small to consider first two terms
after leading one in the expansion of the effective current
\[
\tilde{J}_{Q\mu}^{ab}[A^{(0)}](p)\approx
\tilde{J}_{Q\mu}^{(0)ab}(p) +
\tilde{J}_{Q\mu}^{(1)ab}[A^{(0)}](p) +
\tilde{J}_{Q\mu}^{(2)ab}[A^{(0)}](p).
\]
Substituting the last expression into Eq.\,(\ref{eq:7t}) we obtain the following
after (\ref{eq:7y}) term in the expression for energy loss of energetic parton,
that we write as a sum of two different in structure (and a physical meaning)
addends
\[
\left(\!-\frac{dE^{(1)}}{dx}\right)^{\!{\rm fluct}}=
\left(\!-\frac{dE^{(1)}}{dx}\right)^{\!{\rm sp}} +
\left(\!-\frac{dE^{(1)}}{dx}\right)^{\!{\rm pol}}.
\]
Here, the first term on the right-hand side
\begin{equation}
\left(\!-\frac{dE^{(1)}}{dx}\right)^{\!{\rm sp}}\!=
-\frac{1}{\vert\check{\bf v}\vert}\,
\frac{C_A}{d_A}
\lim\limits_{\tau\rightarrow\infty}
\frac{(2\pi)^4}{\tau}
\!\int\!d{\bf p}d\omega\,
{\rm Im}\,\langle
\tilde{J}^{\ast(1)ab}_{Q\mu}(p)
\!\,^{\ast}\tilde{\cal D}^{\mu\nu}(p)
\tilde{J}^{(1)ab}_{Q\nu}(p)
\rangle
\label{eq:7i}
\end{equation}
defines energy loss due to the process of spontaneous scattering\footnote{
The notion of spontaneous and also stimulated scattering
will be considered in Section 10 in more detail.} of energetic
parton off plasma waves (i.e. plasma excitations lying on mass-shell). The 
second term
\begin{equation}
\left(\!-\frac{dE^{(1)}}{dx}\right)^{\!{\rm pol}}\!=
-\frac{1}{\vert\check{\bf v}\vert}
\,\frac{C_A}{d_A}
\lim\limits_{\tau\rightarrow\infty}
\frac{(2\pi)^4}{\tau}
\!\int\!d{\bf p}d\omega
\,{\rm Im}\biggl\{
\langle
\tilde{J}^{\ast(0)\,ab}_{Q\mu}(p)
\,^{\ast}\tilde{\cal D}^{\mu\nu}(p)
\tilde{J}^{(2)ab}_{Q\nu}(p)
\rangle
\label{eq:7o}
\end{equation}
\[
+\,
\langle
\tilde{J}^{\ast(2)\,ab}_{Q\mu}(p)
\,^{\ast}\tilde{\cal D}^{\mu\nu}(p)
\tilde{J}^{(0)ab}_{Q\nu}(p)
\rangle\biggr\},
\]
as will be shown in Section 9, is different from zero for plasma excitations
lying off mass-shell. The integrand in Eq.\,(\ref{eq:7o}) is proportional to
$\delta(\check{v}\cdot p)$, that gives ground to assign the second term to
polarization loss (\ref{eq:7y}), more exactly, to its correction due to
nonlinear effects of medium. Hereafter the expressions (\ref{eq:7i}) and
(\ref{eq:7o}) for brevity will be called diagonal and off-diagonal contributions
to the energy loss connected with diagonal and off-diagonal terms in a
product of two series (\ref{eq:7w}).

Our main attention is concerned with an analysis of expression of energy loss
due to the processes of spontaneous scattering off plasma waves, i.e. with
expression (\ref{eq:7i}), assuming thereby that this contribution to energy
loss is a main in general dynamics of energy losses in this approximation.
By using the expression (\ref{eq:7r}) of equilibrium soft-gluon propagator,
the diagonal contribution to energy loss similar to (\ref{eq:7y})
reads
\begin{equation}
\left(\!-\frac{dE^{(1)}}{dx}\right)^{\!{\rm sp}}\!=
-\frac{1}{\vert\check{\bf v}\vert}
\,\frac{C_A}{d_A}
\lim\limits_{\tau\rightarrow\infty}
\frac{(2\pi)^4}{\tau}
\!\int\!d{\bf p}d\omega
\,\frac{\omega}{{\bf p}^2}\biggl\{
\frac{p^2}{{\omega}^2}
\,\langle\vert({\bf p}\cdot
\tilde{\bf J}^{(1)ab}_Q (p))\vert^2
\rangle\,{\rm Im}(^{\ast}{\!\Delta}^l(p))
\label{eq:7p}
\end{equation}
\[
+\,
\langle\vert({\bf p}\times
\tilde{\bf J}^{(1)ab}_Q (p))\vert^2
\rangle\, {\rm Im}(^{\ast}{\!\Delta}^t(p))\biggr\}.
\]

Following by common line of this work, the contribution to energy loss caused
by spontaneous scattering off longitudinal plasma waves (plasmons) is of our
interest, i.e. on the right-hand side of Eq.\,(\ref{eq:7p}) we leave only
contribution proportional to ${\rm Im}\,(\!\,^{\ast}{\!\Delta}^l(p)).$ By
using an explicit expression for current $\tilde{J}^{(1)ab}_{Q\mu}(p)$,
Eq.\,(\ref{eq:7e}) and also a definition of a coefficient function
Eq.\,(\ref{eq:3f}), the diagonal contribution can be introduced in the form:
\[
\left(\!-\frac{dE^{(1)l}}{dx}\right)^{\!{\rm sp}}\!=
-\frac{(2\pi)^3}{\vert\check{\bf v}\vert}
\Biggl(\frac{N_c\alpha_s}{2{\pi}^2}\Biggr)^{\!2}\!
\!\int\!dpdp_1\omega
\left(\frac{p^2}{{\omega}^2{\bf p}^2}\right)\!
\left(\frac{p^2_1}{{\omega}^2_1{\bf p}^2_1}\right)
\frac{1}{{\omega}_1^2}\langle{\bf E}^2_l\rangle_{p_1}
\vert p^i\!
\stackrel{\scriptscriptstyle{\!\!\!(1)}}{\displaystyle{K}^{ii_1}}
\!(\check{\bf v}\vert\,p,-p_1)
p_1^{i_1}\vert
\]
\begin{equation}
\times
{\rm Im}\,(\!\,^{\ast}{\!\Delta}^l(p))\,\delta(\check{v}\cdot(p-p_1)),
\label{eq:7a}
\end{equation}
where
\begin{equation}
\stackrel{\scriptscriptstyle{\!\!(1)}}{\displaystyle{K}^{ii_1}}
\!(\check{\bf v}\vert\,p,-p_1) =
\frac{\check{v}^{i}\check{v}^{i_1}}{\check{v}\cdot p_1} +
\!\,^\ast\Gamma^{ii_1j}(p,-p_1,-p+p_1)
\,^\ast\!\tilde{\cal D}^{jj^{\prime}}(p-p_1)\check{v}^{j^{\prime}}.
\label{eq:7s}
\end{equation}
In deriving (\ref{eq:7a}) in the spectral density $I^{jj_1}(p_1)$ we leave
only a longitudinal part (Eq.\,(I.3.16))
\[
I^{jj_1}(p_1) \rightarrow -\,\frac{p_1^2}{\omega_1^2}\,
\frac{{\rm p}^j_1{\rm p}_1^{j_1}}{{\bf p}^2_1}\,I^l(p_1)
\]
and going to an average value of the chromoelectric field squared
\[
I^l(p_1)\rightarrow -\frac{1}{\omega_1^2}\langle{\bf E}^2_l\rangle_{p_1}.
\]
The expression (\ref{eq:7a}) is general, since it takes into account
the availability in the system of plasma fluctuations lying both on plasmon
mass-shell and off-shell. To define energy loss due to the scattering
of hard parton off plasma waves in integrand on the right-hand side of
Eq.\,(\ref{eq:7a}) it should be set
\[
{\rm Im}\,(^{\ast}{\!\Delta}^l(p))\simeq
-\pi {\rm sign}(\omega)\,\delta({\rm Re}\,^{\ast}\!\Delta^{\!-1\,l}(p))=
-\pi {\rm sign}(\omega)\,
\frac{{\rm Z}_l({\bf p})}{2\omega_{\bf p}^l}\,
[\delta(\omega-\omega_{\bf p}^l) +
\delta(\omega+\omega_{\bf p}^l)],
\]
\[
\frac{1}{{\omega}_1^2}\langle{\bf E}^2_l\rangle_{p_1} =
\,\frac{1}{(2\pi)^3}
\frac{{\rm Z}_l({\bf p}_1)}{2\omega_{{\bf p}_1}^l}\,
[N_{{\bf p}_1}^l\delta({\omega}_1-\omega_{{\bf p}_1}^l) +
N_{-{\bf p}_1}\delta(\omega+\omega_{{\bf p}_1}^l)].
\]
Substituting these expressions into (\ref{eq:7a}), integrating over
$d\omega$ and $d\omega_1$ after some algebraic transformations we lead
(\ref{eq:7a}) to the follows form
\begin{equation}
\left(\!-\frac{dE^{(1)l}}{dx}\right)^{\!{\rm sp}} =
\frac{2\pi}{\vert\check{\bf v}\vert}
\left(\frac{N_c\alpha_s}{2{\pi}^2}\right)^2\!\int\!d{\bf p}d{\bf p}_1
{\it w}_2(\check{\bf v}\vert\,{\bf p},{\bf p}_1)
\,\omega_{\bf p}^lN_{{\bf p}_1}^l\delta(\check{v}\cdot(p-p_1))
\label{eq:7d}
\end{equation}
\[
=\frac{\pi}{\vert\check{\bf v}\vert}
\left(\frac{N_c\alpha_s}{2{\pi}^2}\right)^2\!\int\!d{\bf p}d{\bf p}_1
{\it w}_2(\check{\bf v}\vert\,{\bf p},{\bf p}_1)
\Bigl\{\omega_{\bf p}^lN_{{\bf p}_1}^l +
{\omega}^l_{{\bf p}_1}N_{{\bf p}}^l\Bigr\}
\delta(\check{v}\cdot(p-p_1)).
\]
Here, a scattering probability
${\it w}_2(\check{\bf v}\vert\,{\bf p},{\bf p}_1)$ is defined by
Eqs.\,(\ref{eq:3h}), (\ref{eq:3j}),
where the factor $g^4N_c$ enters to one on the right-hand side of
Eq.\,(\ref{eq:7d}), and a replacement ${\bf v}\rightarrow\check{\bf v}$ is made.
In deriving (\ref{eq:7d}) we drop a
term with probability 
${\it w}_2(\check{\bf v}\vert\,{\bf p},-{\bf p}_1)$, 
that defines the process of simultaneous emission (absorption) of
two plasmons by energetic parton for the reason discussed in Section 2.
The last line of equation (\ref{eq:7d}) is a consequence of a property of
probability symmetry (\ref{eq:3h}) over permutations of external soft
momenta
\begin{equation}
{\it w}_2(\check{\bf v}\vert\,{\bf p},{\bf
p}_1)= {\it w}_2(\check{\bf v}\vert\,{\bf p}_1,{\bf p}).
\label{eq:7f}
\end{equation}

Now we turn to analysis of the scattering probability 
${\it w}_2(\check{\bf v}\vert\,{\bf p},-{\bf p}_1)$. 
This expression can be considerably simplify if it will be used a fact that on 
plasmon mass-shell the partial coefficient function
$\stackrel{\scriptscriptstyle{\,(1)}}{\displaystyle{K}}
\!(\check{\bf v}\vert\,p,-p_1)$ satisfies an equality (\ref{eq:6w}). Taking into
account this property and a structure of propagator (\ref{eq:7r}),
a scattering probability can be written in the form more
convenient for subsequent analysis
\begin{equation} {\it
w}_2(\check{\bf v}\vert\,{\bf p},{\bf p}_1) = \left(\frac{{\rm Z}_l({\bf
p})}{2\omega_{\bf p}^l}\right)\!  \left(\frac{{\rm Z}_l({\bf
p}_1)}{2\omega_{{\bf p}_1}^l}\right)\!  \frac{p^2p_1^2}{{\bf p}^2{\bf
p}^2_1} \left\vert\, {\rm T}^C(\check{\bf v}\vert\,{\bf p},{\bf p}_1) +
{\rm T}^{\parallel}(\check{\bf v}\vert\,{\bf p},{\bf p}_1) +
{\rm T}^{\perp}(\check{\bf v}\vert\,{\bf p},{\bf p}_1)\right\vert^2,
\label{eq:7g}
\end{equation}
where we extracted all kinematical factors from the scattering amplitudes
\[
{\rm T}^C(\check{\bf v}\vert\,{\bf p},{\bf p}_1) =
\frac{1}{\check{v}\cdot p_1},\quad
{\rm T}^{\parallel}(\check{\bf v}\vert\,{\bf p},{\bf p}_1)=
\!\,^{\ast}\Gamma^{00j}(p,-p_1,-q)\,
\frac{q^2}{q_0^2}\biggl(\frac{{\rm q}^j{\rm q}^{j^{\prime}}}{{\bf q}^2}
\biggr)\check{v}^{j^{\prime}}
\!\,^{\ast}{\!\Delta}^{\!l}(q),
\]
\begin{equation}
{\rm T}^{\perp}(\check{\bf v}\vert{\bf p},{\bf p}_1) =
\delta\Gamma^{00j}(p,-p_1,-q)
\biggl(\delta^{jj^{\prime}}-\frac{{\rm q}^j{\rm q}^{j^{\prime}}}{{\bf q}^2}
\biggr)
\check{v}^{j^{\prime}}
\!\,^{\ast}{\!\Delta}^{\!t}(q) .
\label{eq:7h}
\end{equation}
Here,
$q\equiv p-p_1=(\omega_{\bf p}^l-\omega_{{\bf p}_1}^l, {\bf p}-{\bf p}_1)$
is the soft energy and momentum transfers. It is convenient to interpret
the terms ${\rm T}^C,\,{\rm T}^{\parallel}$ and ${\rm T}^{\perp}$ by
using a quantum language. The term ${\rm T}^C$ is connected with the
Compton scattering of the soft modes (plasmons) by energetic parton.
${\rm T}^{\parallel}$ defines the scattering of a quantum oscillation
through a longitudinal virtual wave with propagator
$\!\,^{\ast}{\!\Delta}^{\!l}(q)$, where a vertex of a three-wave
interaction is induced by three-gluon HTL-amplitude
$\!\,^{\ast}{\Gamma}^{(3)}$. ${\rm T}^{\perp}$ defines the scattering of a 
quantum oscillation off energetic parton through a transverse virtual wave with 
propagator $\!\,^{\ast}{\!\Delta}^{\!t}(q)$. Here, a vertex of a three-wave 
interaction is induced by HTL-correction $\delta{\Gamma}^{(3)}$ only, since the
contribution of the bare three-gluon vertex drops out.

In the subsequent discussion it can be used a general phylosophy,
proposed by Braaten and Thoma in \cite{bra2} applying to this case.
Introduce an arbitrary momentum scale $\vert{\bf q}^{\ast}\vert$ to
separate the region of soft\footnote{By the fact that we restrict our
consideration to study energy loss defined by scattering on longitudinal
plasma waves, consideration of hard momentum transfer makes no sense.  At the
hard momentum scale plasmon mode is overdamped \cite{leb}.} momentum
transfer $\vert{\bf q}\vert\sim gT$ from ultrasoft one $\vert
{\bf q}\vert \sim g^2T$.  It should be chosen so that $g^2T\ll\vert
{\bf q}^{\ast}\vert\ll gT$, $\vert{\bf q}^{\ast}\vert \sim
g^2T\ln(1/g)$, that is possible in the weak-coupling limit for
$g\rightarrow 0$. The general analysis of contribution to energy loss of
three scattering amplitudes each  on the right-hand side (\ref{eq:7g})
shows that a basic contribution is determined by the last term with its
part, that contains a transverse part of gluon propagator
$\!\,^{\ast}{\!\Delta}^{\!t}(q)$ in a region of momentum transfer
$|{\bf q}|\leq |{\bf q}^{\ast} |.$ This fact caused by two reasons. The first
one is connected with existence of the infrared singularity in integrand
(\ref{eq:7d}), that generated by absence of screening in scalar
propagator $\,^{\ast}{\!\Delta}^t(q)$ for $q_0\ll |{\bf q}|\leq |{\bf q}^{\ast}|$.  
Entering the magnetic screening ``mass''
$\mu$ in the transverse scalar propagator we have logarithmic enhancement
in comparison with other contributions both in a region $|{\bf q}|\leq
|{\bf q}^{\ast}|$ and in region $|{\bf q}|\geq |{\bf q}^{\ast}|$. The
second reason is associated with existence of some effective angle
singularity in a medium induced vertex function
$\delta\Gamma^{00j}(p,-p_1,-q)$ with its convolution with a transverse
projector $(\delta^{jj^{\prime}}-{\rm q}^j{\rm q}^{j^{\prime}}/{\bf q}^2)$.  
These two facts will be considered in detail in the next
Section.

\section{Energy loss caused by scattering off colorless plasmons}
\setcounter{equation}{0}

In a region of small momentum transfer $\vert {\bf q}\vert\leq
\vert {\bf q}^{\ast}\vert$ the approximation
$\omega_{\bf p}^l=\omega_{{\bf p}_1+{\bf q}}^l\simeq
\omega_{{\bf p}_1}^l + {\bf q}\cdot {\bf v}_{{\bf p}_1}^l$ holds with
${\bf v}_{{\bf p}_1}^l = \partial \omega_{{\bf p}_1}^l
/\partial {\bf p}_1$, and such the resonance condition reads
\[
\omega_{\bf p}^l-\omega_{{\bf p}_1}^l -
\check{\bf v}\cdot({\bf p}-{\bf p}_1) \simeq {\bf u}\cdot {\bf q} = 0,
\]
where ${\bf u} \equiv \check{\bf v} - {\bf v}_{{\bf p}_1}^l$ is relative
velocity. At this point, it is convenient to change in Eq.\,(\ref{eq:7d})
the integration variables from ${\bf p},\,{\bf p}_1$ to ${\bf p}_1$ and
${\bf q}$. In kinematical factors on the right-hand side of (\ref{eq:7g})
we can set ${\bf p}\simeq{\bf p}_1$ and by this means instead of (\ref{eq:7d})
we obtain
\begin{equation}
\left(\!-\frac{dE^{(1)l}}{dx}\right)^{\!{\rm sp}}\! \simeq
\frac{2\pi}{\vert\check{\bf v}\vert}
\left(\frac{N_c\alpha_s}{2{\pi}^2}\right)^2\!
\!\int\!d{\bf p}_1d{\bf q}\,
\omega_{{\bf p}_1}^lN_{{\bf p}_1}^l\!
\left(\frac{{\rm Z}_l({\bf p}_1)}{2\omega_{{\bf p}_1}^l}\right)^{\!2}\!
\left(\frac{p_1^2}{{\bf p}^2_1}\right)^{\!2}\!
\left\vert{\rm T}^{\perp}(\check{\bf v}\vert\,{\bf p},{\bf p}_1)\right\vert^2
\delta({\bf u}\cdot {\bf q}).
\label{eq:8q}
\end{equation}
Here, on the right-hand side, by reason outlined in closing previous
Section we leave only the scattering amplitude
${\rm T}^{\perp}(\check{\bf v}\vert\,{\bf p},{\bf p}_1)$
(Eq.\,(\ref{eq:7h})). As it will be shown in subsequent discussion this
amplitude contains no singularities for $\vert {\bf u}\vert =0$ and
$\vert {\bf q}\vert =0$, and therefore $\delta$-function in integrand
(\ref{eq:8q}) can be written in the form
\[
\delta({\bf u}\cdot {\bf q})
=\frac{1}{\vert {\bf u}\vert \vert {\bf q}\vert}\,
\delta({\cos\alpha}),
\]
where $\alpha$ is angle between vectors ${\bf u}$ and ${\bf q}$. For
integration over momentum transfer ${\bf q}$ it is more naturally to
introduce the coordinate system in which axis 0Z is aligned with the relative
velocity ${\bf u}$ (Fig.\ref{FIG6}) for fixed momentum ${\bf p}_1$.
\begin{figure}[hbtp]
\begin{center}
\includegraphics*[height=8cm]{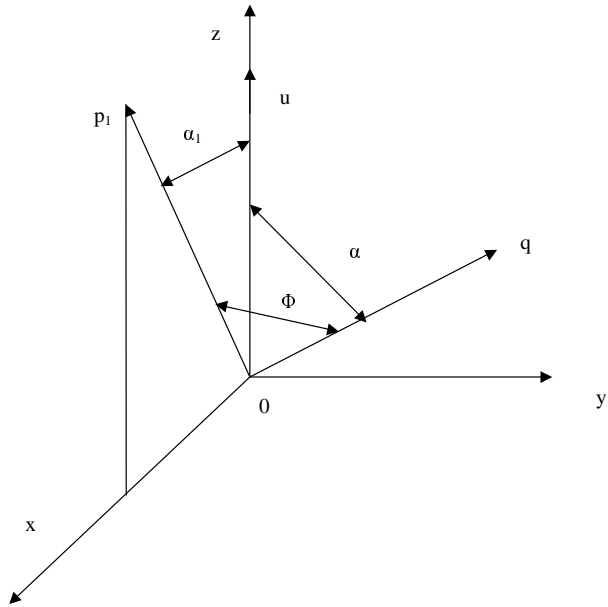}
\end{center}
\caption{\small The coordinate system in ``{\bf q}''-space under fixing
value of vector ${\bf p}_1$.}
\label{FIG6}
\end{figure}
Then the coordinates of vectors ${\bf q}$ and ${\bf p}_1$ are equal to
${\bf q}=(|{\bf q}|,\alpha,\psi),\,{\bf p}_1=(|{\bf p}_1|,{\alpha}_1,0)$.
By $\Phi$ we denote the angle between ${\bf p}_1$ and ${\bf q}$:
$({\bf p}_1\cdot{\bf q})=|{\bf p}_1||{\bf q}|\cos\Phi $.
The angle $\Phi$ can be expressed as
\begin{equation}
\cos\Phi = \sin\alpha\sin{\alpha}_1\cos\psi +
\cos\alpha\cos{\alpha}_1,
\label{eq:8w}
\end{equation}
and the integration measure is $d{\bf q}={\bf q}^2d|{\bf q}|
d\psi\sin\alpha\,d\alpha.$ The integral over polar angle $\alpha$ by virtue
of $\delta$-function determines the value $\alpha=\pi/2$ in the integrand
(\ref{eq:8q}). Futhermore for integration over momentum ${\bf p}_1$ it is
convenient going to coordinate system in which axis 0Z is aligned
with velocity $\check{\bf v}$ of energetic parton, as it was
depicted in Fig.\ref{FIG7}.
\begin{figure}[hbtp]
\begin{center}
\includegraphics*[height=7cm]{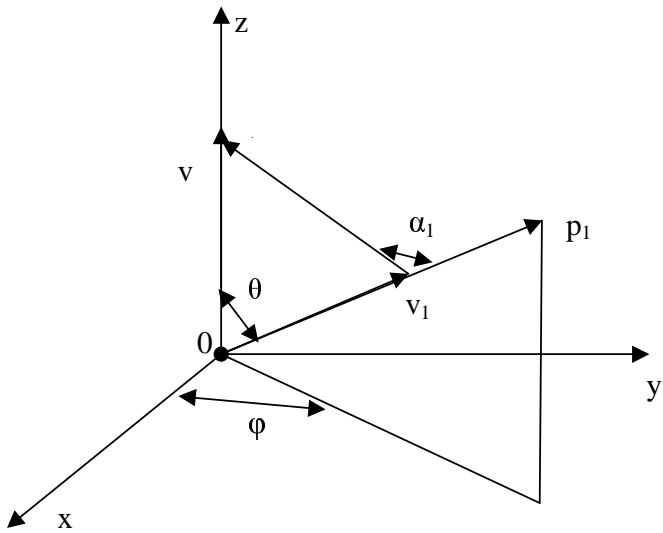}
\end{center}
\caption{\small The coordinate system in ``${\bf p}_1$''-space.}
\label{FIG7}
\end{figure}
Then the coordinate of vector ${\bf p}_1$ is equal
to ${\bf p}_1=(|{\bf p}_1|,\vartheta,\phi)$. The following relations
\[
|{\bf u}|\cos{\alpha}_1 =
|\check{\bf v}|\cos\vartheta - |{\bf v}_{\bf p}^l|,
\]
\[
|{\bf u}|\sin{\alpha}_1 =
|\check{\bf v}|\sin\vartheta
\hspace{1.4cm}
\]
are true. The integration measure with respect to ``${\bf p}_1\!$''-space is
$d{\bf p}_1={\bf p}_1^2d|{\bf q}|d\phi\sin\vartheta d\vartheta.$
Taking into account above-mentioned, the expression (\ref{eq:8q}) can be
presented in the form
\begin{equation}
\left(\!-\frac{dE^{(1)l}}{dx}\right)^{\!{\rm sp}}\!\simeq
\frac{2\pi}{\vert\check{\bf v}\vert}
\left(\frac{N_c\alpha_s}{2{\pi}^2}\right)^2\!\int\limits_{0}^{\infty}
\!d|{\bf p}_1|\, {\bf p}_1^2\,
\omega_{{\bf p}_1}^lN_{{\bf p}_1}^l\!
\left(\frac{{\rm Z}_l({\bf p}_1)}{2\omega_{{\bf p}_1}^l}\right)^{\!2}\!
\left(\frac{p_1^2}{{\bf p}^2_1}\right)^{\!2}
\label{eq:8e}
\end{equation}
\[
\times
\int\limits_0^{2\pi}\!d\phi\!\int\limits_{-1}^{1}\!d\cos\vartheta
\frac{1}{|{\bf u}|}\!\int\limits_0^{|{\bf q}^{\ast}|}\!d|{\bf q}|
|{\bf q}|\!
\int\limits_{-1}^{1}\!d\!\cos\alpha\,\delta(\cos\alpha)
\left\vert{\rm T}^{\perp}(\check{\bf v}\vert\,{\bf p},{\bf p}_1)\right\vert^2.
\]
Running ahead we note that angle singularity mentioned in closing previous
Section is connected with singularity with respect to angle variable
$\alpha_1$. This
singularity has dynamical origin, since it connected with medium induced
three-wave vertex $\delta\Gamma^{(3)}$. The consequence of this angle
singularity will be appearing selected directions in initial isotropic
hot gluon plasma. It is the directions of primary rescattering
of plasmons by hard parton, or directions over which a main energy loss
of parton arise. By virtue of this fact, the basic characteristic of interaction
process of a high energy incident parton with a hot QCD medium, in addition
to integral energy loss (\ref{eq:8e}), is the angular dependence of the energy loss
\[
\left(\!-\frac{dE^{(1)l}}{d\vartheta dx}\right)^{\!{\rm sp}}
\]
with respect to velocity direction $\check{\bf v}$. We turn to derivation and
analysis of this function.

Using an explicit expression for HTL-correction $\delta\Gamma^{00j}$
(Eq.\,(3.30) in work Frenkel and Taylor \cite{bra1}), the initial scattering
amplitude ${\rm T}^{\perp}$ can be presented in an analytic form
\begin{equation}
{\rm T}^{\perp}(\check{\bf v}\vert\,{\bf p},{\bf p}_1) =
\frac{({\bf v}\cdot({\bf n}\times{\bf q}))}{{\bf q}^2{\bf n}^2}\,
\biggl\{\Bigl[
({\bf p}\cdot {\bf q})\omega_1 - \omega({\bf p}_1\cdot{\bf q})\Bigr]
\,\delta\Gamma^{000}(p,-p_1,-q)
\label{eq:8r}
\end{equation}
\[
+\,{\bf q}^2\Pi^{00}(q) + \Bigl[
2{\bf n}^2 + ({\bf p}\cdot {\bf q}){\bf p}^2
- ({\bf p}_1\cdot {\bf q}){\bf p}^2_1\Bigr]\biggl\}
\!\,^{\ast}{\!\Delta}^{\!t}(q)\Bigl|_{\rm on-shell},
\]
where ${\bf n}=({\bf p}\times {\bf p}_1)$, and vertex correction
$\delta\Gamma^{000}$ is
\begin{equation}
\delta\Gamma^{000}(p,-p_1,-q)=-m_D^2
\left\{\omega_1 {\rm M}(p_1,q)-\omega {\rm M}(p,q)
+ \frac{i\pi q_0}{2\sqrt{-\Delta}}\,\theta(-\Delta)\right\},
\label{eq:8t}
\end{equation}
with
\[
{\rm M}(p,q) =
\displaystyle\frac{1}{2\sqrt{-\Delta(p,q)}}
\ln\!\left(\displaystyle\frac{p\cdot q + \sqrt{-\Delta(p,q)}}
{p\cdot q - \sqrt{-\Delta(p,q)}}\right),
\;\; \Delta(p,q)\equiv p^2q^2-(p\cdot q)^2|_{\rm on-shell} < 0,
\]
and finally gluon self-energy $\Pi^{00}$ is
\begin{equation}
\Pi^{00}(q)=m_D^2
\bigg[1 - F\bigg(\frac{q_0}{\vert {\bf q} \vert} \bigg)\bigg]\;,\;
F (x) \equiv \frac{x}{2} \bigg[ \ln \bigg \vert \frac{1 + x}{1 - x}
\bigg \vert - i \pi
\theta ( 1 - \vert x \vert )\bigg],
\label{eq:8y}
\end{equation}
where $q_0\equiv ({\bf v}\cdot{\bf q})$. In approximation of a small momentum
transfer $|{\bf q}|\leq |{\bf q}^{\ast}|$ and large phase velocity of plasmons
$\omega^l_{{\bf p}_1}/|{\bf p}_1|\gg 1$, the expressions in square braces in
Eq.\,(\ref{eq:8r}) can be set equal
\[
({\bf p}\cdot {\bf q})\omega_{{\bf p}_1}^l -
\omega_{\bf p}^l({\bf p}_1\cdot{\bf q})\simeq
{\bf q}^2\omega_{{\bf p}_1}^l, \qquad
2{\bf n}^2 + ({\bf p}\cdot {\bf q}){\bf p}^2
- ({\bf p}_1\cdot {\bf q}){\bf p}^2_1\simeq
3{\bf q}^2{\bf p}_1^2,
\]
and the medium-induced vertex $\delta\Gamma^{000}$ reads
\begin{equation}
\omega_{{\bf p}_1}^l\delta\Gamma^{000}(p,-p_1,-q)\simeq
m_D^2
\left[1 - \frac{q_0}{2|{\bf q}|}\ln\left\vert \frac{|{\bf q}| + q_0}
{|{\bf q}| - q_0}\right\vert - i\pi\,\frac{q_0}{2|{\bf q}|}
\right].
\label{eq:8u}
\end{equation}
As we see from Eq.\,(\ref{eq:8r}), in this approximation the exact
cancellation
of the contribution with $\delta\Gamma^{000}$ with gluon self-energy
(\ref{eq:8y}) is arisen, and therefore expression for scattering amplitude
${\rm T}^{\perp}$ has a simple form
\begin{equation}
{\rm T}^{\perp}(\check{\bf v}\vert\,{\bf p}_1,{\bf q})\simeq
3{\bf p}_1^2\,^{\ast}{\!\Delta}^{\!t}(q)\,
\frac{\check{\bf v}\cdot(({\bf q}\times{\bf p}_1)\times{\bf q})}
{({\bf q}\times{\bf p}_1)^2}.
\label{eq:8i}
\end{equation}
We approximate the transverse scalar propagator $\!\,^{\ast}{\!\Delta}^{\!t}(q)$
by its infrared limit as $q_0\ll |{\bf q}|\rightarrow 0$ over surface
$(\check{\bf v}\cdot{\bf q})=({\bf v}_{{\bf p}_1}^l\cdot{\bf q})$, namely
\begin{equation}
\!\,^{\ast}{\!\Delta}^{\!t}(q)\simeq
\frac{1}{({\bf q}^2 + {\mu}^2) - \displaystyle\frac{i\pi}{2} m_D^2
|{\bf v}_{{\bf p}_1}^l\!|\cos\Phi},
\label{eq:8o}
\end{equation}
where the angle $\Phi$ is derived by relation (\ref{eq:8w}). Here, we
introduce the magnetic ``mass'' $\mu$ to screen the infrared singularity.

After integration over angle $\alpha$ in Eq.\,(\ref{eq:8e}) the expression for
amplitude (\ref{eq:8i}) reads
\begin{equation}
{\rm T}^{\perp}(\check{\bf v}\vert\,{\bf p}_1,{\bf q})\simeq
3{\bf p}_1^2\,^{\ast}{\!\Delta}^{\!t}(q)\,
\frac{|{\bf u}|\cos{\alpha}_1}{1-{\sin}^2\alpha_1{\cos}^2\psi},
\label{eq:8p}
\end{equation}
and in the propagator (\ref{eq:8o}) it is necessary to set
$\cos\Phi=\sin{\alpha}_1\cos\psi$.  The integrand for loss (\ref{eq:8e})
represents a very complicated function of angle $\psi$. However, here,
one can use the fact that propagator (\ref{eq:8o}) represents a
``smooth'' function of angle variables with respect to the function going
from vertex part in Eq.\,(\ref{eq:8p}).  The last one has a singularity
with respect to
variable $\cos{\alpha}_1$. For this reason within accepted computing
accuracy we can factor out the function $|\!\,^{\ast}{\!\Delta}^{\!t}(q)|^2$
from the integral with respect to angle $\psi$, replacing it to average
over $\psi$ on interval $(0,2\pi)$
\begin{equation}
|\!\,^{\ast}{\!\Delta}^{\!t}(q)|^2 \rightarrow
\overline{|\!\,^{\ast}{\!\Delta}^{\!t}(q)|^2} =
\frac{1}{2\pi}\!\int\limits_0^{2\pi}\!
|\!\,^{\ast}{\!\Delta}^{\!t}(q)|^2 d\psi=
\frac{1}{({\bf q}^2 + {\mu}^2)
\sqrt{({\bf q}^2 + {\mu}^2) + \chi_{{\bf p}_1}^2{\sin}^2\alpha_1}}\,,
\label{eq:8a}
\end{equation}
where $\chi_{{\bf p}_1} \equiv (\pi/2)m_D|{\bf v}_{{\bf p}_1}^l\!|$.
The remainning integral over $d\psi$ in a general expression (\ref{eq:8e}) is
simple calculated and by virtue of (\ref{eq:8p}) equals
\[
{\cos}^2\alpha_1\!\int\limits_0^{2\pi}\!
\frac{d\psi}{(1-{\sin}^2\alpha_1{\cos}^2\psi)^2} =
\pi\,\frac{1 + {\cos}^2\alpha_1}{|\cos\alpha_1|}.
\]
We remind that angle $\alpha_1$ is connected with angle $\vartheta$ by means
of equality $|{\bf u}|\cos\alpha_1
= |\check{\bf v}|\cos\vartheta -|{\bf v}_{{\bf p}_1}^l\!|$,
and therefore the last relation has singularity in point
$|{\bf v}_{{\bf p}_1}^l\!|=
|\check{\bf v}|\cos\vartheta$. This, in particular, enables us to approximate
the relative velocity
\[
|{\bf u}|^2 = \check{\bf v}^2 + {\bf v}_{{\bf p}_1}^{l\,2} -
2|\check{\bf v}||{\bf v}_{{\bf p}_1}^l\!|\cos\vartheta\simeq
\check{\bf v}^2{\sin}^2\vartheta.
\]
We set the plasmon number density $N_{{\bf p}_1}^l$ by step function
\[
N_{{\bf p}_1}^l =
\left\{ \begin{array}{ll}
\displaystyle\frac{1}{g^{\rho}}
N_0, & |{\bf p}_1|\leq |{\bf p}_1|_{\rm max}\\
0, & |{\bf p}_1| > |{\bf p}_1|_{\rm max},
\end{array}
\right.
\]
where $N_0$ is a certain constant independing of coupling $g$,
parameter $1\leq\rho\leq 2$. The parameter $\rho $, as it have been
discussed in Section 5, determines a level of soft plasma excitations
(or value of plasmon occupation number). In deciding on value
$|{\bf p}_1|_{\rm max}$ we can guide by the next circumstance.
The leading order dispersion equation (I.2.1) defines a spectrum of longitudinal
oscillations for the entire $(\omega,|{\bf p}|)$-plane. However it was shown
(for QCD case in Refs.\,\cite{leb} and for SQED case in Ref.\,\cite{krae}),
that by virtue of specific behaviour of longitudinal oscillation spectrum
near light-cone $\omega = |{\bf p}|$, consideration next-to-leading terms
in the dispersion equation leads to important qualitative change.
There is only a finite range of $|{\bf p}|$ in which longitudinal plasmons
can exist. Therefore as $|{\bf p}_1|_{\rm max}$ one can choose a value
of plasmon momentum such that modified dispersion curve hit the light-cone.
For pure gluon plasma we have \cite{krae}
\begin{equation}
|{\bf p}_1|_{\rm max}^2 =
3\,\omega_{\rm pl}^2\left[\,
\ln\Biggl(\frac{2\sqrt{6}}{g\sqrt{N_c}}\Biggr)
+\frac{1}{2} - \gamma_E + \frac{\zeta^{\prime}(2)}{\zeta(2)}\,
\right],
\label{eq:8aa}
\end{equation}
where $\gamma_E$ is an Euler's constant and $\zeta$ is the Riemann zeta function.

The chosen approximation for plasmon number density is
very crude, and we pursue only an aim of obtaining explicit analytical
expression for energy loss. The dependence $N_{\bf p}^l$ on ${\bf p}$
is determined by solution of kinetic equation (\ref{eq:2q}), where the
right-hand side represents a sum of collision terms
(\ref{eq:2w})\,--\,(\ref{eq:2r}), (I.2.3)\,--\,(I.2.5) etc. (see, e.g.
for ordinary plasma Refs.\,\cite{tsyt}). In this case as an alternative value
of $|{\bf p}_1|_{\rm max}$, Eq.\,(\ref{eq:8aa}), one can consider
$$
|{\bf p}_1|_{\rm max}^2 \simeq
\biggl(\int\!{\bf p}_1^2 N_{{\bf p}_1}^l
\frac{d{\bf p}_1}{(2\pi)^3}\biggr)\!
\biggm/\!
\biggl(\int\!N_{{\bf p}_1}^l\frac{d{\bf p}_1}{(2\pi)^3}\biggr),
$$
i.e. $|{\bf p}_1|_{\rm max}$ will be dependent on temperature
by very nontrivial manner.

With regard to all above-mentioned approximation and
${\rm Z}^{l}({\bf p}_1)\stackrel{{\bf p}_1^2\rightarrow 0}{\longrightarrow}1$,
we find the next expression for the angular dependence of the energy loss of
a hard parton from Eq.\,(\ref{eq:8e})
\begin{equation}
\left(\!-\frac{dE^{(1)l}}{d\vartheta dx}\right)^{\!{\rm sp}} \simeq
\left(\frac{N_c\alpha_s}{2{\pi}^2}\right)^{\!2}
\frac{9\pi^3}{g^{\rho}}\,\omega_{\rm pl}^3N_0 |\check{\bf v}|\,
{\sin}^3\vartheta
\!\int\limits_{0}^{|{\bf p}_1\!|_{\rm max}}
\!\frac{d|{\bf p}_1|}{\left|
|{\bf v}_{{\bf p}_1}^l| - |\check{\bf v}|\cos\vartheta
\right|}
\!\int\limits_0^{|{\bf q}^{\ast}|}\!d|{\bf q}|\,|{\bf q}|
\,\overline{|\,\!\,^{\ast}{\!\Delta}^{\!t}(q)|^2}.
\label{eq:8s}
\end{equation}
Using expression for mean value of scalar propagator (\ref{eq:8a}), the
integral over $d|{\bf q}|$ is easily calculated and in two limiting cases
it equals
\[
\int\limits_0^{|{\bf q}^{\ast}|}\!d|{\bf q}|\,|{\bf q}|
\,\overline{|\,\!\,^{\ast}{\!\Delta}^{\!t}(q)|^2}
\simeq\frac{1}{2}\!\int\limits_{1/|{\bf q}^{\ast}|^2}^{1/{\mu}^2}\!
\frac{dz}{\sqrt{1 + z^2\chi_{{\bf p}_1}^2}} =
\left\{
\begin{array}{ll}
\displaystyle\frac{1}{2\chi_{{\bf p}_1}}
\ln\!\left(\frac{|{\bf q}^{\ast}|^2}{\mu^2}\right)
&,\, \mu^2\ll\chi_{{\bf p}_1}\\
\displaystyle\frac{1}{2\mu^2}
&,\, \mu^2\gg\chi_{{\bf p}_1}.
\end{array}
\right.
\]
By using this fact, we can approximate this integral by the following
expression
\begin{equation}
\int\limits_0^{|{\bf q}^{\ast}|}\!d|{\bf q}|\,|{\bf q}|
\,\overline{|\,\!\,^{\ast}{\!\Delta}^{\!t}(q)|^2}
\simeq
\theta(\chi_{{\bf p}_1}-\mu^2)\,\frac{1}{2\chi_{{\bf p}_1}}
\ln\!\left(\frac{|{\bf q}^{\ast}|^2}{\mu^2}\right)+
\theta(\mu^2 - \chi_{{\bf p}_1})\,\frac{1}{2\mu^2}.
\label{eq:8d}
\end{equation}
Here, on the right-hand side we leave the terms which are
more singular over
$\mu^2$. Magnetic screening mass in the approximation (\ref{eq:8d})
carries out separation of plasmons over group velocity
${\bf v}_{{\bf p}_1}^l$ on ``fast'' ($\chi_{{\bf p}_1}\geq\mu^2$)
and ``slow'' ones ($\chi_{{\bf p}_1}\leq\mu^2$).

We going in Eq.\,(\ref{eq:8s}) from integrating over $|{\bf p}_1|$ to
integrating over $|{\bf v}_{{\bf p}_1}^l\!|$, setting for small $|{\bf p}_1|$:
$d|{\bf p}_1|\simeq (5/3)\omega_{\rm pl}d|{\bf v}_{{\bf p}_1}^l\!|
\equiv (5/3)\omega_{\rm pl}dv_1$. Then with regard to (\ref{eq:8d})
we will have instead of (\ref{eq:8s})
\begin{equation}
\left(\!-\frac{dE^{(1)l}}{d\vartheta dx}\right)^{\!{\rm sp}} \simeq
\left(\frac{N_c\alpha_s}{2{\pi}^2}\right)^2
G^2|\check{\bf v}|\,{\sin}^3\vartheta
\label{eq:8f}
\end{equation}
\[
\times\left\{
\ln\!\left(\frac{|{\bf q}^{\ast}|^2}{\mu^2}\right)\!\!
\int\limits_{0}^{v_{1{\rm max}}}\!
\theta\Bigl(v_1 - \frac{2\mu^2}{\pi m_D^2}\Bigr)\,
\frac{d v_1}{v_1\left|v_1 - |\check{\bf v}|\cos\vartheta\right|}
+ \frac{\pi m_D^2}{2\mu^2}
\!\int\limits_{0}^{v_{1{\rm max}}}\!
\theta\Bigl(\frac{2\mu^2}{\pi m_D^2}-v_1\Bigr)\,
\frac{d v_1}{\left|v_1 - |\check{\bf v}|\cos\vartheta\right|}
\right\},
\]
\[
\,G^2 \equiv \frac{5\pi^2}{3}\,\frac{1}{g^{\rho}}\,m_D^2N_0,\quad
v_{1\,{\rm max}} = \frac{3|{\bf p}_1|_{\rm max}}{5\omega_{\rm pl}}.
\hspace{3cm}
\]
The integrals on the right-hand side of Eq.\,(\ref{eq:8f}) are well defined
in regions $|\check{\bf v}|\cos\vartheta < 0$ and
$|\check{\bf v}|\cos\vartheta > v_{1\, {\rm max}}$. In the region
$0<|\check{\bf v}|\cos\vartheta < v_{1\, {\rm max}}$ these integrals
contain non-integrating singularity, and here, the procedure of
their extension to finite value is required.

Now we introduce the notation for integrals entering into the right-hand side of
Eq.\,(\ref{eq:8f})
\[
V(z|\,a,b)\equiv
\int\limits_a^b\!\frac{dy}{|y - z|},\; 0\leq a<b<1\,;\quad
W(z|\,a,b)\equiv
\int\limits_a^b\!\frac{dy}{y|y - z|},\; 0<a<b<1,
\]
\[
|z|\leq 1.
\]
We extend the functions $V(z|\,a,b)$ and $W(z|\,a,b)$ based on requirement
of continuous of total loss
$(dE^{(1)l}/dx)^{\rm sp}$ as velocity function of energetic parton
in the neighborhood of value $|\check{\bf v}| = v_{1\,{\rm max}}$.
The result of extension is
\[
V(z|\,a,b)=
\theta(a-z)\,\ln\!\left(\frac{b-z}{a-z}\right) +
\theta(z-a)\theta(b-z)\,\ln\!\left(\frac{1}{(z-a)(b-z)}\right)
\]
\begin{equation}
+\,\theta(z-b)\,\ln\!\left(\frac{z-a}{z-b}\right),
\label{eq:8g}
\end{equation}
\[
W(z|\,a,b)=
\theta(a-z)\,\frac{1}{z}\,\ln\!\left(\frac{a(b-z)}{b(a-z)}\right) +
\theta(z-a)\theta(b-z)\,\frac{1}{z}\,
\ln\!\left(\frac{z-(z-a)^2(b-z)}{a(z-a)(b-z)}\right)
\]
\begin{equation}
+\,\theta(z-b)\,\frac{1}{z}\,\ln\left(\frac{b(z-a)}{a(z-b)}\right).
\label{eq:8h}
\end{equation}
These extended expressions are logarithmic diverge on the endpoint of integration
interval.

The existence of singularities over variable $|\check{\bf v}|\cos\vartheta$
leads to that the angle distribution of energy loss
$(-dE^{(1)l}/d\vartheta dx)^{\rm sp}$ has sharp peak in definite directions
with respect to vector of velocity $\check{\bf v}$, that as was mentioned
above, it can be considered as more probably directions of plasmons reradiation,
after their scattering off energetic parton. If reradiated plasmons are
sufficiently energetic, i.e. $|{\bf v}_{{\bf p}_1}^l\!|\sim v_{1\,{\rm max}}$
then here, one can tell about formation of plasmon jets along these directions,
which after hadronization can be manifested in the typical hadron distribution.

Let us calculate the contribution to
$(-dE^{(1)l}/d\vartheta dx)^{\rm sp}$ of the
region of fast plasmons
$v_{1\,{\rm max}}-\delta < |{\bf v}_{{\bf p}_1}^l\!| < v_{1\,{\rm max}}$,
where parameter $\delta$ is defined in the interval
$0<\delta<v_{1\,{\rm max}} - 2\mu^2/\pi m_D^2$. From the general expression
(\ref{eq:8f}) in this case we have
\[
\left(\!-\frac{dE^{(1)l}}{d\vartheta dx}\right)^{\!{\rm sp}}_{\!{\rm fast}}
\!\simeq
\left(\frac{N_c\alpha_s}{2{\pi}^2}\right)^2
G^2|\check{\bf v}|\,{\sin}^3\vartheta
\ln\!\left(\frac{|{\bf q}^{\ast}|^2}{\mu^2}\right)
W(|\check{\bf v}|\cos\vartheta|\,v_{1{\rm max}} - \delta, v_{1{\rm max}}),
\]
where the function $W$ is defined by Eq.\,(\ref{eq:8h}). In the Fig.\,\ref{FIG8}
the dependence of this expression from $\cos\vartheta$ for $\delta =0,2$ and
0,05; $|\check{\bf v}|=1$ and choice $v_{1\,{\rm max}}\simeq 3/5$, is given.
\begin{figure}[hbtp]
\begin{center}
\includegraphics*[height=7cm]{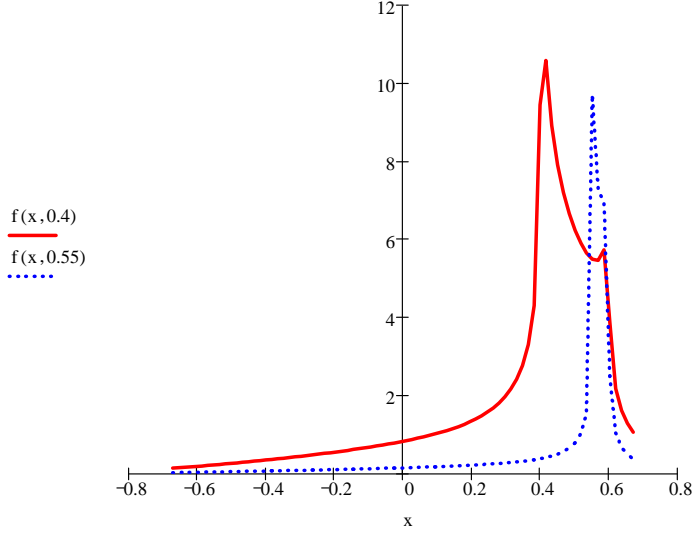}
\end{center}
\caption{\small The dependence
$\left(\!-\frac{dE^{(1)l}}{d\vartheta dx}\right)^{\!{\rm sp}}_{\!{\rm fast}}$
from $\cos\vartheta$ for $\delta =0,2$ (solid) and $\delta =0,05$ (point)
under $|\check{\bf v}|=1$.}
\label{FIG8}
\end{figure}
The typical peculiarity of this dependence is presence of sufficiently sharp
boundary over angle $\vartheta$, for which the loss become essential. This
boundary is determined by $\cos{\vartheta}_0=v_{1{\rm max}}/|\check{\bf v}|$.

The angle distribution of energy loss from whole region of rescattering
plasmons $0<|{\bf v}^l_{{\bf p}_1}\!|<|{\bf v}^l_{{\bf p}_1}\!|_{\rm max}$, is
presented in Fig.\ref{FIG9}. It is seen that main loss is connected with
\begin{figure}[h]
\begin{center}
\includegraphics*[height=7cm]{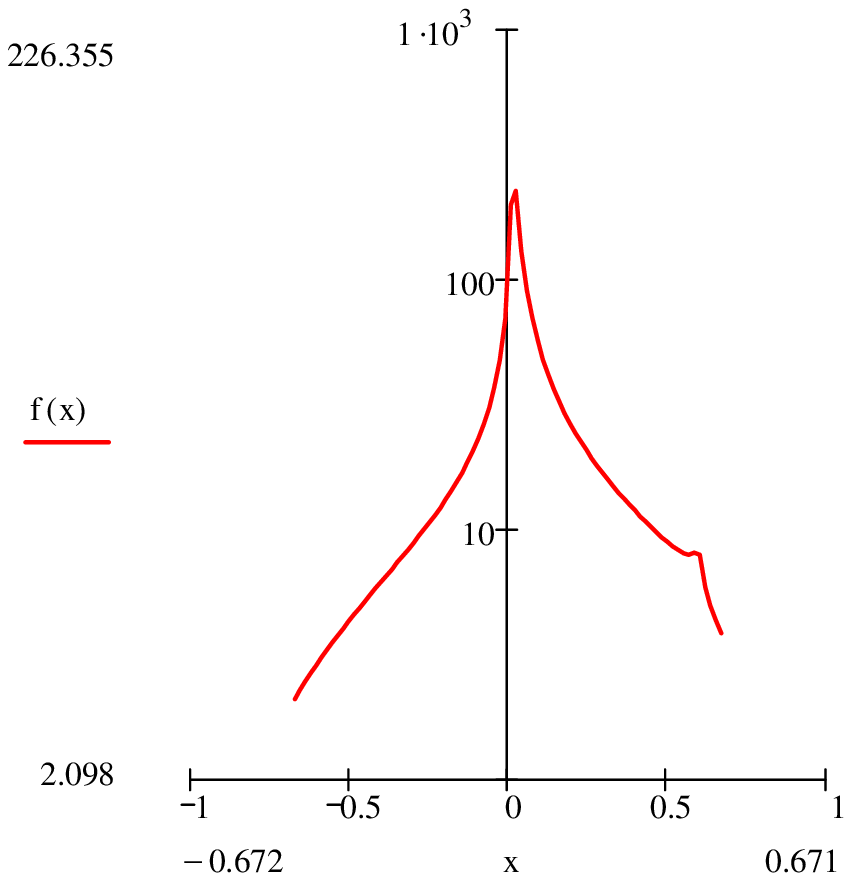}
\end{center}
\caption{\small The dependence
$\left(\!-\frac{dE^{(1)l}}{d\vartheta dx}\right)^{\!{\rm sp}}$
from $\cos\vartheta$ for whole region of rescattering
plasmons, $0<|{\bf v}^l_{{\bf p}_1}\!|<|{\bf v}^l_{{\bf p}_1}\!|_{\rm max}$
under $|\check{\bf v}|=1$.}
\label{FIG9}
\end{figure}
creation of ``slow'' plasmons $|{\bf v}^l_{{\bf p}_1}\!|\sim 0$ and lie in
region $\vartheta \sim \pi/2$.

Now we calculate the total energy loss. Integrating the expression
(\ref{eq:8f}) over angle $\vartheta$ and using (\ref{eq:8g}), (\ref{eq:8h})
we will have
\[
\left(\!-\frac{dE^{(1)l}}{dx}\right)^{\!{\rm sp}}\! =\;
\left(\!-\frac{dE^{(1)l}}{dx}\right)^{\!{\rm sp}}_{\!v_1<2\mu^2\!/\pi m_D^2}
+\,\;\;
\left(\!-\frac{dE^{(1)l}}{dx}\right)^{\!{\rm sp}}_{\!v_1>2\mu^2\!/\pi m_D^2}
\]
\begin{equation}
\equiv
\left(\frac{N_c\alpha_s}{2{\pi}^2}\right)^{\!2}\!
\frac{G^2}{|\check{\bf v}|^2}\,\frac{\pi m_D^2}{2\mu^2}\!
\int\limits_{-|\check{\bf v}|}^{|\check{\bf v}|}\!dz
\,(|\check{\bf v}|^2 - z^2)\,
V\Bigl(z\Bigm|\,0,\frac{2\mu^2}{\pi m_D^2}\Bigr)
\label{eq:8j}
\end{equation}
\[
+\,
\left(\frac{N_c\alpha_s}{2{\pi}^2}\right)^{\!2}\!
G^2\ln\!\left(\frac{|{\bf q}^{\ast}|^2}{\mu^2}\right)\!
\int\limits_{-|\check{\bf v}|}^{|\check{\bf v}|}\!dz
\,(|\check{\bf v}|^2 - z^2)
\,W\Bigl(z\Bigm|\,\frac{2\mu^2}{\pi m_D^2},\frac{3}{5}\Bigr),
\]
where in the last line we set $v_{1\,{\rm max}}=3/5$. Using an explicit
expression for function $V$ (Eq.\,(\ref{eq:8g})), we derive in limit
$\mu^2\rightarrow 0$ an asymptotic expression for contribution from
``slow'' plasmons
\begin{equation}
\left(\!-\frac{dE^{(1)l}}{dx}\right)^{\!{\rm sp}}_{\!v_1<2\mu^2\!/\pi m_D^2}
\simeq\;
\left(\frac{N_c\alpha_s}{2{\pi}^2}\right)^{\!2}\!
4G^2\ln\!\left(\frac{\pi m_D^2}{2\mu^2}\right)\!+
{\rm const} + O(\mu^2) +\ldots\,.
\label{eq:8k}
\end{equation}
Similar calculations result in asymptotic expression for the second
contribution on the right-hand side of (\ref{eq:8j})
\[
\left(\!-\frac{dE^{(1)l}}{dx}\right)^{\!{\rm sp}}_{\!v_1>2\mu^2\!/\pi m_D^2}
\simeq\;
\left(\frac{N_c\alpha_s}{2{\pi}^2}\right)^{\!2}\!
G^2\ln\!\left(\frac{|{\bf q}^{\ast}|^2}{\mu^2}\right)
\left\{\frac{3}{2}\,
{\ln}^2\!\left(\frac{\pi m_D^2}{2\mu^2}\right) +
\ln\!|\check{\bf v}|
\ln\!\left(\frac{\pi m_D^2}{2\mu^2}\right) \right.
\]
\begin{equation}
\left.
+\,\ln\left(\frac{|\check{\bf v}|}{v_{1{\rm max}}}\right)
\ln\!\left(\frac{v_{1{\rm max}}\pi m_D^2}{2\mu^2}\right)
\theta(|\check{\bf v}| - v_{1{\rm max}}) +
{\rm const} + O(\mu^2) + \ldots\,\right\}.
\label{eq:8l}
\end{equation}
We see from Eq.\,(\ref{eq:8k}) and (\ref{eq:8l}) that a main contribution
to energy loss in the limit $\mu^2\rightarrow 0$ is connected with a first
term in braces on the right-hand side of Eq.\,(\ref{eq:8l}). Thus leading
contribution to parton energy loss, caused by spontaneous scattering off
plasmons in hot gluon plasma is defined by expression
\begin{equation}
\left(\!-\frac{dE^{(1)l}}{dx}\right)^{\!{\rm sp}}
\simeq
\frac{5N_0N_c^3}{24\pi^2}\,
g^{2-\rho}\alpha_s^2T^2
\ln\!\left(\frac{|{\bf q}^{\ast}|^2}{\mu^2}\right)
{\ln}^2\!\left(\frac{\pi m_D^2}{2\mu^2}\right).
\label{eq:8z}
\end{equation}
For highly excited plasma, when $\rho\simeq 2$, suppression
in coupling constant of this process of energy loss as compared with
polarization loss \cite{thom}, is not arisen. Moreover here, we have
two logarithmic enhacements. The first of them is connected with absence
of sreening in scalar propagator $\!\,^{\ast}\!\Delta^{\!t}(q)$, that
gives a factor $\ln(|{\bf q}^{\ast}|^2\!/\mu^2)$. The second one is
associated with singularity of integrand of Eq.\,(\ref{eq:8e}) with
respect to angle variable $\alpha_1$ in point $\alpha_1=\pi/2$. This
singularity sets a singular curve $v_1 - |\check{\bf v}|\cos\vartheta =
0$ in variables space $(\vartheta,v_1)$, whose consequence is appearing
of logarithmic singularities with respect to angle $\vartheta$ in points
$\vartheta\sim\pi/2$ and $\vartheta\sim\vartheta_0=\arccos\,(|\check{\bf
v}|/v_{1{\rm max}})$.  They lead to a factor ${\ln}^2(\pi m_D^2/2\mu^2)$
in expession for loss (\ref{eq:8z}). It can be assumed that this angle 
singularity is associated with the massless basic constituents of a hot gluon 
plasma, i.e. with massless of hard transverse gluons. The inclusion
of the asymptotic thermal mass $m_{g,\,\infty}^2 = g^2T^2N_c/6$ for hard thermal
gluons and use of the improved three-gluon HTL amplitude 
$\!\,^{\ast}\Gamma_{\mu\mu_1\mu_2}(p,-p_1,-p_2)$ \cite{flech} 
can removes this type of singularity. This hypothesis requires separate and 
more careful consideration.

\section{Off-diagonal contribution to energy loss}
\setcounter{equation}{0}

In this Section we briefly analyze on a qualitative level a role of
off-diagonal term (\ref{eq:7o}) in the process of energy loss of energetic parton.
Gailitis and Tsytovich \cite{gai} have pointed to the necessity of
considering this contribution within usual plasma, and here,
we extend the conclusions of this work to non-Abelian case.

Using the explicit expression for effective
current $\tilde{J}_{Q\mu}^{(2)ab}(p)$ (\ref{eq:7e}) and leaving only
longitudinal parts in the propagator and correlation function we lead to the
expression (instead of (\ref{eq:7o}))
\begin{equation}
\left(\!-\frac{dE^{(1)l}}{dx}\right)^{\!{\rm pol}}\! =
\frac{(2\pi)^3}{\vert\check{\bf v}\vert}
\Biggl(\frac{N_c\alpha_s}{2{\pi}^2}\Biggr)^{\!2}\!
\frac{1}{d_AN_c}
\!\int\!dpdp_1\,{\omega}^2
\left(\frac{p^2}{{\omega}^2{\bf p}^2}\right)\!
\left(\frac{p^2_1}{{\omega}^2_1{\bf p}^2_1}\right)
\frac{1}{{\omega}_1^2}\langle{\bf E}^2_l{\rangle}_{p_1}
\label{eq:9q}
\end{equation}
\[
\times\,
{\rm Im}\,(\!\,^{\ast}{\Delta}^l(p))\,
{\rm Re}\,\Bigl(K^{abba\,ii_1i_1^{\prime}}(\check{\bf v}\vert\,p,-p_1,p_1)
p^ip_1^{i_1}p_1^{i_1^{\prime}}\Bigr)
\,\delta(\check{v}\cdot p).
\]

The explicit form of the coefficient function
$K_{\mu \mu_1 \mu_2}^{aa_1a_2}(\check{\bf v}\vert\,p,-p_1,-p_2)$
in linear approximation in a color charge $Q_0^b$ of parton
is defined by expression (\ref{eq:4i}). With the help of (\ref{eq:4i})
we obtain the required expression, where
\begin{equation}
K^{abba\,ii_1i_1^{\prime}}({\check{\bf v}\vert\,p,-p_1,p_1)=
{\rm Tr}\,(T^aT^a)\Bigl\{
\stackrel{\scriptscriptstyle{\!\!\!\!\!\!\!\!\!\!(1)}}
{\displaystyle{K}^{ii_1i_1^{\prime}}}
\!(\check{\bf v}\vert\,p,-p_1,p_1) +
\stackrel{\scriptscriptstyle{\!\!\!\!\!\!\!\!\!\!(1)}}
{\displaystyle{K}^{ii_1^{\prime}i_1}}}
\!(\check{\bf v}\vert\,p,p_1,-p_1)\Bigr\} .
\label{eq:9w}
\end{equation}
The special feature of the last expression is the condition that in spite of the fact
that every term in braces on the right-hand side of
equation (\ref{eq:9w}) contains the singularities of the
$1/(\check{v}\cdot p),\,^{\ast}\!\tilde{\cal D}_{\mu \mu^{\prime}}(0)$
etc. types, however in a sum they are reduced.  Therefore the expression
(\ref{eq:9w}) is well-definite and equals
\[
K^{abba\,ii_1i_1^{\prime}}({\check{\bf v}}\vert\,p,-p_1,p_1)=
N_cd_A\biggl\{
\!\,^{\ast}\tilde{\Gamma}^{ii_1ji_1^{\prime}}(p,p_1,-p,-p_1)
\,^{\ast}\tilde{\cal D}^{jj^{\prime}}\!(p)\check{v}^{j^{\prime}}
\]
\begin{equation}
-\,\check{v}^{i_1^{\prime}}
\biggl(\,\frac{\check{v}^{i}\check{v}^{i_1}}{(\check{v}\cdot p_1)^2} +
\frac{1}{(\check{v}\cdot p_1)}
\!\,^\ast\Gamma^{ii_1j}(p,-p_1,-p+p_1)
\,^\ast\tilde{\cal D}^{jj^{\prime}}\!(p-p_1)\check{v}^{j^{\prime}}
\label{eq:9e}
\end{equation}
\[
-\,
\frac{1}{(\check{v}\cdot p_1)}
\!\,^\ast\Gamma^{ii_1j}(p,p_1,-p-p_1)
\,^\ast\tilde{\cal D}^{jj^{\prime}}\!(p+p_1)\check{v}^{j^{\prime}}\biggr)
\biggr\}.
\]
Futhermore we add off-diagonal contribution (\ref{eq:9q}) with expression
for polarization loss (\ref{eq:7y}). After some regrouping of the terms this
sum reads
\[
\left(\!-\frac{dE^{(0)l}}{dx}\right)^{\!{\rm pol}} +
\left(\!-\frac{dE^{(1)l}}{dx}\right)^{\!{\rm pol}} =
\Lambda_1 + \Lambda_2.
\]
Here, on the right-hand side the function $\Lambda_1$ is
\begin{equation}
{\Lambda}_1 =
-\frac{1}{\vert\check{\bf v}\vert}
\Biggl(\frac{N_c\alpha_s}{2{\pi}^2}\Biggr)\!
\int\! dp\,\Biggl(\frac{\omega\,p^2}{{\bf p}^2}\Biggr)
{\rm Im}\,(\!\,^{\ast}{\!\Delta}^l(p))
\biggl\{1 + {\rm Re}\,(\Pi^{(1)l}(p)\!\,^{\ast}\!{\Delta}^l(p))\biggr\}
\delta(\check{v}\cdot p),
\label{eq:9r}
\end{equation}
where
\begin{equation}
\Pi^{(1)l}(p)\equiv -(2\pi)^3\!
\Biggl(\frac{N_c\alpha_s}{2{\pi}^2}\Biggr)\!
\left(\frac{p^2}{{\omega}^2{\bf p}^2}\right)\!
\int\!dp_1
\!\left(\frac{p^2_1}{{\omega}^2_1{\bf p}^2_1}\right)
\frac{1}{{\omega}_1^2}\langle{\bf E}^2_l\rangle_{p_1}
\label{eq:9t}
\end{equation}
\[
\times
\Bigl[\!\,^{\ast}\tilde{\Gamma}^{ii_1jj_1}(p,p_1,-p,-p_1)
p^ip_1^{i_1}p^jp_1^{j_1}\Bigr],
\]
and the function ${\Lambda}_2$ has the form
\begin{equation}
{\Lambda}_2\equiv
-\frac{(2\pi)^3}{\vert\check{\bf v}\vert}
\Biggl(\frac{N_c\alpha_s}{2{\pi}^2}\Biggr)^{\!2}\!
\!\int\!dpdp_1\,{\omega}^2\!
\left(\frac{p^2}{{\omega}^2{\bf p}^2}\right)\!
\left(\frac{p^2_1}{{\omega}^2_1{\bf p}^2_1}\right)
\frac{1}{{\omega}_1^2}\langle{\bf E}^2_l\rangle_{p_1}
(\check{\bf v}\cdot{\bf p}_1)
\label{eq:9y}
\end{equation}
\[
\times\,
{\rm Im}\,(\!\,^{\ast}{\!\Delta}^l(p))\,
{\rm Re}\,\Biggl\{
\frac{\check{v}^{i}\check{v}^{i_1}}{(\check{v}\cdot p_1)^2} +
\frac{1}{(\check{v}\cdot p_1)}
\,^\ast\Gamma^{ii_1j}(p,-p_1,-p+p_1)
\,^\ast\tilde{\cal D}^{jj^{\prime}}\!(p-p_1)\check{v}^{j^{\prime}}
\]
\[
-\,
\frac{1}{(\check{v}\cdot p_1)}
\,^\ast\Gamma^{ii_1j}(p,p_1,-p-p_1)
\,^\ast\tilde{\cal D}^{jj^{\prime}}\,(p+p_1)\check{v}^{j^{\prime}}\biggr\}
p^ip_1^{i_1}
\delta(\check{v}\cdot p).
\]

Let us analyze the expression ${\Lambda}_1$. It can be shown that the entering
function $\Pi^{(1)l}(p)$ represents the correction to longitudinal part of
gluon self-energy in HTL-approximation (i.e. to $\Pi^l(p)$) taking into
account change
of dielectric properties of hot gluon plasma by the action of the processes of
nonlinear interaction (longitudinal) plasma excitations among themselves.
For low excited state of plasma ($\delta = 1/2$ for estimation (\ref{eq:7u})),
corresponding to level of thermal fluctuation, correction $\Pi^{(1)l} (p)$
is suppressed by more power of $g$ in comparison with $\Pi^l (p)$. In the
limiting case of a strong field $(\delta = 0)$ the function $\Pi^{(1)l} (p)$
(and also corrections of higher power in $\langle {\bf E}_l^2 \rangle_p$) is the same
order in $g$ as gluon self-energy in HTL-approximation, and therefore considering
influence of nonlinear plasma processes on the energy loss of energetic
parton becomes necessary.

One can define an effective scalar propagator $\!\,^{\ast}\!\tilde{\Delta}^l(p)$,
which takes into account additional contributions, considering nonlinear effects
of plasmon self-interactions. Here, we have
\begin{equation}
\!\,^{\ast}\!\tilde{\Delta}^{\!-1l}(p)\equiv
\!\,^{\ast}\!{\Delta}^{\!-1l}(p) -
\sum_{n=1}^{\infty}\Pi^{(n)l}(p),
\label{eq:9u}
\end{equation}
where
\[
\Pi^{(n)l}(p)\equiv -(2\pi)^3
\Biggl(\frac{N_c\alpha_s}{2{\pi}^3}\Biggr)^{\!n}\!
\left(\frac{p^2}{{\omega}^2{\bf p}^2}\right)\!\int\!
\prod_{i=1}^n dp_i
\left(\frac{p^2_i}{{\omega}^2_i{\bf p}^2_i}\right)
\frac{1}{{\omega}_i^2}\langle{\bf E}^2_l\rangle_{p_i}
\]
\[
\times\biggl[
\,^{\ast}\tilde{\Gamma}^{ii_1\ldots i_njj_1\ldots j_n}(p,p_1,\ldots,p_n,
-p_1,\ldots,-p_n)p^ip_1^{i_1}\ldots p_n^{i_n}p^jp_1^{j_1}\ldots p_n^{j_n}
\biggr].
\]
The algorithm of calculation of the effective subamplitudes
$\!\,^{\ast}\tilde{\Gamma}^{(n)}$ entering in integrand, was defined in
\cite{mark1}. For low excited state, when
$\!\,^{\ast}\!{\Delta}^{\!-1l}(p)\gg\Pi^{(1)l}(p)\gg\Pi^{(2)l}(p)\ldots$
from the equation (\ref{eq:9u}) to the first order in
$\langle{\bf E}^2_l\rangle_{p_i}$ we obtain an approximation
\[
\!\,^{\ast}\!\tilde{\Delta}^{\!l}(p) \simeq
\!\,^{\ast}\!{\Delta}^{\!l}(p) +
\!\,^{\ast}\!{\Delta}^{\!l}(p)\,
\Pi^{(1)l}(p)\,^{\ast}\!{\Delta}^{\!l}(p),
\]
whose imaginary part is equal to
\begin{equation}
{\rm Im}\,(\!\,^{\ast}\!\tilde{\Delta}^{\!l}(p))\simeq
{\rm Im}\,(\!\,^{\ast}\!{\Delta}^{\!l}(p))
\biggl\{1 + {\rm Re}\,(\Pi^{(1)l}(p)\!\,^{\ast}\!{\Delta}^l(p))\biggr\}+
{\rm Re}\,(\!\,^{\ast}\!{\Delta}^{\!l}(p))
{\rm Im}\,(\Pi^{(1)l}(p)\!\,^{\ast}\!{\Delta}^l(p)),
\label{eq:9i}
\end{equation}
correspondingly. Here, the first term on the right-hand side is precisely
coincident  with appropriate function in integrand of (\ref{eq:9r}). If the
second term is absent on the right-hand side of Eq. (\ref{eq:9i}), than the
contribution $\Lambda_1$ to the energy loss can be interpretated as polarization
loss with regard to change of dielectric property of medium induced by
plasmon self-nteraction. This can be effective present as a replacement
of the HTL resummed longitudinal propagator, entering into expression
for polarization loss (\ref{eq:7y}) by an effective one
\[
^{\ast}\!{\Delta}^{\!l}(p)\rightarrow\, ^{\ast}\tilde{\!\Delta}^{\!l}(p).
\]
However the existence of the second term on the right-hand side of Eq.
(\ref{eq:9i}) gives no way completely restricted ourselves by such simple
transition. The physical meaning of this fact is not clear.

Let us analyzed the contribution $\Lambda_2$ (Eq.\,(\ref{eq:9y})). This contribution
(as previous one) is not equil to zero only for plasma excitations
lying off
mass-shell. From the expression for basic diagonal contribution to energy loss
(\ref{eq:7a}), (\ref{eq:7s}) we see that it contains the singularities
off-plasmon shell in the forms $1/(\check{v} \cdot p_1)^2$ and $1/(\check{v}
\cdot p_1)$, when frequency and momentum of plasma excitations approach to
"Cherenkov cone"
\[
(\check{v} \cdot p_1) \rightarrow 0.
\]
The existence of these singularities causes the large rotation velocity of
color vector $Q^a$ of energetic parton in random plasma field. However these
singularities are exactly compensated by analogous ones in expressions for energy
loss $\Lambda_2$, Eq. (\ref{eq:9y}). It is not difficult to check  this by
correlating the expressions (\ref{eq:7a}) and (\ref{eq:9y}).
Thus the complete sum of all contributions to energy loss of energetic parton
for scattering off plasma excitations lying off-plasmon shell is a
finite value, but as supposed it is sufficiently small in comparison with energy loss for
scattering off plasma excitations lying on-plasmon shell to neglect it.

\section{The Fokker-Planck equation for beam of energetic partons}
\setcounter{equation}{0}

The goal of this Section is a derivation of the Fokker-Planck equation,
describing an evolution of a distribution function $f({\bf k}, x, Q)$
for a beam
of energetic color partons\footnote{We will denote the momentum of energetic
partons of beam by the same symbol ${\bf k}$ as momentum of hard
thermal particles of hot medium (unlike velocity), lest not introduce a new
notation.} due to the processes of energy loss and diffusion in momentum
space for scattering off soft longitudinal plasma excitations. Let us
supposed that particles number density is so small that one can
neglect by partons interaction in beam among themselves. One of the
consequences of derivation of this equation having relevance to the subject of
our study will be obtaining more complete expression for energy loss and
also a possible gain of energy of isolated energetic parton of beam
which follows from well-known connection of the drag coefficient
${\rm A}^{i}({\bf k})$ with expression for energy loss $dE/dx$ \cite{svet},
\cite{must1}. 

First of all we consider a more general problem for initial formulation,
which in principle gives a possibility to consider a diffusion of beam of
energetic color particles in an effective color space on an equal footing
with diffusion in momentum space. We assume that a time-space dependent
external color current $j_{\mu}^{{\rm ext}\,a} (x)$ starts acting on the
system. In this case the soft-gauge field develops an expectation value
$\langle A_{\mu}^a(x)\rangle \equiv {\cal A}_{\mu}^a(x)\neq 0$, and the number
density of the plasmons $N_{\bf p}^{l ab}$ acquires a nondiagonal color structure.
The equation describing the time-space evolution of the distribution
function $f({\bf k}, x, Q)$ in first approximation in a number of emitted
and absorbed plasmons reads
\[
\left(\frac{\partial}{\partial t}+
\check{\bf v}\cdot \frac{\partial}{\partial {\bf x}} +
gQ^a\biggl({\bf {\cal E}}^a(x) + (\check{\bf v}\times{\bf {\cal B}}^a(x))
\biggr)
\cdot \frac{\partial}{\partial{\bf k}}\right)\!f({\bf k},x,Q)
\]
\begin{equation}
+\,gf^{abc}\check{v}^{\mu}{\cal A}_{\mu}^a(x)\frac{\partial}{\partial Q^b}
(Q^c f({\bf k},x,Q))
\label{eq:10q}
\end{equation}
\[
=-\,\frac{1}{2}\!\int\!
\frac{d{\bf p}}{(2\pi)^3}\frac{d{{\bf p}_1}}{(2\pi)^3}\,
{\it w}_2^{\{a^{\prime}\},\{a\}}(\check{\bf v}({\bf k})\vert\,{\bf p},
{\bf p}_1)
\]
\[
\times
\Bigl\{ N_{\bf p}^{l\,a^{\prime}a}
(N_{{\bf p}_1}^l + 1)^{a_1a_1^{\prime}} f({\bf k},x,Q)- (N_{\bf p}^{l}
+1)^{aa^{\prime}}N_{{\bf p}_1}^{la_1^{\prime}a_1} f({\bf k} +
{\bf q},x,Q) \Bigr\}
\]
\[
-\,\frac{1}{2}\!\int\!
\frac{d{\bf p}}{(2\pi)^3}\frac{d{{\bf p}_1}}{(2\pi)^3}\,
{\it w}_2^{\{a\},\{a^{\prime}\}}(\check{\bf v}({\bf k})\vert\,
{\bf p}_1,{\bf p})
\]
\[
\times\Bigl\{ N_{{\bf p}_1}^{l\,aa^{\prime}}
(N_{\bf p}^l + 1)^{a_1^{\prime}a_1} f({\bf k},x,Q)- (N_{{\bf p}_1}^{l}
+1)^{a^{\prime}a}N_{\bf p}^{la_1a_1^{\prime}} f({\bf k} - {\bf q},x,Q)
\Bigr\}.
\]
Here, we use a multiindex notation $\{a\} = (a, a_1)$. In deriving
collision integrals on the right-hand side of Eq.\,(\ref{eq:10q}) we take into
account a property of scattering probability relative to permutation of
the external soft momenta ${\bf p}$ and ${\bf p}_1$:
\begin{equation}
{\it w}_2^{\{a^{\prime}\},\{a\}}(\check{\bf v}({\bf k})\vert\,{\bf p},
{\bf p}_1)
= {\it w}_2^{\{a\},\{a^{\prime}\}}(\check{\bf v}({\bf k}+
{\bf q})\vert\,{\bf p}_1,{\bf p}).
\label{eq:10w}
\end{equation}
This property of the scattering probability is a generalization of the property
(\ref{eq:7f}), since the last one is valid only in the framework of
HTL-approximation and for condition $\vert{\bf q}\vert\ll\vert{\bf
k}\vert$.  The equality (\ref{eq:10w}) is a consequence of detailed
balancing principle and therefore is exact. The existence of two
collision terms on the right-hand side of Eq.\,(\ref{eq:10q}) takes into
account two possible processes of exchange of the beam state of energetic
color particles.

In the general case the scattering probability in integrand on the right-hand side
of Eq. (\ref{eq:10q}) also depends on a color vector $Q=(Q^b)$. Suppose that this
dependence is analytic, i.e. the function ${\it w}_2^{\{a^{\prime}\};\{a\}}$
can be represented as an expansion in powers of $Q$ and a first term in this
expansion can be written in the form
\[
{\it w}_2^{\{a^{\prime}\},\{a\}}(\check{\bf v}({\bf k})\vert\,
{\bf p},{\bf p}_1)=
\Bigl(T^{aa_1b}(\check{\bf v}({\bf k})\vert\,{\bf p},-{\bf p}_1)Q^b\Bigr)
\Bigl(T^{\ast\,a^{\prime}a_1^{\prime}b^{\prime}}
(\check{\bf v}({\bf k})\vert\,{\bf p},-{\bf
p}_1)Q^{b^{\prime}}\Bigr)
\]
\begin{equation}
\times
2\pi\,\delta(E_{\bf k}-E_{{\bf k} + {\bf q}} + \omega_{\bf p}^l -
\omega_{{\bf p}_1}^l).
\label{eq:10e}
\end{equation}
Here, $E_{\bf k}$ is energy of hard parton and
$T^{a a_1 b}(\check{\bf v}({\bf k})\vert\,{\bf p},-{\bf p}_1)=-if^{a a_1 b}
T(\check{\bf v}({\bf k})\vert\,{\bf p}, - {\bf p}_1)$ is interaction
matrix element. The scalar function $T(\check{\bf v}({\bf k})\vert\,
{\bf p}, - {\bf p}_1)$ is reduced to the function (\ref{eq:3j}) in
HTL-approximation and small momentum transfers.

Furthermore we restricted our consideration to study of the scattering process
of a bundle hard partons off colorless part of soft-gluon excitations,
i.e. set on the right-hand side of Eq.\,(\ref{eq:10q}) $N_{\bf p}^{l a a^{\prime}} =
\delta^{a a^{\prime}} N_{\bf p}^l$. If one neglect by influence of a mean
field (setting it a small), then kinetic equation (10.1) can be written
as equation for momentum of zeroth order in color charge of distribution
function
\[
f({\bf k}, x) = \int\! dQ f({\bf k}, x, Q).
\]
With regard to
a structure of scattering probability (\ref{eq:10e}), after some
algebraic transformation and regrouping the terms, initial equation
(\ref{eq:10q}) can be introduced in the form more convenient for subsequent
study
\begin{equation}
\left(\frac{\partial}{\partial t}+
\check{\bf v}\cdot \frac{\partial}{\partial {\bf x}}\right)\!f({\bf k},x) =
\left(\frac{\partial f}{\partial t}\right)^{\!{\rm sp}}\!+
\left(\frac{\partial f}{\partial t}\right)^{\!{\rm st}}\!,
\label{eq:10r}
\end{equation}
where
\begin{equation}
\left(\frac{\partial f}{\partial t}\right)^{\!{\rm sp}}\!\equiv -\pi
\left(\frac{N_c\alpha_s}{2{\pi}^2}\right)^{\!2}\!\left(\frac{C_2}{N_c}\right)\!
\int\!d{\bf p}d{\bf p}_1\Bigl\{
{\it w}_2(\check{\bf v}({\bf k})\vert\,{\bf p},{\bf p}_1)
\Bigl(
N_{\bf p}^l f({\bf k},x) -  N_{{\bf p}_1}^l f({\bf k}+{\bf q},x)
\Bigr)
\label{eq:10t}
\end{equation}
\[
+\,{\it w}_2(\check{\bf v}({\bf k})\vert\,{\bf p}_1,{\bf p})
\Bigl(
N_{{\bf p}_1}^l f({\bf k},x) - N_{\bf p}^l f({\bf k} - {\bf q},x)
\Bigr)\Bigr\}
\delta(E_{\bf k}-E_{{\bf k} + {\bf q}} + \omega_{\bf p}^l -
\omega_{{\bf p}_1}^l).
\]
is collision term linear in plasmon number density $N_{\bf p}^l$ corresponding
to spontaneous scattering processes\footnote{Here, we follows by terminology
accepted in \cite{och}.}, and
\begin{equation}
\left(\frac{\partial f}{\partial t}\right)^{\!{\rm st}}\!
\equiv -\pi
\left(\frac{N_c\alpha_s}{2{\pi}^2}\right)^{\!2}\!
\left(\frac{C_2}{N_c}\right)\!
\int\!d{\bf p}d{\bf p}_1\Bigl\{
{\it w}_2(\check{\bf v}({\bf k})\vert\,{\bf p},{\bf p}_1)\,
(f({\bf k},x) - f({\bf k}+{\bf q},x))
\label{eq:10y}
\end{equation}
\[
+\,
{\it w}_2(\check{\bf v}({\bf k})\vert\,{\bf p}_1,{\bf p})\,
(f({\bf k},x) - f({\bf k} - {\bf q},x))
\Bigr\}
N_{\bf p}N_{{\bf p}_1}
\delta(E_{\bf k}-E_{{\bf k} + {\bf q}} + \omega_{\bf p}^l -
\omega_{{\bf p}_1}^l)
\]
is collision term quadratic in $N_{\bf p}^l$ corresponding to stimulated 
scattering process. Here, a constant $C_2 \equiv Q^a Q^a$ is the second order
Casimir.  We have $C_2 = C_A$
for beam of high energy gluons and $C_2 = C_F$ for quarks. In the scattering
probability
$w_2(\check{\bf v}({\bf k})\vert\,{\bf p},{\bf p}_1) \equiv
\vert T(\check{\bf v}({\bf k})\vert\,{\bf p},-{\bf p}_1)\vert^2$
we take out a coupling constant and isotopic factors in a coefficient before
integral, and the $\delta$-function expressing the energy conservation -- in
integrand.

Let us consider approximation, when exchange of plasmon momentum in scattering
process is much smaller than momentum of energetic parton from beam, that is
$\vert {\bf q} \vert = \vert {\bf p} - {\bf p}_1\vert\ll\vert{\bf k}
\vert$.  It enables us to expand the distribution function in powers of
$\vert {\bf q} \vert$:
\begin{equation}
f({\bf k}\pm{\bf q},x)\simeq
f({\bf k},x) \pm
\Biggl({\bf q}\cdot\frac{\partial}{\partial {\bf k}}\Biggr)
f({\bf k},x) +\frac{1}{2!}\,\Biggl({\bf q}\cdot\frac{\partial}
{\partial {\bf k}}\Biggr)^{\!2}\! f({\bf k},x).
\label{eq:10u}
\end{equation}
The expansion of scattering probability ${\it w}_2$ requires a larger accuracy.
Let us consider equality ${\it w}_2(\check{\bf v}({\bf k})\vert\,{\bf
p}_1,{\bf p})= {\it w}_2(\check{\bf v}({\bf k}-{\bf q})\vert\,{\bf
p},{\bf p}_1)$, which is a consequence of (\ref{eq:10w}). We expand the
right-hand side in a small parameter $|{\bf q}|/|{\bf k}|$:
\begin{equation}
{\it w}_2(\check{\bf v}({\bf k})\vert\,{\bf p}_1,{\bf p})=
{\it w}_2(\check{\bf v}({\bf k})\vert\,{\bf p},{\bf p}_1) -
\Biggl({\bf q}\cdot\frac{\partial}{\partial {\bf k}}\Biggr)
{\it w}_2(\check{\bf v}({\bf k})\vert\,{\bf p},{\bf p}_1) \,+\ldots\,.
\label{eq:10i}
\end{equation}
The expression ${\it w}_2(\check{\bf v}({\bf k})\vert\,{\bf p},{\bf p}_1)$ is
also required to expand in a small parameter $|{\bf q}|/|{\bf k}|$:
\begin{equation}
{\it w}_2(\check{\bf v}({\bf k})\vert\,{\bf p},{\bf p}_1) =
{\it w}_2^{(0)}(\check{\bf v}({\bf k})\vert\,{\bf p},{\bf p}_1) +
{\it w}_2^{(1)}(\check{\bf v}({\bf k})\vert\,{\bf p},{\bf p}_1) +\ldots\,.
\label{eq:10o}
\end{equation}

Substituting the last expansion in the either side of Eq.\,(\ref{eq:10i}),
we obtain
\begin{equation}
{\it w}_2^{(0)}(\check{\bf v}({\bf k})\vert\,{\bf p},{\bf p}_1) =
{\it w}_2^{(0)}(\check{\bf v}({\bf k})\vert\,{\bf p}_1,{\bf p}),
\label{eq:10p}
\end{equation}
\begin{equation}
{\it w}_2^{(1)}(\check{\bf v}({\bf k})\vert\,{\bf p},{\bf p}_1) =
{\it w}_2^{(1)}(\check{\bf v}({\bf k})\vert\,{\bf p}_1,{\bf p}) +
{\bf q}\cdot\frac{\partial
{\it w}_2^{(0)}(\check{\bf v}({\bf k})\vert\,{\bf p},{\bf p}_1)}
{\partial{\bf k}},
\label{eq:10a}
\end{equation}
etc. The principle term of expansion
${\it w}_2^{(0)}(\check{\bf v}({\bf k})\vert {\bf p}, {\bf p}_1)$
represents the scattering probability in quasiclassical approximation with
neglecting of recoil of hard parton. Therefore this function is identified
with a scattering probability in HTL-approximation (Eqs. (\ref{eq:7g}),
(\ref{eq:7h})), having the same property of symmetry relative to permutation
of soft external momenta (Eq. (\ref{eq:7f})) as (\ref{eq:10p}).

At first we consider expansion of collision term defining a process of stimulated
scattering. Substituting an expansion (\ref{eq:10u}), (\ref{eq:10o}) into
(\ref{eq:10y}) and using the properties (\ref{eq:10p}) and (\ref{eq:10a}), we
obtain that $(\partial f / \partial t)^{\rm st}$ can be introduced in the form 
of a Fokker-Planck containing the scattering probability in HTL-approximation 
only
\[
\left(\frac{\partial f}{\partial t}\right)^{\!{\rm st}}
\simeq\frac{\partial}{\partial k^i}\Bigl({\rm A}^{{\rm st},\,i}
({\bf k})f\Bigr) +\frac{{\partial}^2}{\partial k^i\partial k^j}
\Bigl({\rm B}^{{\rm st},\,ij}({\bf k})f\Bigr),
\]
where the drag and diffusion coefficients respectively are
\begin{equation}
{\rm A}^{{\rm st},\,i}({\bf k})= -\pi
\left(\frac{N_c\alpha_s}{2{\pi}^2}\right)^{\!2}\!
\left(\frac{C_2}{N_c}\right)\!
\int\!
q^i\biggl({\bf q}\cdot\frac{\partial}{\partial {\bf k}}\,
{\it w}_2^{(0)}(\check{\bf v}({\bf k})\vert\,{\bf p},{\bf p}_1)
\biggr)
N_{\bf p}N_{{\bf p}_1}\,\delta(\check{v}\cdot(p-p_1))
d{\bf p}d{{\bf p}_1},
\label{eq:10s}
\end{equation}
\[
{\rm B}^{{\rm st},\,ij}({\bf k})=
\pi\left(\frac{N_c\alpha_s}{2{\pi}^2}\right)^{\!2}\!
\left(\frac{C_2}{N_c}\right)\!
\int\!
q^iq^j
{\it w}_2^{(0)}(\check{\bf v}({\bf k})\vert\,{\bf p},{\bf p}_1)
N_{\bf p}N_{{\bf p}_1}\,
\delta(\check{v}\cdot(p-p_1))
d{\bf p}d{{\bf p}_1}.
\hspace{1.6cm}
\]
In the case of collision term, associated with spontaneous scattering
process (Eq.\,(\ref{eq:10t})) a situation is more complicated. If we perform
calculations similar to previous case then we will have
\[
\left(\frac{\partial f}{\partial t}\right)^{\!{\rm sp}}
\simeq\frac{\partial}{\partial k^i}\Bigl({\rm A}^{{\rm sp},\,i}
({\bf k})f\Bigr)
+\frac{{\partial}^2}{\partial k^i\partial k^j}
\Bigl({\rm B}^{{\rm sp},\,ij}({\bf k})f\Bigr) +
\Phi^{(1)}[f],
\]
where the drag and diffusion coefficient are
\[
{\rm A}^{{\rm sp},\,i}({\bf k})=
-\pi
\left(\frac{N_c\alpha_s}{2{\pi}^2}\right)^{\!2}\!
\left(\frac{C_2}{N_c}\right)\!
\int\!q^i{\it w}_2^{(0)}(\check{\bf v}({\bf k})\vert\,{\bf p},{\bf p}_1)
(N_{\bf p}^l-N_{{\bf p}_1}^l)\,\delta(\check{v}\cdot(p-p_1))
d{\bf p}d{{\bf p}_1}
\]
\begin{equation}
-\,\frac{\pi}{2}
\left(\frac{N_c\alpha_s}{2{\pi}^2}\right)^{\!2}\!
\left(\frac{C_2}{N_c}\right)\!
\int\!q^i
\biggl({\bf q}\cdot\frac{\partial}{\partial {\bf k}}
\,{\it w}_2^{(0)}(\check{\bf v}({\bf k})\vert\,{\bf p},{\bf p}_1)
\biggr)
(N_{\bf p}^l+N_{{\bf p}_1}^l)
\,\delta(\check{v}\cdot(p-p_1))d{\bf p}d{{\bf p}_1},
\label{eq:10d}
\end{equation}
\[
\]
\[
{\rm B}^{{\rm sp},\,ij}({\bf k})=
\frac{\pi}{2}
\left(\frac{N_c\alpha_s}{2{\pi}^2}\right)^{\!2}\!
\left(\frac{C_2}{N_c}\right)\!
\int\!q^iq^j
{\it w}_2^{(0)}(\check{\bf v}({\bf k})\vert\,{\bf p},{\bf p}_1)
(N_{\bf p}^l+N_{{\bf p}_1}^l)\,
\delta(\check{v}\cdot(p-p_1))d{\bf p}d{{\bf p}_1},
\]
and the function $\Phi^{(1)}[f]$ is defined by expression
\[
\Phi^{(1)}[f]=
\frac{\pi}{2}
\left(\frac{N_c\alpha_s}{2{\pi}^2}\right)^{\!2}\!
\left(\frac{C_2}{N_c}\right)\!
\int\Bigr\{{\it w}_2^{(1)}(\check{\bf v}({\bf k})\vert\,{\bf p},{\bf p}_1)+
{\it w}_2^{(1)}(\check{\bf v}({\bf k})\vert\,{\bf p}_1,{\bf p})\Bigr\}
(N_{\bf p}^l - N_{{\bf p}_1}^l)
\]
\[
\times\Biggl({\bf q}\cdot\frac{\partial f}{\partial {\bf k}}\Biggr)
\,\delta(\check{v}\cdot(p-p_1))d{\bf p}d{{\bf p}_1}.
\]
The existence of the last term ${\Phi}^{(1)}[f]$ containing a correction to
quasiclassical scattering probability shows that for collision term connected
with spontaneous scattering process in limit of a small momentum transfer and
accepted calculation accuracy, it is impossible to close on quasiclassical
description level. In other words, it is not sufficient to use the properties
(\ref{eq:10p}) and (\ref{eq:10a}) to remove by the next term in expansion
of the scattering probability (\ref{eq:10o}) considering in fact an effect
of quantum response of hard parton.

Nevetheless, having an explicit expression for coefficient functions
(\ref{eq:10s}), (\ref{eq:10d}) we can write the expression for full drag
coefficient
\[
{\rm A}^{i}({\bf k})=
{\rm A}^{{\rm sp},\,i}({\bf k}) + {\rm A}^{{\rm st},\,i}({\bf k})
\]
and in so doing to define the energy loss by using relation \cite{svet},
\cite{must1}
\[
\frac{dE^{l}}{dx} =
-\frac{1}{|{\bf k}|}\,({\bf k}\cdot{\bf A}({\bf k})) \equiv
-\frac{1}{|\check{\bf v}|}\,(\check{\bf v}\cdot {\bf A}({\bf k}))
\]
\[
= -\frac{\pi}{|\check{\bf v}|}
\left(\frac{N_c\alpha_s}{2{\pi}^2}\right)^{\!2}\!
\left(\frac{C_2}{N_c}\right)\!
\int\!d{\bf p}d{{\bf p}_1}
{\it w}_2^{(0)}(\check{\bf v}({\bf k})\vert\,{\bf p},{\bf p}_1)
\Bigl\{\omega_{\bf p}^lN_{{\bf p}_1}^l
+ \omega_{{\bf p}_1}^lN_{\bf p}^l\Bigr\}
\,\delta(\check{v}\cdot(p-p_1))
\]
\[
+\,\frac{\pi}{|\check{\bf v}|}
\left(\frac{N_c\alpha_s}{2{\pi}^2}\right)^{\!2}\!
\left(\frac{C_2}{N_c}\right)\!
\int\!d{\bf p}d{{\bf p}_1}
{\it w}_2^{(0)}(\check{\bf v}({\bf k})\vert\,{\bf p},{\bf p}_1)
\Bigl\{\omega_{\bf p}^lN_{\bf p}^l
+ \omega_{{\bf p}_1}^lN_{{\bf p}_1}\Bigr\}
\,\delta(\check{v}\cdot(p-p_1))
\]
\[
+\,\frac{\pi}{|\check{\bf v}|}
\left(\frac{N_c\alpha_s}{2{\pi}^2}\right)^{\!2}\!
\left(\frac{C_2}{N_c}\right)\!
\int\!d{\bf p}d{{\bf p}_1}
\biggl({\bf q}\cdot\frac{\partial}{\partial {\bf k}}
{\it w}_2^{(0)}(\check{\bf v}({\bf k})\vert\,{\bf p},{\bf p}_1)\biggr)
\]
\[
\times ({\omega}_{\bf p}^l - {\omega}_{{\bf p}_1}^l)
\biggl\{\frac{1}{2}(N_{\bf p}^l+N_{{\bf p}_1}^l)
+ N_{\bf p}^lN_{{\bf p}_1}^l\biggr\}\,
\delta(\check{v}\cdot(p-p_1)).
\]
The first term on the right-hand side exactly reproduces the expression
for energy loss (\ref{eq:7d}) used in previous Sections for $C_2 = C_A$.
The second term defines acceleration obtained by hard color particle
for Compton plasmon scattering. As it is seen from this expression in fact
it equals the product of total section of this scattering process in HTL
approximation by mean energy of longitudinal QGP excitations. It is
necessary to take into account this contribution in complete balance of
energy loss. The third term is formally suppressed in comparison with the
first factor $\vert {\bf q} \vert / \vert {\bf k}\vert \ll 1$. However it
can be expected, that for highly excited state of system, when the
plasmon number density is very large, the term in integrand proportional to
$N_{\bf p}^l N_{{\bf p}_1}^l$ (stimulated scattering process), can give
appreciable contribution to general expression for energy loss. By virtue
of the fact that this contribution is indefinit in a general case then
here, both drag and acceleration of partons from beam are possible.  In
particular it depends on concrete form of the plasmon number density
$N_{\bf p}^l$ as a function of soft momentum ${\bf p}$.

\section{\bf Conclusion}
\setcounter{equation}{0}

In conclusion we would like to dwell on some important moments which
remain to be touched upon in this work and on the whole concerned
with a mechanism of energy loss studied in previous sections.

In studies of plasmon-hard particle scattering processes and energy losses
their related, we consider that a mean-gluon field
${\cal A}_{\mu}^a (x)$ in a system equils zero or is so weak that it can
be neglected by its influence on dynamics of the processes in
QGP\footnote{The exception here, is only the beginning of Section 10. The
kinetic equation (\ref{eq:10q}) written there represents a simplest
example of that as far as a dynamics of a system for
${\cal A}_{\mu}^a (x)\neq 0$
becomes more complicated and intricated.}.  This in particular leads to the
fact that we have considered interaction of energetic color particle (or
thermal test particle) with colorless part of soft-gluon excitations
only. However, opposite case -- very strong mean gluon field
that is to be necessary account exactly in all orders
of perturbation theory, is of great interest. In this case, for example,
coefficient functions
$K_{\mu\mu_1\cdots\mu_s}^{a a_1\ldots a_s}({\bf v}\vert\,{\bf p},-{\bf p}_1,
\ldots,-{\bf p}_s)$ under the integral sign of effective currents
(\ref{eq:7e}) depend by themselves on ${\cal A}_{\mu}^a (x)$, i.e. they
can contain
insertions of an external or self-consisting mean field, in general, of arbitrary
order. The interest to such setting up a problem is specified by that it is
simpler model for research of jet quenching phenomenon in heavy ion
collisions, whereby high-$p_t$ partons get depleted through strong (classical)
color field.The latter is encountered in {\it the Color Glass Condensate} (CGC)
\cite{ian}. About ten years ago Mc.Lerran and Venugopalan \cite{mcl} suggest
that the prompt phase of heavy ion collisions is CGC, and therefore influence
of CGC on energy losses of color partons can be very essential. The example of
considering CGC in calculation of energy loss has been given in Introduction.
This is a mechanism of synchrotron-like radiation of an ultrarelativistic
charge in QCD going through a constant chromomagnetic field, proposed by
Shuryak and Zahed \cite{shur1}.

Another important moment that was not touched in this work is concerned with
consideration of contributions to energy losses of terms higher in powers of
the field in expansion of the effective current (\ref{eq:7w}), $s\geq 3$. In
principal, in limiting case of a strong random field
$\vert A_{\mu}^{(0)a}(p)\vert\sim 1/g (gT)^3$ (Eq.\,(\ref{eq:7u})) account
must be taken of the whole of series in expansion of the effective current,
since formally each term of the series becomes of the same order as a first
term $\tilde{J}_{Q \mu}^{(0)}$.
A preliminary analysis of contribution to energy loss of higher scattering
processes shows that in each subsequent `diagonal' contribution
$(-dE^{(n)}/dx)^{\rm sp},\,n\geq 2$, a term having practically the same
peculiarity as $(-dE^{(1)}/dx)^{\rm sp}$ can be separated. However, here, we come
up against the further problem not allowing directly summarized all relevant
terms. These contributions contain the pathological terms generated by product
of two $\delta$-functions of
$[\delta ({\rm Re}\,^{\ast}\!\Delta^{-1t}(p + p_1))]^2$ type etc.
This strongly remind a situation with that one runs in out of equilibrium
thermal field theory connected with so-called {\it pinch singularity}
\cite{alth}, with the distinction that argument of $\delta$-function contains
inverse resumming propagator instead of bare one. For this reason a problem of
contribution to energy loss of higher spontaneous scattering processes requires
separate and careful study.

A futher remark is concerned with existence of
soft-gluon transverse excitations in the system. Entering the transverse
oscillations in consideration result in appearing additional
channels of energy losses of energetic particle similar to ones that we
study for longitudinal oscillations. Besides the
scattering processes off pure transverse plasma waves here, it is also
possible the scattering processes resulting in change of polarization
type of soft excitations (so-called {\it a conversion processes} of plasmons to
transverse waves\footnote{For example, on Fig.\,\ref{FIG1} and
Fig.\,\ref{FIG2} incoming soft-gluon line have a longitudinal
polarization, whereas outgoing soft gluon line -- transverse ones.}) and
more complicated processes being a combination of usual scattering
process and the fusion process of two plasmons into one transverse
eigenwave, as depicted in Fig.\,\ref{FIG10}.
\begin{figure}[h]
\begin{center}
\includegraphics*[height=4.5cm]{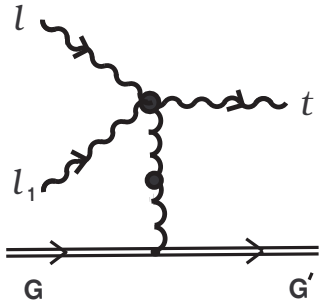}
\end{center}
\caption{\small A scattering-fusion process of two soft-gluon longitudinal
plasma excitations off energetic color particle with transition to one
soft-gluon transverse plasma excitation.}
\label{FIG10}
\end{figure}
These processes also need to be taken into account for total balance
of energy losses. However, here, a new specific effect, namely for transverse
mode is appeared. This is a possibility of large energy-momentum transfer
from ultrarelativistic particle for scattering off plasmons with
its conversion to transverse quantum of oscillations (see the last
footnote).  The energy of secondary quantum may be close to energy of
initial hard particle.  This is connected with that unlike longitudinal
excitations, transverse ones have no restriction to maximum
possible value of momentum of (\ref{eq:8aa}) type.  However when energy of
secondary quantum is close to energy of fast particle, then
quantum effects become essential.  Here, it is necessary to use quantum
expressions for scattering processes on gluons of Compton's type with
change of incoming line of vacuum gluon by plasma gluon with
longitudinal polarization, and a factor provides renormalization of a
gluon wave function by thermal effects. The scattering on longitudinal
gluons with conversion to hard transverse radiation produces additional
mechanism imposed upon gluon bremsstrahlung. The processes of such a kind
for usual (Abelian) plasma were first considered by Gailitis and
Tsytovich in Ref.\,\cite{gai1}.

Finally, a last remark that should be mentioned, is concerned with energy loss
of {\it color dipole}, next in order of complexity to color object after
single color-charge particle. Classical color dipole can be considered as
`toy' model for heavy quarkonia ($c\bar{c}$ or $b\bar{b}$) propagating through
QGP, and its energy loss may be relevant for the distortion of
heavy-quarkonium production by plasma formation. Within linear response theory
this problem was first considered by Koike and Matsui in Ref.\,\cite{thom}.
However an attempt of direct extension of the result obtained by Koike and
Matsui to the case of the scattering of color dipole off soft-gluon excitations
of QGP encounter specific difficulties. In fact we can easily define expression
for ``starting'' current generated by the color pairs. For a neutral pair, in the
approximation `point' dipole we have
\begin{equation}
\tilde{J}_{Q\bar{Q}}^{(0)a\mu}(p) =
\frac{g}{(2\pi)^3}\,\tilde{v}^{\mu}\delta(\tilde{v}\cdot p)
\,\frac{1}{2}
\biggl\{\frac{\sqrt{1-{\tilde{\bf v}}^2}}{{\tilde{\bf v}}^2} \,
({\bf p}\cdot\tilde{\bf v})({\bf d}^a\cdot\tilde{\bf v})+
\frac{1}{{\tilde{\bf v}}^2}\,
({\bf p}\times\tilde{\bf v})({\bf d}^a\times\tilde{\bf v})
\biggr\},
\label{eq:11q}
\end{equation}
where ${\bf d}^a=(Q^a_{10} -Q^a_{20}){\bf r}_0$ is the color dipole moment with
fixed separation ${\bf r}_0$ of color charges in a intrinsic frame of reference,
and $\tilde{\bf v}$ is a velocity of dipole. Substituting (\ref{eq:11q})
into expression for energy loss (\ref{eq:7q}), previously replacing
\[
\int\!dQ \rightarrow \int\!dQ_1dQ_2,\quad
E_{Q}^{ai}(p) \rightarrow
E_{Q\bar{Q}}^{ai}(p) = -i\omega\, ^\ast{\cal D}^{i\nu}(p)
\tilde{J}_{Q\bar{Q}\nu}^{\,a}[A^{(0)}](p),
\]
we derive polarization energy loss for color dipole. To take
into account scattering process for color dipole off soft plasma waves
and energy loss associated with this process, we must calculate the next
term in the expansion of effective color-dipole current
$\tilde{J}_{Q\bar{Q}}^{\,a \mu}[A^{(0)}](p)$, linear over soft fluctuation
field of system $A_{\mu}^{(0)a}$. The problem lies in the fact that not only
a field of incident wave, but also field of second charge of pair are
acted on each of color charge vectors of dipole. This circumstance very complicates and tangles
the problem, compared to single color charge (see, e.g. work of Khriplovich
\cite{khrip} on construction of a field produced by two fixed Yang-Mills
point charges), and therefore demands development of more refined method for
calculating effective color-dipole current, that is out of the scope of the
present study.

\section*{\bf Acknowledgments}
This work was supported by the Russian Foundation for Basic Research
(project no 03-02-16797) and Grant INTAS (project no 2000-15).


\end{document}